\documentclass[final,3p,times]{elsarticle}
\usepackage{amsmath}
\usepackage{lipsum}
\usepackage{braket}
\usepackage{bm}
\usepackage{cases}
\usepackage{mathtools}
\usepackage{dsfont}
\usepackage{siunitx}
\usepackage[linktocpage=true]{hyperref}
\usepackage{verbatim}
\hypersetup{
    colorlinks,
    citecolor=blue,
    filecolor=black,
    linkcolor=blue,
    urlcolor=black
}
\setcitestyle{square}
\numberwithin{equation}{section}
\newcommand{\pvec}[1]{\vec{#1}\mkern2mu\vphantom{#1}}

\renewcommand{\vec}{\bm}
\renewcommand{\a}{\alpha}
\renewcommand{\b}{\beta}
\renewcommand{\c}{\gamma}
\renewcommand{\d}{\delta}
\newcommand{\e}{\epsilon}
\newcommand{\f}{\varphi}
\renewcommand{\t}{\tau}
\newcommand{\m}{\mathcal}

\renewcommand{\epsilon}{\varepsilon}
\renewcommand{\rho}{\varrho}
\DeclarePairedDelimiter{\abs}{\lvert}{\rvert}

\def\orcid#1{\href{https://orcid.org/#1}{\!\includegraphics[keepaspectratio,width=0.7em]{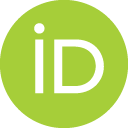}}}

\journal{Annals of Physics}

\begin{document}

\begin{frontmatter}

\title{Expansion of one-, two- and three-body matrix elements on a generic spherical basis for nuclear \emph{ab initio} calculations}

\author[1]{A. Scalesi
\orcid{0000-0002-7526-4110} }
\author[2,3]{C. Barbieri \orcid{0000-0001-8658-6927} }
\author[3]{E. Vigezzi \orcid{0000-0002-2683-6442}}
\affiliation[1]{organization={IRFU, CEA, Université Paris-Saclay},%Department and Organization
            %addressline={}, 
            city={Gif-sur-Yvette},
            postcode={91191}, 
            country={France}}
\affiliation[2]{organization={Dipartimento di Fisica "Aldo Pontremoli", Università degli studi di Milano},%Department and Organization
            addressline={Via Celoria 16}, 
            city={Milano},
            postcode={20133}, 
            %state={},
            country={Italy}}
\affiliation[3]{organization={INFN, Sezione di Milano},%Department and Organization
            addressline={Via Celoria 16}, 
            city={Milano},
            postcode={20133}, 
            %state={},
            country={Italy}}

\begin{abstract}
\emph{Ab initio} studies of atomic nuclei are based on Hamiltonians including one-, two- and three-body operators with very complicated structures.
Traditionally, matrix elements of such operators are expanded on a Harmonic Oscillator single-particle basis, which allows for a simple separation of the center-of-mass motion from the intrinsic one.
A few recent investigations have showed that the use of different single-particle bases can bring significant advantages to numerical nuclear structure computations.
In this work, the complete analytical expression of the Hamiltonian matrix elements expanded on a generic spherical basis is presented for the first time.
This will allow systematic studies aimed at the determination of optimal nuclear bases.
\end{abstract}


\end{frontmatter}

\section{Introduction}
\label{intro}

Much of our understanding of complex quantum mechanical systems is based on the capability to represent them as being composed of simpler (or more elementary) particles that are subject to specific mutual interactions. This is referred to as the \emph{quantum many-body problem}. Techniques for finding accurate predictions for such systems have found applications across several fields of Science, ranging from the modelling of crystals in solid state physics, to molecules for quantum chemistry, to cold gasses, all the way to the equation of state for neutron star matter and the unstable isotopes created during nucleosynthesis.

A vast range of different computational quantum many-body methods has been applied across all of these disciplines. Yet, almost all of these start by representing each constituent as an independent particle in a one-body Hilbert space with a complete basis. The correlations due to the inter-particle interactions are then added afterwards in order to construct the true many-body quantum states of the system. The key ingredient for a simulation to be successful is to find a proper choice of the single-particle basis and compute correctly the matrix elements of the Hamiltonian operator in such basis. For example, the realization that Gaussian states can approximate accurately the electron cuspid while also accelerating the computation of Coulomb matrix elements has enabled the whole development of Quantum Chemistry~\cite{PopleNobelLecture, pople}.

For nuclear physics things are more complicated because the nuclear interaction has a very complex operator structure, with non-local terms and non-trivial radial expressions. The only possible simplifications come from exploiting spherical symmetry but no other straightforward analytical properties of the Hamiltonian can be employed to accelerate the computation of matrix elements. 
Modern applications use harmonic oscillator (HO) wave functions as the best surrogate to represent nuclear single-particle states~\cite{Tajima04} while being able to exploit the symmetries of the HO under the transformation between the center-of-mass and laboratory frames. Different strategies can be used \emph{a posteriori} to accelerate the convergence of bulk properties of nuclei with respect to the size of the model space compared to the HO basis. For example, a Hartree-Fock (HF) computation defines a better reference state and a single-particle basis optimized in energy. 
The Self Consistent Green's Function (SCGF) method has employed the so-called Optimised Reference State (OpRS)~\cite{Barbieri2022Gkadc3} to handle non-soft interactions since~Refs.~\cite{Barbieri2009Ni56,Lecluse2017,McIlroy2018}.
More recently, natural orbitals (NAT) obtained from the correlated density matrix~\cite{NAT1,NAT2} have proven to enable a more efficient convergence than HF bases. The OpRS basis is built by reproducing the first moments of the spectral distributions, hence, it is tuned to optimize both the correlated density matrix~\cite{Barbieri2022Gkadc3} and the energy sum rules.
Yet both the NAT and OpRS have been so far based on a preliminary expansion on the HO basis and require handling nuclear matrix elements in a sufficiently large oscillator space.

HO wave functions are known to bring the wrong asymptotic behavior, which negatively affects for example the study of weakly bound nuclei.
A few past investigations in nuclear physics have focused on single-particle bases different than the HO~\cite{negoita, 
CoulombSturmian,puddu} and new classes of candidates for nuclear wave functions have been addressed by Ref.~\cite{BulgacForbes}.
Though, all these studies are limited and the corresponding proposed bases have never really been competitive with the HO basis.
It is then an important open issue to investigate more systematically whether the use of bases different from the conventional HO can lead to significant improvements in the computation of nuclear properties.

Know-how for handling matrix elements in generic bases is fragmented into several publications that appeared across past years but it mostly covers generic two-particle interactions. 
To the best of our knowledge, analytic expressions for the two- and three-body matrix elements of a realistic nuclear Hamiltonian suitable for \emph{ab initio} calculations in a generic (possibly spin- and isospin-dependent) single-particle basis are still not available. 
The present paper aims to fill such gap, setting as a starting point the matrix elements of the interaction in Jacobi coordinates \cite{MachleidtBonn, Hebeler2015} which is the usual form adopted by \emph{ab initio} practitioners. Expressions of such matrix  elements are derived in a generic spherical basis, making use of the Wong-Clement (WC) \cite{WC} coefficients in place of the usual Moshinsky brackets \cite{Moshinsky}.
Known material and new developments are included in a self-contained manuscript that incorporates all the necessary elements for computing matrix elements. Although the focus is on modern nuclear interactions, including spin-isospin degrees of freedom, the formulae reported in this work could be easily adapted to other rotationally invariant quantum many-body systems (and even specialized to simpler Hamiltonians as needed).

The structure of the manuscript is the following: Sec.~\ref{definitions} presents the main conventions and definitions used, as well as the basic ingredients necessary for getting started with the calculation of matrix elements.
Sec.~\ref{1b}, \ref{2b} and \ref{3b} focus respectively on the matrix elements relative for one-, two- and three-body operators.
The matrix elements of two- and three-body interactions are obtained as a change of basis from two- and three-body matrix elements in momentum-space.
Finally Sec.~\ref{conclusions} draws some final conclusions and future perspectives.
All the fine details regarding specific topics as the calculation of particular coefficients are left in the Appendix. 

\section{Basic ingredients for the calculation of matrix elements}
\label{definitions}
\subsection{The nuclear Hamiltonian}
The goal of \emph{ab initio} methods \cite{MBPT,FrontSCGF,CCHagen,IMSRGHer} is to solve the non-relativistic many-body Schr\"{o}dinger equation
\begin{equation}
H\ket{\Psi_n} = E_n\ket{\Psi_n},
\end{equation}
where $H$ represents the intrinsic Hamiltonian of the system
\begin{equation}
H = T + V + W + \dots
\label{eqn:Hh}
\end{equation}
and $T$, $V$ and $W$ represent respectively the intrinsic kinetic energy, the two-body potential and the three-body potential. Higher-body operators of the Hamiltonian will be ignored in this work.
The operators in Eq.~\eqref{eqn:Hh} are assumed to be invariant under Galilean boosts.

The intrinsic kinetic energy is conveniently written as the difference between the kinetic energy expressed in the laboratory frame and its center-of-mass contribution:
\begin{equation}
T = T^{lab} - T^{cm},
\label{eqn:tlabtcm}
\end{equation}
with
\begin{equation}
T^{lab} = \sum_i \dfrac{\vec{p}_i^2}{2m_N}.
\end{equation}
$T$ is a two-body operator that can be expressed as
\begin{equation}
T = \dfrac{1}{2m_N\hat{A}}\sum_{i<j}(\vec{p}_i -\vec{p}_j)^2,
\label{eqn:Tint_2b}
\end{equation}
where $\hat{A}$ represents the particle-number operator\footnote{While operators are denoted in this paper without the `hat' symbol, a different notation is used for the particle-number operator to allow a clear distinction from the number of particles $A$ of the system, consistently with \cite{CoM_roth}.} and $m_N$ represents the average mass of the nucleon.
Following \cite{CoM_roth,IntrOp}, Eq.~\eqref{eqn:Tint_2b} can be re-written by using the identity
\begin{equation}
\sum_{i<j}(\vec{p}_i-\vec{p}_j)^2 = \sum_{i<j}(\vec{p}_i^2 + \vec{p}_j^2 - 2\vec{p}_i\cdot\vec{p}_j) = (\hat{A}-1)\sum_i\vec{p}_i^2 - 2\sum_{i<j}\vec{p}_i\cdot\vec{p}_j,
\end{equation}
which leads to a alternative expression for the intrinsic kinetic energy
\begin{equation}
T = \dfrac{\hat{A}-1}{\hat{A}}\sum_i\dfrac{\vec{p}_i^2}{2m_N} - \dfrac{1}{m_N\hat{A}}\sum_{i<j}\vec{p}_i\cdot\vec{p}_j \,.
\label{eqn:Tint_1b2b}
\end{equation}
Eq.~\eqref{eqn:tlabtcm} is then recovered by identifying the center-of-mass correction of the kinetic energy with
\begin{equation}
T^{cm} = \dfrac{1}{\hat{A}}\sum_i\dfrac{\vec{p}_i^2}{2m_N} + \dfrac{1}{m_N\hat{A}}\sum_{i<j}\vec{p}_i\cdot\vec{p}_j,
\label{eqn:Tcm}
\end{equation}
which is the sum of a one-body and a two-body operator.
Note that Eqs.~\eqref{eqn:Tint_2b} and~\eqref{eqn:Tint_1b2b} are equivalent only in virtue of the presence of the particle number operator~$\hat{A}$~\cite{CoM_roth}. In most practical applications this is simply replaced with the number of particles $A$ so that the equivalence between the two relations is preserved only for wave function based methods that are formulated in the $A$-body Hilbert space, such as Full Configuration Interaction (FCI) or Quantum Monte Carlo (QMC) in coordinate space.  For methods formulated in Fock space--for example, the SCGF that explores intermediate configuration states by particle attachment and removal--the two forms of the intrinsic kinetic energy with a fixed value of $A$ are no longer exactly equivalent.  In practice, it is found that Eq.~\eqref{eqn:Tint_1b2b} is the most accurate choice in presence of explicit truncations of the many-body expansion.  The calculation of the matrix elements of the kinetic operator is then reduced to the calculation of the one-body $\vec{p}_i^2$ and the two-body $\vec{p}_i\cdot\vec{p}_j$ matrix elements.
Thus, the complete intrinsic Hamiltonian reads
\begin{equation}
\label{eqn:Hint_1stQuant}
\begin{split}
H &= \dfrac{\hat{A}-1}{\hat{A}}\sum_i\dfrac{\vec{p}_i^2}{2m_N} - \dfrac{1}{m_N\hat{A}}\sum_{i<j}\vec{p}_i\cdot\vec{p}_j + \sum_{i<j}V_{ij} + \sum_{i<j<k}W_{ijk}\\
&= \dfrac{\hat{A}-1}{\hat{A}} T^{lab} + \sum_{i<j}(V_{ij} - \widetilde{T}^{cm}_{ij}) + \sum_{i<j<k}W_{ijk}\,,
\end{split}
\end{equation}
where $\widetilde{T}^{cm}_{ij}$ is the two-body component of Eq.~\eqref{eqn:Tcm}.

\subsection{Single-particle basis}
In this work, Greek letters ($\a$) will be used as collective indices for the quantum numbers defining the single-particle basis and the corresponding letter $a$ and $\tilde{a}$ for specific subgroups of them. For spherical basis states with both spin ($\hat{\vec s}$) and isospin ($\hat{\vec t}$) degrees of freedom, one has 
\begin{subequations}
\begin{align}
\tilde{a} &\equiv (n_\a, l_\a, j_\a) = (n_\a, \pi_\a, j_\a)\,,\\
a &\equiv (\tilde{a}, \tau_\a) = (n_\a, l_\a, j_\a, \tau_\a)\,,\\
\alpha &\equiv (a, m_\a) = (n_\a, l_\a, j_\a, \tau_\a, m_\a)\,,
\label{eq:label_sp_all}
\end{align}
\end{subequations}
where $n_\a$, $l_\a$, $\pi_\a$, $j_\a$, $\tau_\a$ and $m_\a$ represent respectively the principal quantum number, the orbital angular momentum, the parity, the total angular momentum (resulting from the sum of the orbital angular momentum and the spin, $\hat{\vec{j}}=\hat{\vec{l}}+\hat{\vec{s}}$), the projection of the isospin and the projection of the total angular momentum along the quantization axis of the quantum state. The parity is implied by the angular momentum ($\pi_\a=(-1)^{l_\a}$). Likewise, the magnitudes of spin and isospin are fixed as they are intrinsic properties of particles ($s=t=1/2$ for protons and neutrons).
The coupling of $l_\a$ and $s_\a$ to give $j_\a$ will be indicated with the notation $(l_\a s_\a)j_\a$ and will be always implicitly assumed in the rest of this work when a total angular momentum $j$ is used.
The projection of the isospin encodes informations about the charge of the nucleon (neutron or proton).
Many-body states built by taking the quantum number $m_\a$ explicitly into account are said to be in ‘$m$-scheme'.
Likewise, for Nuclear Physics applications one refers to \emph{isospin} or to \emph{proton-neutron} (p-n) scheme (or equivalently to ‘$T$-coupled’ or ‘$T$-decoupled’ scheme) when nucleons are coupled respectively to the total isospin ($\hat{\vec{T}}=\hat{\vec{t}}_1+\hat{\vec{t}}_2$) or when the isospin-projections $\tau_1$ and $\tau_2$ are kept as an explicit label for the single-particle states. In the latter case, one keeps track of the total number of protons and neutron separately.
In this work, $J$-coupled $T$-decoupled matrix elements will always be considered. Expressions for cases different from nuclear physics, such as electron or quantum gasses where no isospin is present, can be obtained by specializing our results to systems with all polarized isospins (for example assigning $\tau=+1/2$ to all the particles).

To describe fermions in Fock space, consider a set of anti-commuting creation ($c_{\alpha}^\dagger$) and annihilation ($c_{\alpha}$) operators
\begin{equation}
\{c_{\alpha}, c_{\beta}\} = 0, \quad\{c_{\alpha}^\dagger, c_{\beta}^\dagger\} = 0, \quad\{c_{\alpha}, c_{\beta}^\dagger\} = \delta_{\alpha\beta},
\end{equation}
where the latter notation represents the product of Kronecker deltas on all relevant quantum numbers,
\begin{equation}
\label{eqn:coll_deltas}
\delta_{\alpha\beta} = \delta_{n_\a n_\b}\delta_{l_\a l_\b}\delta_{j_\a j_\b}\delta_{m_\a m_\b}\delta_{\tau_\a\tau_\b}.
\end{equation}
One-, two- and three-nucleon states are defined as
\begin{subequations}
\begin{align}
\ket{\alpha} &\equiv c_{\alpha}^\dagger\ket{0},\\
\ket{\alpha\beta} &\equiv c_{\alpha}^\dagger c_{\beta}^\dagger\ket{0},\\
\ket{\alpha\beta\gamma} &\equiv c_{\alpha}^\dagger c_{\beta}^\dagger c_{\gamma}^\dagger\ket{0},
\end{align}
\end{subequations}
where $\ket{0}$ is the vacuum state.
A generic $n$-body operator $O^{[n]}$ can be expressed as a function of creation and annihilation operators:
\begin{equation} 
O^{[n]} \equiv \biggl(\dfrac{1}{n!}\biggr)^2\sum_{\alpha'_1\dots \alpha'_n}\sum_{\alpha_1\dots \alpha_n} O_{{\alpha}'_1\dots \alpha'_n\, \alpha_1\dots \alpha_n} c^\dagger_{\alpha'_1}\dots c^\dagger_{\alpha'_n} c_{\alpha_n}\dots c_{\alpha_1},
\end{equation}
where the object $O_{\alpha'_1\dots \alpha'_n \alpha_1\dots \alpha_n}$ is the properly antisymmetrized matrix element $\braket{\alpha'_1\dots \alpha'_n | O^{[n]} | \alpha_1\dots \alpha_n}_A$.
The Hamiltonian in Eq.~\eqref{eqn:Hint_1stQuant} in second quantization reads
\begin{equation}
\begin{split}
H = \sum_{\alpha\delta} \left(1-\dfrac{1}{A}\right) T^{lab}_{\alpha\delta} c^\dagger_{\alpha}c_{\delta} + \biggl( \dfrac{1}{2!} \biggr) \sum_{\alpha\beta\delta\epsilon} (V_{\alpha\beta\,\delta\epsilon} - T^{cm}_{\alpha\beta\,\delta\epsilon}) c^\dagger_{\alpha}c^\dagger_{\beta}c_{\epsilon}c_{\delta} + \biggl( \dfrac{1}{3!} \biggr) \sum_{\alpha\beta\gamma\delta\epsilon\varphi} W_{\alpha\beta\gamma\,\delta\epsilon\varphi} c^\dagger_{\alpha}c^\dagger_{\beta}c^\dagger_{\gamma}c_{\varphi}c_{\epsilon}c_{\delta}.
\end{split}
\label{eqn:H}
\end{equation}

\subsubsection*{\texorpdfstring{$J$}{J}-coupled scheme}
The invariance under rotations of the Hamiltonian in Eq.~\eqref{eqn:H} allows to reduce the number of matrix elements to be computed for calculations in a fixed model space. To exploit this symmetry, the following two- and three-body states can be defined
\begin{equation}
\ket{(ab) JM_J}_N \equiv \dfrac{1}{\sqrt{1+\delta_{ab}}} \sum_{m_\a m_\b} \braket{j_\a m_\a j_\b m_\b | JM_J} \ket{\alpha\beta},
\label{eqn:2bnorm}
\end{equation}
\begin{equation}
\ket{[(ab)J_{ab}c] J_{tot}M_{J_{tot}}} \equiv \sum_{m_\a m_\b m_\c}\braket{j_\a m_\a j_\b m_\b | J_{ab} M_{J_{ab}}} \braket{J_{ab} M_{J_{ab}} j_\c m_\c | J_{tot} M_{J_{tot}}} \ket{\alpha\beta\gamma},
\label{eqn:tbnnorm}
\end{equation}
where the subscript $N$ indicates that the two-body state is normalized and the delta function 
\begin{equation}
\delta_{ab} = \delta_{n_\a n_\b}\delta_{l_\a l_\b}\delta_{j_\a j_\b}\delta_{\tau_\a \tau_\b} \,
\end{equation}
is analogous to~Eq.~\eqref{eqn:coll_deltas}. 
The normalization of the three-body state is cumbersome and is typically not included in the computed matrix elements.
The matrix elements of spherically symmetric operators among the $J$-scheme states~\eqref{eqn:2bnorm}-\eqref{eqn:tbnnorm} drop the dependence on the $M$ quantum number. As a consequence, less states are required to represent a given model space and the computational time as well as the memory required to store two- and three-body matrix elements are reduced.

A generic $J$-coupled two-body state can be antisymmetrized as follows:
\begin{equation}
\ket{(ab) JM_J}_A = \ket{(ab) JM_J} - (-1)^{j_\a+j_\b-J}\ket{(ba) JM_J} \, ,
\label{eqn:jasymm}
\end{equation}
while the antisymmetrization of $J$-coupled three-body states is more complicated and will be discussed in details in Sec.~\ref{3b} and \ref{T23}.

\subsection{The single-particle wave function}
The nuclear single-particle isospin-dependent wave functions in coordinate space (CS) and momentum space (MS) take the form
\begin{equation}
\Psi_{nlm_l\tau}(\vec{r}) \equiv \braket{\vec{r} | nlm_l\tau} \equiv \phi_{nl\tau}(r)Y_{lm_l}(\hat{r}),
\label{eqn:psi_sp_spher_CS}
\end{equation}
\begin{equation}
\widetilde\Psi_{nlm_l\tau}(\vec{k}) \equiv \braket{\vec{k} | nlm_l\tau} \equiv \tilde\phi_{nl\tau}(k) Y_{lm_l}(\hat{k}).
\label{eqn:psi_sp_spher_MS}
\end{equation}
where $\phi_{nl\tau}$ represents the isospin-dependent radial wave function and $Y_{lm_l}$ is a \emph{spherical harmonic}, whose argument is a solid angle $\Omega = (\vartheta, \varphi)$.
Note that the radial wave functions can be easily generalized to spin-dependent wave functions $\phi_{nlj\tau}(r)$ and $\tilde\phi_{nlj\tau}(k)$.
The total wave functions in CS and MS are related by a \emph{Fourier transform} and equivalently their radial wave functions are related through Hankel transforms
\begin{equation}
\phi_{nl\tau}(r) \equiv \braket{r | nl\tau} = \sqrt{\dfrac{2}{\pi}} \int_0^{+\infty} dk\,k^2\,j_l(kr)\hat\phi_{nl\tau}(k),
\label{eqn:momtor}
\end{equation}
\begin{equation}
\hat\phi_{nl\tau}(k) \equiv \braket{k | nl\tau} = \sqrt{\dfrac{2}{\pi}} \int_0^{+\infty} dr\,r^2\,j_l(kr)\phi_{nl\tau}(r),
\label{eqn:coordtor}
\end{equation}
where $j_l(kr)$ represents the \emph{spherical Bessel function of the first kind}.
It is important to notice that Eq.~\eqref{eqn:coordtor} does not directly produce the radial part of the Fourier transformed state~\eqref{eqn:psi_sp_spher_MS} but there is an additional complex phase involved (see~\ref{HOmomcoord} for further details):
\begin{equation}
\tilde\phi_{nl\tau}(k) = (-i)^l \, \hat\phi_{nl\tau}(k) \,.
\label{eqn:Hank_phase}
\end{equation}
One may chose to disregard this phase and use Eqs.~\eqref{eqn:momtor} and~\eqref{eqn:coordtor} to define the CS and MS states [that is, $\tilde\phi_{nl\tau}(k) \equiv \hat\phi_{nl\tau}(k)$]. This is convenient because both radial parts can be kept real, however, it must be bore in mind that Eqs.~\eqref{eqn:psi_sp_spher_CS} and ~\eqref{eqn:psi_sp_spher_MS} are no longer exact Fourier transforms of each other in this case: matrix elements computed in the CS basis may acquire additional phases in the MS one and vice versa. Furthermore, the radial functions can have oscillating asymptotic behaviours with changing number of nodes--that is when varying the principal quantum number $n$. In this work, the general case where $\phi_{nl\tau}(r)$ and $\tilde\phi_{nl\tau}(k)$ take complex values is considered.
Leaving the imaginary phase aside, the following convention is assumed:
\begin{equation}
\phi_{nl\tau}(r) > 0, \qquad\text{for}\,\,r \to 0,
\end{equation}
\begin{equation}
\hat\phi_{nl\tau}(k) \sim (-1)^{n}, \qquad\text{for}\,\,k \to 0.
\end{equation}

\subsection{Beyond the Harmonic Oscillator basis}
The HO basis has been widely employed in \emph{ab initio} nuclear structure. The reasons are multiple: the HO potential seems to be a proper choice to describe well-bound closed-shell systems, which are the first systems that have been studied in \emph{ab initio} theory.
The specific analytical expression (Eqs.~\eqref{eqn:phir} and~\eqref{eqn:phik2}) of the HO wave functions allows for various simplifications: first of all, it allows for an exact separation of the intrinsic and center-of-mass motions, which as discussed in Secs.~\ref{Sec:2b_force} and~\ref{3b} is necessary to go from the laboratory frame to the intrinsic one.
Second, the specific transformation between center-of-mass ($\vec{r}$, $\vec{R}$) and single-particle ($\vec{r}_1$, $\vec{r}_2$) coordinates:
\begin{equation}
\begin{pmatrix}
\vec{r}\\
\vec{R}
\end{pmatrix}
=
\begin{pmatrix}
\sqrt{\frac{d}{d+1}} & -\sqrt{\frac{1}{d+1}}\\
\sqrt{\frac{1}{d+1}} & \sqrt{\frac{d}{d+1}}
\end{pmatrix}
\begin{pmatrix}
\vec{r}_1\\
\vec{r}_2
\end{pmatrix},
\label{eqn:hotransf}
\end{equation}
(where $d = m_1/m_2$ for a two-body system) can be carried out in the HO basis by the so called Moshinsky brackets ($\braket{nN(lL)\lambda|n_1n_2(l_1l_2)\lambda}_d$) \cite{Moshinsky},
whose implementation is computationally very convenient.
Furthermore, these coefficients are diagonal among major oscillator shells so that the relation $2n+l+2N+L = 2n_1+l_1+2n_2+l_2$ is always valid. This property is referred to as the \emph{conservation of the energy} for the Moshinsky brackets and it allows to greatly simplify the calculation based on HO states given the large number of vanishing coefficients it implies.
Finally, since HO wave-functions are isospin-independent, the number of matrix elements to be stored can be reduced exploiting this symmetry.

The formalism developed in this paper allows to go beyond several limitations imposed by the HO basis. First of all, isospin-dependent bases can be employed, so that one can exploit the advantages of expanding matrix elements on two different bases to tackle for instance neutron-rich exotic systems in which the radial density behavior of neutrons and protons can be very different. Second, the Moshinsky brackets that are employed in the case of the HO wave functions are substituted by the WC brackets $\braket{rR(lL)\lambda|n_1n_2(l_1l_2)\lambda}$ \cite{WC}, which are at the heart of this paper and can be applied to a completely generic basis. This will allow breaking the constraints that presently link \emph{ab initio} methods in nuclear physics to the HO basis.
These coefficients allow for a change of coordinates that generalizes Eq.~\eqref{eqn:hotransf}, as follows:
\begin{equation}
\begin{pmatrix}
\vec{r}_1\\
\vec{r}_2
\end{pmatrix}
=
\begin{pmatrix}
s_1 & t_1\\
s_2 & t_2
\end{pmatrix}
\begin{pmatrix}
\vec{r}\\
\vec{R}
\end{pmatrix},
\label{eqn:wctransf}
\end{equation}
with $s_1, t_1, s_2, t_2 \in \mathbb{R}$. The matrix constituted by these coefficients will be referred to as \emph{matrix of Wong-Clement coefficients}. It is straightforward to invert Eq.~\eqref{eqn:wctransf} to obtain a generalization of Eq.~\eqref{eqn:hotransf}.
The WC bracket is different compared to the Moshinsky bracket also in the type of states that are coupled, since it can act directly on MS momenta or CS positions, allowing to use single-particle wave functions in either space. On the other hand, the presence of these continuous variables complicates the computation of WC coefficients. A complete description of these coefficients is given in~\ref{coeff}.

\section{One-body matrix elements}
\label{1b}
\subsection{Laboratory-frame kinetic energy}
Consider the matrix elements of the kinetic energy operator in the laboratory frame
\begin{equation}
\braket{\a | T^{lab} | \b} = \delta_{m_\a m_\b} \braket{a | \dfrac{p^2}{2m_N} | b} = \delta_{m_\a m_\b} \delta_{j_\a j_\b}\delta_{l_\a l_\b}\delta_{\tau_\a \tau_\b} t_{n_\a n_\b}^{l_\a \tau_\a},
\label{eqn:tsmall}
\end{equation}
where the deltas select subblocks $t_{n_\a n_\b}^{l_\a \tau_\a}$ of non-vanishing matrix elements of $T^{lab}$. These matrix elements are easily computed for MS radial wave functions,
\begin{equation}
t_{n_\a n_\b}^{l \tau} = -\dfrac{\hbar^2}{2m_N} \int_0^{+\infty} dp\,p^4\, \tilde\phi_{n_\a l \tau}(p) \, \tilde\phi_{n_\b l \tau}(p),
\end{equation}
with $l=l_\a=l_\b$ and $\tau=\tau_\a=\tau_\b$.
The one-body matrix element in Eq.~\eqref{eqn:tsmall} can be re-expressed as a function of CS single-particle wave functions by exploiting the representation of the momentum operator in CS
\begin{equation}
\vec{p} \to -i\hbar\vec{\nabla}
\end{equation}
and the expansion of the \emph{Laplace operator} in spherical coordinates
\begin{equation}
\Delta(r) = \nabla^2(r) = \biggl[ \dfrac{d^2}{dr^2} + \dfrac{2}{r}\dfrac{d}{dr} - \dfrac{L^2}{\hbar^2 r^2} \biggr].
\end{equation}
The matrix elements of the operator $t$ are easily re-expressed in CS
\begin{equation}
t_{n_\a n_\b}^{l \tau} = -\dfrac{\hbar^2}{2m_N} \int_0^{+\infty} dr\,r^2\,\phi_{n_\a l \tau}(r) \biggl[ \dfrac{d^2}{dr^2} + \dfrac{2}{r}\dfrac{d}{dr} - \dfrac{l(l+1)}{r^2} \biggr] \phi_{n_\b l \tau}(r),
\end{equation}
where the completeness relation
\begin{equation}
\int dr r^2 \ket{r}\bra{r} = 1
\label{eqn:compl1}
\end{equation}
and the orthonormality relation
\begin{equation}
\braket{r|r'} = \dfrac{\delta(r-r')}{r^2}
\label{eqn:orth1}
\end{equation}
for the radial state $\ket{r}$ hold. Equations analogous to~\eqref{eqn:compl1} and~\eqref{eqn:orth1} stand for the MS state $\ket{p}$.

\section{Two-body matrix elements}
\label{2b}
In this section, working relations for computing two-body matrix elements in $J$-coupled $T$-decoupled scheme are obtained. The $J$-coupled scheme minimizes the number of matrix elements to be considered, while the $T$-decoupling allows to take into account for possible isospin dependence in the single-particle wave functions.

\subsection{Center-of-mass correction of the kinetic energy}
\label{sec:2b_KE}
\vspace{0.2cm}
\noindent
\emph{Expression with MS wave function}
\vspace{0.1cm}
\\
The calculation of the center-of-mass correction of the kinetic energy involves evaluating of the matrix elements of the two-body operator $\vec{p}_1\cdot\vec{p}_2$ \cite{CoulombSturmian, ShellModelGenBas}, where the subscripts 1 and 2 denote the two particles. By means of Eqs.~\eqref{eqn:r_sph} and \eqref{eqn:prod_sph}, one can  write this operator as a sum of spherical components, $[\vec{p}_1\cdot\vec{p}_2]_\mu$:
\begin{equation}
\vec{p}_1\cdot\vec{p}_2 = \sum_\mu [\vec{p}_1\cdot\vec{p}_2]_\mu = \dfrac{4\pi}{3} \sum_\mu (-1)^\mu p_1p_2 Y_{1\mu}(\hat{p}_1)Y_{1-\mu}(\hat{p}_2) \, .
\label{eqn:pipj}
\end{equation}
The radial and angular parts of Eq.~\eqref{eqn:pipj} decouple easily in $m$-scheme. Thus,
\begin{equation}
\begin{split}
\braket{\a\b | [\vec{p}_1\cdot\vec{p}_2]_\mu | \c \d} = \dfrac{4\pi}{3}(-1)^\mu \braket{n_\a l_\a \tau_\a, n_\b l_\b \tau_\b | p_1p_2 | n_\c l_\c \tau_\c, n_\d l_\d \tau_\d}\\
\quad\times\braket{j_\a m_\a, j_\b m_\b | Y_{1\mu}(\hat{p}_1)Y_{1-\mu}(\hat{p}_2) | j_\c m_\c, j_\d m_\d}.
\end{split}
\end{equation}
The radial part is further separable in one-body matrix elements
\begin{equation}
\braket{n_\a l_\a \tau_\a, n_\b l_\b \tau_\b | p_1p_2 | n_\c l_\c \tau_\c, n_\d l_\d \tau_\d} = \braket{n_\a l_\a \tau_\a | p | n_\c l_\c \tau_\c} \braket{n_\b l_\b \tau_\b | p | n_\d l_\d \tau_\d}\,,
\end{equation}
where each contribution is expressed as a radial integral
\begin{equation}
\braket{n_\a l_\a \tau_\a | p_1 | n_\c l_\c \tau_\c} = \int_0^{+\infty}dp\,p^3\, \tilde\phi_{n_\a l_\a \tau_\a}^*(p) \, \tilde\phi_{n_\c l_\c \tau_\c}(p)\, .
\label{eqn:com_mom}
\end{equation}
Similarly, the angular bracket is separable in two one-body contributions
\begin{equation}
\braket{j_\a m_\a, j_\b m_\b | Y_{1\mu}(\hat{p}_1)Y_{1-\mu}(\hat{p}_2) | j_\c m_\c, j_\d m_\d} = \braket{(l_\a s_\a)j_\a m_\a | Y_{1\mu} | (l_\c s_\c)j_\c m_\c} \braket{(l_\b s_\b)j_\b m_\b | Y_{1-\mu} | (l_\d s_\d)j_\d m_\d}.
\label{eqn:Yp1Yp2_mSch}
\end{equation}
To compute the one-body expectation value of a spherical harmonic we convert it from $j$-scheme to $ls$-scheme
\begin{equation}
\begin{split}
\braket{(l_\a s_\a)j_\a m_\a | Y_{1\mu} | (l_\c s_\c)j_\c m_\c} =&
\sum_{m_{l_\a}m_{s_\a}m_{l_\c}m_{s_\c}}
\braket{l_\a m_{l_\a}s_\a m_{s_\a} | j_\a m_\a} \braket{l_\c m_{l_\c}s_\c m_{s_\c} | j_\c m_\c} \, \braket{l_\a m_{l_\a}s_\a m_{s_\a} | Y_{1\mu} | l_\c m_{l_\c}s_\c m_{s_\c}} \\
 =& \sum_{m_{l_\a}m_{s_\c}m_{l_\c}}
\braket{l_\a m_{l_\a}s_\c m_{s_\c} | j_\a m_\a} \braket{l_\c m_{l_\c}s_\c m_{s_\c} | j_\c m_\c} \braket{l_\a m_{l_\a} | Y_{1\mu} | l_\c m_{l_\c}} \, ,
\end{split}
\label{eqn:jmY}
\end{equation}
where we use the fact that $Y_{1\mu}$ does not act on spin and $\braket{s_\a m_{s_\a} | s_\c m_{s_\c}}=\delta_{s_\a s_\c}\delta_{m_{s_\a} m_{s_\c}}$. Furthermore, all spins have the same value for a system of identical fermions.
The angular bracket in Eq.~\eqref{eqn:jmY} is given by Eq.~\eqref{eqn:gaunt}:
\begin{equation}
\braket{l_\a m_{l_\a} | Y_{1\mu} | l_\c m_{l_\c}} = \sqrt{\dfrac{3}{4\pi}} \dfrac{\hat{l}_\c}{\hat{l}_\a}\braket{l_\c010 | l_\a0} \braket{l_\c m_{l_\c}1\mu | l_\a m_{l_\a}}\,,
\end{equation}
where we use the notation of Eq.~\eqref{eqn:def_hatj}.
Eq.~\eqref{eqn:3j6j} can then be applied to simplify the summation over triple Clebsch-Gordan coefficients
\begin{equation}
\begin{split}
\braket{(l_\a s_\a)j_\a m_\a | Y_{1\mu} |(l_\c s_\c) j_\c m_\c}
=& \sqrt{\dfrac{3}{4\pi}} \dfrac{\hat{l}_\c}{\hat{l}_\a} \braket{l_\c010 | l_\a0}
\sum_{m_{l_\a}m_{s_\c}m_{l_\c}} \, \braket{l_\c m_{l_\c} s_\c m_{s_\c} | j_\c m_\c} \, \braket{l_\a m_{l_\a} s_\c m_{s_\c} | j_\a m_\a}\,\braket{l_\c m_{l_\c} 1	\mu | l_\a m_{l_\a}}\\
=& \sqrt{\dfrac{1}{4\pi}} \braket{l_\c0l_\a0 | 10}(-)^{j_\c+s_\c} \hat{j}_\c\hat{l}_\a\hat{l}_\c
\braket{j_\c  m_\c  1 \mu | j_\a m_\a}
\begin{Bmatrix}
l_\c & l_\a & 1\\
j_\a & j_\c & s_\c
\end{Bmatrix} \,.
\end{split}
\label{eqn:trcg}
\end{equation}
Eq.~\eqref{eqn:trcg} is valid for fermions with $s_\a=s_\c$.  For the special case of spin $s=1/2$, such as electrons or nucleons, Eq.~\eqref{eqn:3j6j2} allows further simplifications:
\begin{equation}
\begin{split}
&\braket{l_\c 0l_\a 0|10}
\begin{Bmatrix}
l_\c & l_\a & 1\\
j_\a & j_\c & 1/2
\end{Bmatrix}\\
&\qquad\quad= \dfrac{[1+(-1)^{l_\a+l_\c+1}]}{2} \dfrac{(-1)^{j_\c-j_\a+1+l_\a-l_\c} }{\hat{l}_\a \, \hat{l}_\c} \braket{j_\c \dfrac{1}{2} j_\a -\dfrac{1}{2} | 10}\\
&\qquad\quad= \pi(l_\c, 1, l_\a) \dfrac{(-1)^{j_\c - j_\a + 1 + l_\a - l_\c} }{ \hat{l}_\a \, \hat{l}_\c} (-1)^{j_\c-1/2} \sqrt{\dfrac{3}{2j_\a+1}}\braket{j_\c \dfrac{1}{2}  1 0 | j_\a \dfrac{1}{2} }\,,
\end{split}
\end{equation}
where the definition 
\begin{equation}
\pi(l_1,l_2,\dots) \equiv \dfrac{1}{2}[1+(-1)^{l_1+l_2+\dots}]
\end{equation}
has been used to enforce the parity constraint $l_\a = l_\c \pm 1$.
Putting together all above results, one can confirm that the matrix element~\eqref{eqn:jmY} satisfies the Wigner-Eckart theorem
\begin{equation}
\begin{split}
\braket{(l_\a s_\a)j_\a m_\a | Y_{1\mu} | (l_\c s_\c)j_\c m_\c} =& \braket{j_\c  m_\c 1 \mu | j_\a m_\a}
\braket{(l_\a s_\a) j_\a || Y_{1} || (l_\c s_\c) j_\c} 
\end{split}
\label{eqn:jYj}
\end{equation}
and identifies the reduced matrix elements for $s=1/2$ fermions and integer angular momentum with
\begin{equation}
\braket{(l_\a s_\a) j_\a || Y_{1} || (l_\c s_\c) j_\c} \equiv  \sqrt{\dfrac{3}{4\pi}} \dfrac{\hat{j}_\c}{\hat{j}_\a} (-1)^{j_\a-j_\c+1}\braket{j_\c\dfrac{1}{2}10 | j_\a \dfrac{1}{2}}\pi(l_\c, 1, l_\a)\,.
\label{eqn:ljY}
\end{equation}
The second contribution in Eq.~\eqref{eqn:Yp1Yp2_mSch} is analogous but with the substitution $\mu\rightarrow-\mu$. By summing over $\mu$ and coupling to total angular momentum J we obtain:
\begin{equation}
\begin{split}
& \sum_\mu(-1)^\mu \braket{ab; J | Y_{1\mu}(\hat{p}_1)Y_{1 -\mu}(\hat{p}_2) | cd; J} = \\
&\qquad \quad  = \sum_{\substack{m_\a \, m_\b  \, \mu \\ m_\c \, m_\d}} (-1)^\mu \braket{j_\a m_\a j_\b m_\b | J M_J} \braket{j_\c m_\c j_\d m_\d | J M_J} \braket{\a \b | Y_{1\mu}(\hat{p}_1)Y_{1 -\mu}(\hat{p}_2) | \c \d} \\
&\qquad \quad = \hat{j}_\a\hat{j}_\b \sum_{\substack{m_\a \, m_\b  \, \mu \\ m_\c \, m_\d \, M_J}}  (-1)^{j_\b+j_\c+J} \braket{(l_\a s_\a) j_\a || Y_{1} || (l_\c s_\c) j_\c} \braket{(l_\b s_\b) j_\b || Y_{1} ||(l_\d s_\d) j_\d} \, (-1)^{j_\d+1+j_\c+m_\d+\mu-m_\c}\\
&\qquad \qquad \times
\begin{pmatrix}
j_\a & J & j_\b\\
-m_\a & M_J & -m_\b
\end{pmatrix}
\begin{pmatrix}
j_\a & 1 & j_\c\\
-m_\a & \mu & -m_\c
\end{pmatrix}
\begin{pmatrix}
j_\d & J & j_\c\\
-m_\d & M_J & -m_\c
\end{pmatrix}
\begin{pmatrix}
j_\d & 1 & j_\b\\
m_\d & -\mu & -m_\b
\end{pmatrix}  \\
&\qquad \quad= (-1)^{j_\b+j_\c+J} \, \hat{j}_\a \hat{j}_\b \,
\begin{Bmatrix}
j_\a & j_\b & J\\
j_\d & j_\c & 1
\end{Bmatrix} \,
\braket{(l_\a s_\a) j_\a || Y_{1} || (l_\c s_\c) j_\c}\braket{(l_\b s_\b) j_\b || Y_{1} || (l_\d s_\d) j_\d}\, ,
\end{split}
\label{eqn:p1p2_mtsel}
\end{equation}
where Eq.~\eqref{eqn:2bnorm} has been used without normalization and we have exploited the symmetry properties of the $3j$ symbols and Eq.~\eqref{eqn:6jfrom3j}.
Eventually, the antisymmetrized and normalized matrix elements of product of momenta $\vec{p}_1\cdot\vec{p}_2$ can be expressed as
\begin{equation}
%\begin{gathered}
\braket{ab; J | \vec{p}_1\cdot\vec{p}_2 | cd; J}_{AN} = \dfrac 1{\sqrt{1+\delta_{ab}}} \dfrac 1{\sqrt{1+\delta_{cd}}} \biggl[ \braket{ab; J | \vec{p}_1\cdot\vec{p}_2 | cd; J} - (-1)^{j_\c+j_\d-J}\braket{ab; J| \vec{p}_1\cdot\vec{p}_2 | dc; J} \biggr] \,.
\label{eqn:asme}
\end{equation}
Each two-body contribution is conveniently written as
\begin{equation}
\braket{ab; J | \vec{p}_1\cdot\vec{p}_2 | cd; J} = (-1)^{j_\b+j_\c+J}\, \hat{j}_\a \hat{j}_\b \,
\begin{Bmatrix}
j_\a & j_\b & J\\
j_\d & j_\c & 1
\end{Bmatrix}
\braket{a||\vec{p}_1||c}\braket{b||\vec{p}_2||d},
\label{eqn:pipjrel}
\end{equation}
where the one-body reduced brackets, which include both the radial and angular one-body contributions, are defined by
\begin{equation}
\braket{a||\vec{p}||b} \equiv \biggl[ \int_0^{+\infty} dp\,p^3\, \tilde\phi_{n_\a l_\a \tau_\a}^*(p)\, \tilde\phi_{n_\b l_\b \tau_\b}(p) \biggr] \braket{(l_\a s_\a) j_\a || Y_1 || (l_\b s_\b) j_\b}.
\label{eqn:adoubpb}
\end{equation}
The separability of the two-body matrix elements in Eq.~\eqref{eqn:pipjrel} into a product of two one-body matrix elements makes the computational time required for their calculation scale the same as one-body matrix elements.

\vspace{0.2cm}
\noindent
\emph{Expression with CS wave function}
\vspace{0.1cm} \\
For single-particle wave functions expressed in CS, the center-of-mass correction of the kinetic energy can be computed from Eq.~\eqref{eqn:asme} through a preliminary Hankel transform [c.f.r. Eq.~\eqref{eqn:coordtor}] of the radial part in MS (and keeping track of the complex phase~\eqref{eqn:Hank_phase}, see~\ref{HOmomcoord}). In most cases this implies performing additional computations to evaluate the Hankel transform which might introduce numerical noise.
Another option is to represent the momentum operator in CS without manipulating the wave functions. In this case, Eq.~\eqref{eqn:com_mom} and the reduced matrix elements~\eqref{eqn:adoubpb} need to be substituted with appropriate CS integrals. We discuss the relevant formulae in the following.

Consider the product of momenta in coordinate space
\begin{equation}
\vec{p}_1\cdot\vec{p}_2 = -\hbar^2 \vec{\nabla}_1\cdot\vec{\nabla}_2 = -\hbar^2\sum_{\mu}(-1)^\mu \nabla_{1\mu}\nabla_{1-\mu}.
\end{equation}
The spherical components of the gradient operator have a rather complicated form~\cite{Var}
\begin{equation}
\nabla_{1\mu} = \sqrt{\dfrac{4\pi}{3}} \biggl\{ Y_{1\mu}\dfrac{d}{dr} - \dfrac{i}{r}[Y_1\times L_1]_{1\mu} \biggr\} \,,
\label{eqn:nab}
\end{equation}
which no longer factorizes in a radial and angular part. However, it still is a rank-1 tensor operator and it satisfies the Wigner-Eckart theorem for the matrix elements among spherical angular states:
\begin{equation}
\braket{n_\a l_\a m_{l_\a} | \nabla_{1\,\mu} | n_\b l_\b  m_{l_\b}} = \braket{n_\a l_\a \tau_\a|| \vec{\nabla} || n_\b l_\b \tau_\b}\braket{l_\b m_{l_\b} 1 \mu | l_\a m_{l_\a}}\,.
\end{equation}
The computation of such reduced matrix elements is lengthy but straightforward and is discussed in detail in Ref.~\cite{Var} (Sec.~13.2.4). One has
\begin{equation}
\braket{n_\a l_\a \tau_\a|| \vec{\nabla} || n_\b l_\b \tau_\b} = \sqrt{\dfrac{l_\b+1}{2l_\a+1}} A_{n_\a l_\a \tau_\a}^{n_\b l_\b \tau_\b} \delta_{l_\a, l_\b+1} - \sqrt{\dfrac{l_\b}{2l_\a+1}}B_{n_\a l_\a \tau_\a}^{n_\b l_\b \tau_\b} \delta_{l_\a, l_\b-1},
\end{equation}
with the definitions
\begin{subequations}
\begin{align}
A_{n_\a l_\a \tau_\a}^{n_\b l_\b \tau_\b} &\equiv \int_0^{+\infty} dr\,r^2\,\phi^*_{n_\a l_\a \tau_\a}(r)\biggl[ \dfrac{d}{dr} - \dfrac{l_\b}{r}\biggr] \phi_{n_\b l_\b \tau_\b}(r),\\
B_{n_\a l_\a \tau_\a}^{n_\b l_\b \tau_\b} &\equiv \int_0^{+\infty} dr\,r^2\,\phi^*_{n_\a l_\a \tau_\a}(r)\biggl[ \dfrac{d}{dr} + \dfrac{l_\b+1}{r}\biggr] \phi_{n_\b l_\b \tau_\b}(r).
\end{align}
\end{subequations}
To find the reduced matrix elements in a $J$-coupled single-particle states we proceed in analogy with Eqs.~\eqref{eqn:jmY}--\eqref{eqn:trcg} and write:
\begin{equation}
\begin{split}
\braket{\a| \nabla_{1\,\mu} | \c}
%\braket{n_\a(l_\a s_\a) j_\a  m_\a \t_\a| \nabla_{1\,\mu} | n_\c (l_c\ s_\c) j_\c m_\c \t_\c}
 &= \delta_{\t_\a \t_\c} \braket{n_\a l_\a \tau_\a || \vec{\nabla}|| n_\c l_\c \tau_\c} \sum_{m_{l_\a}m_{s_\c}m_{l_\c}} \braket{l_\a m_{l_\a} s_\c m_{s_\c} | j_\a m_\a} \braket{l_\c m_{l_\c} s_\c m_{s_\c} | j_\c m_\c} \braket{l_\c m_{l_\c}1\mu | l_\a m_{l_\a}}\\
 &= \delta_{\t_\a \t_\c} \braket{n_\a l_\a \tau_\a || \vec{\nabla} || n_\c l_\c \tau_\c} (-1)^{l_\a + j_\c + s_\c + 1}  \, \hat{j}_\c \, \hat{l}_\a \,
 \braket{ j_\c m_\c 1 \mu  | j_\a m_\a}
\begin{Bmatrix}
 j_\c &  1   & j_\a\\
 l_\a & s_\c & l_\c
\end{Bmatrix} \, ,
\end{split}
\end{equation}
so that we have
\begin{equation}
\braket{ a || \vec{\nabla} || c} = \delta_{\t_\a \t_\c} 
 \braket{n_\a l_\a \tau_\a || \nabla_1 || n_\c l_\c \tau_\c}\,  F_{l_\a l_\c s_\c}^{j_\a j_\c}
\end{equation}
with the definition of the quantity
\begin{equation}
F_{l_\a l_\c s_\c}^{j_\a j_\c} \equiv (-1)^{l_\a + j_\c + s_\c + 1} \hat{j}_\c\hat{l}_\a
\begin{Bmatrix}
 j_\c &  1   & j_\a\\
 l_\a & s_\c & l_\c
\end{Bmatrix} \, .
\end{equation}

Performing a coupling to $J$-scheme as for Eq.~\eqref{eqn:p1p2_mtsel}, we find the final expression for the (unnormalized and non-antisymmetric) matrix element:
\begin{equation}
\begin{split}
\braket{ab; J | \vec{p}_1\cdot\vec{p}_2 | cd; J} = & -\hbar^2 \sum_\mu (-1)^\mu \braket{ab; J | \nabla_{1\mu}\nabla_{1-\mu} | cd; J} \\
=& -\hbar^2 \braket{ a || \vec{\nabla}_1 || c} \braket{ b || \vec{\nabla}_1 || d}
(-1)^{j_\b + j_\c + J} \, \hat{j}_\a \, \hat{j}_\b \,
\begin{Bmatrix}
j_\a & j_\b & J\\
j_\d & j_\c & 1
\end{Bmatrix}\, ,\\
=& -\hbar^2 \braket{n_\a l_\a \tau_\a || \nabla_1 || n_\c l_\c \tau_\c} \braket{n_\b l_\b \tau_\b || \nabla_1 || n_\d l_\d \tau_\d} F_{l_\a l_\c s_\c}^{j_\a j_\c}F_{l_\b l_\d s_\d}^{j_\b j_\d} 
(-1)^{j_\b + j_\c + J} \, \hat{j}_\a \, \hat{j}_\b \,
\begin{Bmatrix}
j_\a & j_\b & J\\
j_\d & j_\c & 1
\end{Bmatrix}\, .
\end{split}
\label{eqn:ppij}
\end{equation}
The matrix elements in Eq.~\eqref{eqn:ppij} can be normalized and antisymmetrized according to Eq.~\eqref{eqn:asme}.

\subsection{Center-of-mass correction of the radius}
In this subsection, the matrix elements of the operator $\vec{r}_1\cdot\vec{r}_2$ are obtained. This operator does not enter directly the Hamiltonian but it is used for the calculation of nuclear radii. Given the symmetry between the expressions of the momentum operator in coordinate space and the radius operator in momentum space,
\begin{equation}
\vec{p} \to -i\hbar\vec{\nabla}_r\,, \qquad \vec{r} \to i\hbar\vec{\nabla}_p\,,
\end{equation}
it is straightforward to adapt the equations obtained for the center-of-mass correction of the kinetic energy in the previous section to the case of the center-of-mass correction of the radius. The equations for the expectation value of the operator $\vec{p}_1\cdot\vec{p}_2$ expanded on the MS wave functions (Eqs.~\eqref{eqn:asme}, \eqref{eqn:pipjrel}, \eqref{eqn:adoubpb}) become the equations for the expectation value of the operator $\vec{r}_1\cdot\vec{r}_2$ expanded on CS wave functions:
\begin{align}
%\begin{gathered}
\braket{ab; J | \vec{r}_1\cdot\vec{r}_2 | cd; J}_{AN} ={}& \dfrac1{\sqrt{1+\delta_{ab}}} \dfrac1{\sqrt{1+\delta_{cd}}} \biggl[ \braket{ab; J | \vec{r}_1\cdot\vec{r}_2 | cd; J} - (-1)^{j_\c+j_\d-J}\braket{ab; J| \vec{r}_1\cdot\vec{r}_2 | dc; J} \biggr]\,, \\
%\end{gathered}  \\
%\end{equation}
%
%\begin{equation}
\braket{ab; J | \vec{r}_1\cdot\vec{r}_2 | cd; J} ={}& (-1)^{j_\b+j_\c+J}\hat{j}_\a\hat{j}_\b
\begin{Bmatrix}
j_\a & j_\b & J\\
j_\d & j_\c & 1
\end{Bmatrix}
\braket{a||\vec{r}||c}\braket{b||\vec{r}||d}\,,\\
%\end{equation} 
%
%\begin{equation}
\braket{a||\vec{r}||b} ={}& \biggl[ \int_0^{+\infty} dr\,r^3\,\phi_{n_\a l_\a \tau_\a}^*(r)\, \phi_{n_\b l_\b \tau_\b}(r) \biggr] \braket{l_\a j_\a || Y_1 || l_\b j_\b}\, .
\end{align}

The same applies for the expectation value of the operator $\vec{r}_i\cdot\vec{r}_j$ expanded on MS wave functions:
\begin{equation}
\braket{ab; J | \vec{r}_1\cdot\vec{r}_2 | cd; J} = -\hbar^2\braket{n_\a l_\a \tau_\a || \vec{\nabla}_p || n_\c l_\c \tau_\c} \braket{n_\b l_\b \tau_\b|| \vec{\nabla}_p || n_\d l_\d \tau_\d} F_{l_\a l_\c s_\c}^{j_\a j_\c}F_{l_\b l_\d s_\d}^{j_\b j_\d} (-1)^{j_\b + j_\c + J }
\begin{Bmatrix}
j_\a & j_\b & J\\
j_\d & j_\c & 1
\end{Bmatrix}\,,
\end{equation}

\begin{equation}
\braket{n_\a l_\a \tau_\a|| \vec{\nabla}_p || n_\b l_\b \tau_\b} \equiv \sqrt{\dfrac{l_\b+1}{2l_\a+1}} \widetilde{A}_{n_\a l_\a \tau_\a}^{\,n_\b l_\b \tau_\b} \delta_{l_\a, l_\b+1} - \sqrt{\dfrac{l_\b}{2l_\a+1}}\widetilde{B}_{n_\a l_\a \tau_\a}^{\,n_\b l_\b \tau_\b} \delta_{l_\a, l_\b-1}\,,
\end{equation}

\begin{subequations}
\begin{align}
\widetilde{A}_{n_\a l_\a \tau_\a}^{\,n_\b l_\b \tau_\b} &\equiv \int_0^{+\infty} dp\,p^2\,\tilde\phi^*_{n_\a l_\a \tau_\a}(p)\biggl[ \dfrac{d}{dp} - \dfrac{l_\b}{p}\biggr] \tilde\phi_{n_\b l_\b \tau_\b}(p)\,,\\
\widetilde{B}_{n_\a l_\a \tau_\a}^{\,n_\b l_\b \tau_\b} &\equiv \int_0^{+\infty} dp\,p^2\,\tilde\phi^*_{n_\a l_\a \tau_\a}(p)\biggl[ \dfrac{d}{dp} + \dfrac{l_\b+1}{p}\biggr] \tilde\phi_{n_\b l_\b \tau_\b}(p)\,,
\end{align}
\end{subequations}
where the coefficients $\widetilde{A}_{n_\a l_\a \tau_\a}^{\,n_\b l_\b \tau_\b}$ and $\widetilde{B}_{n_\a l_\a \tau_\a}^{\,n_\b l_\b \tau_\b}$ are defined in terms of integrals over momenta.

\subsection{Coulomb interaction}
At the short distance scales typical of hadronic interactions, the strength of the electromagnetic force becomes important even for nuclear systems. Hence, the two-body part of Hamiltonian~\eqref{eqn:H} must contain a Coulomb interaction in the form
\begin{equation}
V(\abs{\vec{r}_1 - \vec{r}_2}) = \dfrac{\alpha\, Z_1\, Z_2}{\abs{\vec{r}_1 - \vec{r}_2}},
\label{eqn:coul}
\end{equation}
where $\alpha$ is the \emph{fine structure constant} and $Z_i$ is the charge number of the $i$-th particle.
The inverse distance term in Eq.~\eqref{eqn:coul} can be expanded by a \emph{multiple decomposition} \cite{CoulombComm}:
\begin{equation}
\dfrac{1}{\abs{\vec{r}_1 - \vec{r}_2}} = \sum_\lambda \dfrac{r_<^\lambda}{r_>^{\lambda+1}} P_\lambda ( \cos\theta),
\label{eqn:1sr}
\end{equation}
where $\theta$ is the angle between the vectors $\vec{r}_1$ and $\vec{r}_2$,  $P_\lambda(x)$ is a Legendre Polynomial of degree $\lambda$ and the convention
\begin{subequations}
\begin{align}
r_< &\equiv \min(r_1, r_2),\\
r_> &\equiv \max(r_1, r_2)
\end{align}
\end{subequations}
is used.
The Legendre polynomial can be expanded in spherical harmonics:
\begin{equation}
P_\lambda ( \cos\theta) = \sum_{\mu=-\lambda}^{+\lambda} \dfrac{4\pi}{2\lambda+1} Y_{\lambda \mu}(\hat{r}_1)Y_{\lambda \mu}^*(\hat{r}_2),
\end{equation}
such that Eq.~\eqref{eqn:1sr} becomes
\begin{equation}
\dfrac{1}{\abs{\vec{r}_1 - \vec{r}_2}} = \sum_{\lambda} \dfrac{4\pi}{2\lambda+1} \dfrac{r_<^\lambda}{r_>^{\lambda+1}} \sum_{\mu} (-1)^\mu \, Y_{\lambda\,\mu}(\hat{r}_1)Y_{\lambda\,-\!\mu}(\hat{r}_2) \,,
\label{eq:1overR_op}
\end{equation}
where the property $Y_{\lambda \mu}^*(\hat{r}) = (-1)^\mu Y_{\lambda \,-\!\mu}(\hat{r})$ of spherical harmonics has been employed.

Unlike the case of the center-of-mass corrections discussed in previous subsections, the radial part of the operator~\eqref{eq:1overR_op} does not separate in two one-body contributions. However, such factorization still holds for the angular terms.
Thus, the matrix elements among single-particle basis states become
\begin{equation}
\begin{split}
\braket{\a \b | \dfrac{1}{\abs{\vec{r}_1 - \vec{r}_2}} | \c \d}
=& \delta_{\t_\a \t_\c} \delta_{\t_\b \t_\d} \, \sum_{\lambda\,\mu} \dfrac{4\pi}{2\lambda+1}
\braket{n_\a l_\a \tau_\a, n_\b l_\b \tau_\b | \dfrac{r_<^\lambda}{r_>^{\lambda+1}} | n_\c l_\c \tau_\c, n_\d l_\d \tau_\d} \\
& \qquad \times
\braket{(l_\a s_\a) j_\a m_\a | \, Y_{\lambda \,\mu} \, | (l_\c s_\c) j_\c m_\c}
\braket{(l_\b s_\b) j_\b m_\b | \, Y_{\lambda \,-\!\mu} \, | (l_\d s_\d) j_\d m_\d} \,,
\end{split}
\label{eqn:rrpar}
\end{equation}
where the radial integral can be written explicitly as
\begin{equation}
\begin{split}
\braket{n_\a l_\a \tau_\a, n_\b l_\b \tau_\b | \dfrac{r_<^\lambda}{r_>^{\lambda+1}} | n_\c l_\c \tau_\c, n_\d l_\d \tau_\d} =
\int_0^{+\infty} dr_1 \, \phi^*_{n_\a l_\a \tau_\a}(r_1) \, & \phi_{n_\c l_\c \tau_\c}(r_1)  
 \biggl[\dfrac{1}{r_1^{\lambda-1}} \int_0^{r_1}  dr_2 \, \phi^*_{n_\b l_\b \tau_\b}(r_2)\,r_2^{\lambda+2}\,\phi_{n_\d l_\d \tau_\d}(r_2) \\
 &
+ r_1^{\lambda+2} \int_{r_1}^{+\infty}  dr_2 \, \phi^*_{n_\b l_\b \tau_\b}(r_2) \,\dfrac{1}{r^{\lambda-1}_2}\,\phi_{n_\d l_\d \tau_\d}(r_2)  \biggr] \,.
\end{split}
\label{eqn:radcoul}
\end{equation}
Note that no charge term is yet included in the operator~\eqref{eq:1overR_op} so that Eq.~\eqref{eqn:rrpar} is diagonal in isospin. Nevertheless, the integral~\eqref{eqn:radcoul} still depend of isospin through the radial parts of the single-particle basis functions.
The angular matrix elements in Eq.~\eqref{eqn:rrpar} obey the Wigner-Eckart theorem
\begin{equation}
\begin{split}
\braket{(l_\a s_\a) j_\a m_\a | \, Y_{\lambda\,\mu} \, | (l_\c s_\c) j_\c m_\c} =
\braket{j_\c m_\c \, \lambda \mu | j_\a m_\a} 
\braket{(l_\a s_\a) j_\a || \, Y_\lambda \, || (l_\c s_\c) j_\c }
\end{split}
\label{eqn:Ylm_WE}
\end{equation}
and their derivation follows closely the one shown in Sec.~\ref{sec:2b_KE} for MS. The reduced matrix element is found to be
\begin{equation}
\begin{split}
\braket{(l_\a s_\a) j_\a || \, Y_\lambda \, || (l_\c s_\c) j_\c }
=& \delta_{s_\a\,s_\c} \, \sqrt{\dfrac{1}{4\pi}} \braket{l_\c 0 l_\a 0 | \lambda 0}(-)^{j_\c+s_\c}  \hat{j}_\c \hat{l}_\a \hat{l}_\c
\begin{Bmatrix}
l_\c & l_\a & \lambda \\
j_\a & j_\c & s_\c
\end{Bmatrix} \,,
\end{split}
\label{eqn:j_Yl_j_RME}
\end{equation}
which generalises Eq.~\eqref{eqn:trcg} to spherical harmonic operators of rank $\lambda$ and particles of generic spin $s$. This result can also be specialised to spin $s=1/2$ fermions using Eq.~\eqref{eqn:3j6j2}:
\begin{equation}
\begin{split}
\braket{(l_\a \frac12) j_\a || \, Y_\lambda \, || (l_\c \frac 12) j_\c }
&= (-1)^\lambda \sqrt{\dfrac{2\lambda+1}{4\pi}} \braket{j_\a\dfrac{1}{2} \lambda 0 | j_\c \dfrac{1}{2}}\pi(l_\a, \lambda, l_\c)\,.
\\ &= (-1)^{j_\a-j_\c+\lambda} \sqrt{\dfrac{2\lambda+1}{4\pi}} \dfrac{\hat{j}_\c}{\hat{j}_\a} \braket{j_\c\dfrac{1}{2} \lambda 0 | j_\a \dfrac{1}{2}}\pi(l_\a, \lambda, l_\c)\,.
\label{eqn:j_Yl_j_RME_s1h}
\end{split}
\end{equation}
The general $J$-coupled and $T$-decoupled  matrix element is then calculated as follows
\begin{equation}
\begin{split}
\braket{ab; J | \dfrac{1}{\abs{\vec{r}_1 - \vec{r}_2}} | cd; J} ={}& \sum_{m_\a \, m_\b \, m_\c \, m_\d} \braket{j_\a m_\a j_\b m_\b | JM_J}\braket{j_\c m_\c j_\d m_\d | JM_J} \braket{\a \b | \dfrac{1}{\abs{\vec{r}_1 - \vec{r}_2}} | \c \d}\\
={}& \delta_{\t_\a \t_\c} \delta_{\t_\b \t_\d} \, \hat{j}_\a \hat{j}_\b \, \sum_\lambda \dfrac{4\pi}{2\lambda+1}
 (-1)^{j_\b+j_\c+J + 2\lambda} \braket{n_\a l_\a \tau_\a, n_\b l_\b \tau_\b | \dfrac{r_<^\lambda}{r_>^{\lambda+1}} | n_\c l_\c \tau_\c, n_\d l_\d \tau_\d} \\
& \quad \times
\braket{(l_\a s_\a) j_\a || \, Y_\lambda\, || (l_\c s_\c) j_\c}
\braket{(l_\b s_\b) j_\b || \, Y_\lambda\, || (l_\d s_\d) j_\d}
\begin{Bmatrix}
j_\a & j_\b & J\\
j_\d & j_\c & \lambda
\end{Bmatrix} \, .
\end{split}
\label{eqn:Coulomb_final}
\end{equation}
A simpler relation can be found for nucleons by substituting Eq.~\eqref{eqn:j_Yl_j_RME_s1h} directly into the latter relation. One finds
\begin{equation}
\begin{split}
\braket{ab; J | \dfrac{1}{\abs{\vec{r}_1 - \vec{r}_2}} | cd; J} ={}&
\delta_{\t_\a \t_\c} \delta_{\t_\b \t_\d} \, \hat{j}_\a \hat{j}_\b \, \sum_\lambda (-1)^{j_\b+j_\c+J} \braket{n_\a l_\a \tau_\a, n_\b l_\b \tau_\b | \dfrac{r_<^\lambda}{r_>^{\lambda+1}} | n_\c l_\c \tau_\c, n_\d l_\d \tau_\d} \\
& \quad \times
\pi(l_\a, \lambda, l_\c) \pi(l_\b, \lambda, l_\d)
\braket{j_\a\dfrac{1}{2} \lambda 0 | j_\c \dfrac{1}{2}}
\braket{j_\b\dfrac{1}{2} \lambda 0 | j_\d \dfrac{1}{2}}
\begin{Bmatrix}
j_\a & j_\b & J\\
j_\d & j_\c & \lambda
\end{Bmatrix} \, ,
\end{split}
\label{eqn:Coulomb_final_s1h}
\end{equation}
which is valid for spin $s=1/2$ fermions.
The final normalized and antisymmetrized matrix elements are found substituting the latter relations into Eq.~\eqref{eqn:asme}.

\subsection{Two-body interaction}
\label{Sec:2b_force}
The matrix elements of a two-body interaction entering Eq.~\eqref{eqn:H} are intended in the laboratory frame so that they must be expressed either with respect to products of single-particle orbits or between the corresponding $J$-coupled states~\eqref{eqn:2bnorm}. Invariance under Galilean translation  and rotational symmetry make it easier to model many-body interactions in terms of degrees of freedom for the relative motion. For Nuclear physics this is normally done using relative two-body coordinates, or Jacobi coordinates for three or more nucleons.

Let define the two-body relative and center-of-mass momenta $\vec{p}$ and $\vec{P}$:
\begin{equation}
\begin{pmatrix}
\vec{p}\\
\vec{P}
\end{pmatrix}
=
\begin{pmatrix}
\frac{1}{2} & -\frac{1}{2}\\
1 & 1
\end{pmatrix}
\begin{pmatrix}
\vec{k}_1\\
\vec{k}_2
\end{pmatrix},
\label{eqn:2bcoord}
\end{equation}
where $\vec{k}_1$ and $\vec{k}_2$ represent the momenta of the two particles.
A general two-body interaction can be expanded in terms of the momenta ($\vec{k}_1$, $\vec{k}_2$) as
\begin{equation}
V = \int d\vec{k}_1d\vec{k}_2d\vec{k}'_1d\vec{k}'_2\,\ket{\vec{k}_1\vec{k}_2}\braket{\vec{k}_1\vec{k}_2 | V | \vec{k}'_1\vec{k}'_2}\bra{\vec{k}'_1\vec{k}'_2}\delta{\biggl( \sum_{i=1}^2(\vec{k}_i-\vec{k}'_i) \biggr)}.
\end{equation}
Equivalently, the expansion can be performed with respect to the momenta ($\vec{p}$, $\vec{P}$) and reads
\begin{equation}
\begin{split}
V &= \int d\vec{p}d\vec{P}d\vec{p}'d\vec{P}'\, \ket{\vec{p}\vec{P}}\braket{\vec{p}\vec{P} | V | \vec{p}'\vec{P}'} \bra{\vec{p}'\vec{P}'} \delta{(\vec{P}-\vec{P}')}\\
&= \int d\vec{p}d\vec{p}'\, \ket{\vec{p}}\braket{\vec{p} | V | \vec{p}'} \bra{\vec{p}'} \otimes \mathds{1}_{\vec{P}}\,,
\label{eqn:Vcom}
\end{split}
\end{equation}
where $\braket{\vec{p} | V | \vec{p}'}$ is independent of $\vec{P}$ for a translationally invariant interaction and the center-of-mass contribution has been isolated in the term
\begin{equation}
\mathds{1}_{\vec{P}} \equiv \int d\vec{P}\,\ket{\vec{P}}\bra{\vec{P}}.
\end{equation}
In practice, the most generic two-nucleon potential that is employed in \emph{ab initio} simulations
can be expanded on a partial-wave basis as
\begin{equation}
\braket{\vec{p} | V | \vec{p}'} \longrightarrow \braket{p(lS)j, TM_T | V | p'(l'S')j, TM_T},
\label{eqn:2bmom}
\end{equation}
where $p$, $l$, $S$, $j$, $T$ ad $M_T$ represent respectively the modulus of $\vec{p}$, the relative orbital angular momentum, the total spin, the relative total angular momentum, the total isospin and the projection of the total isospin along the quantization axis of the two-body system. The projection $m$ of the relative total angular momentum along the quantization axis does not appear in Eq.~\eqref{eqn:2bmom} in virtue of spherical symmetry. Note that Eq.~\eqref{eqn:2bmom} is also diagonal in $M_T$ in virtue of the conservation of electric charge. For most of the nuclear forces that are employed in nuclear physics, including earlier high-precision potentials~\cite{AV18,MachleidtBonn} and those based on Chiral Effective Field Theory (ChEFT)~\cite{RevModPhysChEFT,Machleidt2011PhysRep,PiarulliFornt}%Soma2020LNL,HutherInt,NNLOsat,MagicInt}
, the matrix elements are also diagonal in $T$.

As in the case of the center-of-mass corrections discussed above, the goal is to calculate the $J$-coupled $T$-decoupled matrix elements:
\begin{equation}
\braket{ab; JM_J | V | cd; JM_J} \equiv \braket{n_\a l_\a j_\a \tau_\a, n_\b l_\b j_\b \tau_\b; JM_J | V | n_\c l_\c j_\c \tau_\c, n_\d l_\d j_\d \tau_\d; JM_J}.
\label{eqn:2blab}
\end{equation}
Differently from what seen in the previous sections, the matrix elements of the two-body interaction discussed here are not calculated from scratch since for most applications they are provided in a partial wave basis. Hence, one need to perform a change of basis from Eq.~\eqref{eqn:2bmom} to Eq.~\eqref{eqn:2blab}. This change of basis is the topic of the rest of this section.
We demonstrate it by focusing on the sole isospin-independent part of the two-body states, $\ket{n_\a l_\a j_\a, n_\b l_\b j_\b; JM_J}$, since the isospin couplings enter only the final part of the transformation. In the following, the various intermediate transformations required are listed.
\vspace{0.2cm}
\\
\underline{STEP 1}
\vspace{-0.1cm}
\begin{equation}
\ket{n_\a l_\a j_\a, n_\b l_\b j_\b; JM_J} \to
\ket{n_\a n_\b [ (l_\a l_\b)\lambda, (s_\a s_\b)S] JM_J}
\end{equation}
A change from $J$- to $ls$-coupling is operated:
\begin{equation}
\ket{n_\a l_\a j_\a, n_\b l_\b j_\b; JM_J} = 
\sum_{\lambda \, S}\hat{j}_\a\hat{j}_\b\hat{\lambda}\hat{S}
\begin{Bmatrix}
l_\a & s_\a & j_\a\\
l_\b & s_\b & j_\b\\
\lambda & S & J
\end{Bmatrix}
\ket{n_\a n_\b [(l_\a l_\b)\lambda, (s_\a s_\b)S] JM_J},
\end{equation}
where $\lambda$ is the orbital angular momentum that couples $l_\a$ and $l_\b$, while $S$ is the total spin that couples $s_\a$ and $s_\b$.
\vspace{0.2cm}
\\
\underline{STEP 2}
\vspace{-0.1cm}
\begin{equation}
\ket{n_\a n_\b [(l_\a l_\b)\lambda, S] JM_J} \to
\ket{pP [(lL)\lambda, S] JM_J}
\end{equation}
Two different transformations are performed here with one step: the change of coordinate systems from single-particle to relative and center-of-mass coordinates represented in Eq.~\eqref{eqn:2bcoord}, and the integration of the single-particle momenta, to give a state with single-particle quantum numbers $n_\a$ and $n_\b$:
\begin{equation}
\ket{n_\a n_\b (l_\a l_\b)\lambda} = \sum_{l\,L}\int dpdP\,p^2P^2\,\braket{pP(lL)\lambda|n_\a n_\b (l_\a l_\b) \lambda}_{\tau_\a \tau_\b} \ket{pP (lL)\lambda},
\label{eqn:2bwceq}
\end{equation}
where $l$ and $L$ are the relative and center-of-mass orbital angular momenta of the two-body system. The intermediate bracket represents the WC bracket discussed in Ref.~\cite{WC} and~\ref{coeff_WC}. In the specific case of the transformation between momenta given by Eq.~\eqref{eqn:2bcoord}, the bracket appearing in Eq.~\eqref{eqn:2bwceq} is associated with the inverse relation:
\begin{equation}
\begin{pmatrix}
\vec{k}_1\\
\vec{k}_2
\end{pmatrix}
=
\begin{pmatrix}
1 & \frac{1}{2}\\
-1 & \frac{1}{2}
\end{pmatrix}
\begin{pmatrix}
\vec{p}\\
\vec{P}
\end{pmatrix} \, ,
\end{equation}
so that the WC coefficients (analogous to Eq.~\eqref{eqn:wctransf}) read
\begin{equation}
\begin{pmatrix}
s_1 & t_1\\
s_2 & t_2
\end{pmatrix}_{MS}
=
\begin{pmatrix}
1 & \frac{1}{2}\\
-1 & \frac{1}{2}
\end{pmatrix}.
\label{eqn:2bwc}
\end{equation}
\vspace{0.2cm}
\\
\underline{STEP 3}
\vspace{-0.1cm}
\begin{equation}
\ket{pP [(lL)\lambda, S] JM_J} \to
\ket{pP[(lS)j,L]JM_J}
\end{equation}
A change of coupling is required to couple the relative angular momentum to the total spin:
\begin{equation}
\ket{pP [(lL)\lambda, S] JM_J} = \sum_j \hat{\lambda}\hat{j}(-1)^{j+L+S+\lambda}
\begin{Bmatrix}
j & L & J\\
\lambda & S & l
\end{Bmatrix}
\ket{pP[(lS)j,L]JM_J},
\label{eqn:sumj}
\end{equation}
where $j$ is the relative total angular momentum of the two particles.
\vspace{0.2cm}
\\
\underline{STEP 4}
\vspace{-0.1cm}
\begin{equation}
\ket{pP[(lS)j,L]JM_J} \to
\ket{p(lS)jm, PLm_L}
\end{equation}
A decoupling of the center-of-mass angular momentum $L$ from the total $J$ is used to exploit the independency of the two-body potential from the center-of-mass variables
\begin{equation}
\ket{pP[(lS)j,L]JM_J} = \sum_{m\,m_L} \braket{jmLm_L | JM_J}
\ket{p(lS)jm, PLm_L},
\label{eqn:pPJ}
\end{equation}
where $m$ and $m_L$ represent respectively the projection of $j$ and $L$ along the quantization axis.

The two-body potential is independent of the center-of-mass variables because of traslational invariance, see Eq.~\eqref{eqn:Vcom}. Furthermore, spherical symmetry implies that Eq.~\eqref{eqn:2bmom} it is independent of $m$ and couples only states with the same total angular \hbox{momentum $j$:}
\begin{equation}
\begin{gathered}
\braket{p(lS)jm, PLm_L | V | p'(l'S')j'm', P'L'm'_L} = \dfrac{\delta(P-P')}{P^2}\delta_{LL'}\delta_{m_Lm'_L} \delta_{jj'}\delta_{mm'} \braket{p(lS)j| V | p'(l'S')j}.
\end{gathered}
\end{equation}
Hence, set aside the isospin degrees of freedom, the expectation value of the two-body potential over the states of Eq.~\eqref{eqn:pPJ} leads to
\begin{equation}
\begin{split}
&\braket{pP[(lS)j,L]JM_J | V | p'P'[(l'S')j',L']J'M'_J}\\
&\qquad\qquad\qquad=\sum_{m\,m_L\,m'\,m'_L} \braket{jmLm_L | JM_J} \braket{j'm'L'm'_L | J'M'_J} \braket{p(lS)jm, PLm_L | V | p'(l'S')j'm', P'L'm'_L}\\
&\qquad\qquad\qquad= \dfrac{\delta(P-P')}{P^2}\delta_{LL'}\delta_{jj'}
\delta_{JJ'}\delta_{M_JM'_J}
\braket{p(lS)j| V | p'(l'S')j},
\end{split}
\end{equation}
where at the last step the orthonormality of the Clebsch-Gordan coefficients (Eq.~\eqref{eqn:CGort}) has been used.
At a last step we still need to recouple the isospin quantum numbers according to Eq.~\eqref{eqn:2bmom}. 
\vspace{0.2cm}
\\
\underline{STEP 5}
\vspace{-0.1cm}
\begin{equation}
\ket{t_\a\tau_\a, t_\b\tau_\b} \to
\ket{(t_\a t_\b)TM_T}
\end{equation}
The total isospin is recoupled
\begin{equation}
\ket{t_\a\tau_\a, t_\b\tau_\b} = \sum_{TM_T}\braket{t_\a \tau_\a t_\b \tau_\b | TM_T} \ket{TM_T}
,
\end{equation}
and the only value of $M_T$ that is giving a non-vanishing contribution to the summation is given by the Clebsch-Gordan coefficient through Eq.~\eqref{eqn:msumrule}: $M_T=\tau_\a+\tau_\b$.
After gathering all the above transformations, the complete expression for the change of basis reads
\begin{equation}
\begin{split}
\ket{n_\a l_\a j_\a \tau_\a, n_\b l_\b j_\b \tau_\b; JM_J} =& \sum_{\lambda\, S\,l\,L\,j\,T\,M_T}\int dpdP\,p^2P^2\,(-1)^{j+L+S+\lambda}\hat{j}_\a\hat{j}_\b\hat{\lambda}^2\hat{S}\hat{j}
\begin{Bmatrix}
l_\a & s_\a & j_\a\\
l_\b & s_\b & j_\b\\
\lambda & S & J
\end{Bmatrix}\\
&\quad\times
\braket{pP(lL)\lambda|n_\a n_\b(l_\a l_\b)\lambda}_{\tau_\a \tau_\b}
\begin{Bmatrix}
j & L & J\\
\lambda & S & l
\end{Bmatrix}
\braket{t_\a \tau_\a t_\b \tau_\b | TM_T}
\ket{p(lS)j, TM_T}.
\end{split}
\label{eqn:Tc2b}
\end{equation}
Let us now define the collective index\footnote{To keep with standard notation in the literature, the Greek letter $\alpha$ is used to label the angular and isospin quantum numbers of the relative motion, and similarly for three-body relative states in Eq.~\eqref{eqn:3bmomc}. This is not to be confused with the labels for single-particle states~\eqref{eq:label_sp_all}. The two cases should be distinguished easily from the context.}:
\begin{equation}
\alpha \equiv \{l, S, j, T, M_T\}.
\label{eqn:twobas}
\end{equation}
The coefficient that expresses the probability amplitude for the change of basis presented in this section is called \emph{$T$-coefficient} and can be obtained multiplying Eq.~\eqref{eqn:Tc2b} by the state $\ket{p\alpha, PL}$
\begin{equation}
\begin{gathered}
T^{2B}
%= T_{p\alpha, PL}^{ab; JM_J}
\equiv 
\braket{p\alpha, PL | ab; JM_J}_N =
\dfrac{1}{\sqrt{1+\delta_{ab}}} \sum_\lambda(-1)^{j+L+S+\lambda}\hat{j}_\a\hat{j}_\b\hat{\lambda}^2\hat{S}\hat{j}\\
\times
\begin{Bmatrix}
l_\a & s_\a & j_\a\\
l_\b & s_\b & j_\b\\
\lambda & S & J
\end{Bmatrix}
\braket{pP(lL)\lambda|n_\a n_\b(l_\a l_\b)\lambda}_{\tau_\a \tau_\b}
\begin{Bmatrix}
j & L & J\\
\lambda & S & l
\end{Bmatrix}
\braket{t_\a \tau_\a t_\b \tau_\b | TM_T},
\end{gathered}
\label{eqn:tlst_N}
\end{equation}
where the term $1/\sqrt{1+\delta_{ab}}$ is the normalization factor for the two-body $J$-coupled state shown introduced in Eq.~\eqref{eqn:2bnorm}.
The final matrix elements must be antisymmetrized using the two-body antisymmetrization operator,
\begin{equation}
\ket{\a\b}_A \equiv \sqrt{2!}\mathcal{A}\ket{\a\b} = \dfrac{1}{\sqrt{2}}(\ket{\a\b} - \ket{\b\a}).
\label{eq:twob_antisymm}
\end{equation}
The transposition operator is defined by $T_{12}\ket{\a\b} = \ket{\b\a}$. When acting on the coupled state $\ket{ab; JM_J}$ it brings in a phase $(-1)^{j_\a+j_\b-J}$ arising from the Clebsch-Gordan coefficient needed to uncouple the angular momenta. Similarly, inverting the states $a$ and $b$ in Eq.~\eqref{eqn:twobas} introduces a phase $(-1)^{t_\a+t_\b-T}$ from the isospin coupling, a phase $(-1)^{l_\a+l_\b+s_\a+s_\b+j_\a+j_\b+\lambda+S+J}$ from the $9j$ symbol and a phase $(-1)^{\lambda-L}$ from the WC bracket. The latter can be obtained from the symmetry relations listed in~\cite{WC}. 
In the assumption of spin and isospin $s=t=1/2$ and using parity conservation, all the above phases simplify to $(-1)^{l+S+T}$, 
so that the $T$-coefficient for the antisymmetrized two-body state is given by
\begin{equation}
\begin{split}
\braket{p\alpha, PL | ab; JM_J}_{A\,N} 
=& \bra{p\alpha, PL} \dfrac{1}{\sqrt{2}}[\ket{ab; JM_J} - (-1)^{j_\a+j_\b-J} \ket{ba; JM_J}] \\
=& \bra{p\alpha, PL} \dfrac{1}{\sqrt{2}}[\ket{ab; JM_J} - (-1)^{l+S+T} \ket{ab; JM_J}] \\
=& \sqrt{2} \, \pi(l,S,T) \braket{p\alpha, PL | ab; JM_J}_N.
\end{split}
\label{eqn:tlst_AN}
\end{equation}
Finally, the complete formula for change of basis for two-body matrix elements reads
\begin{equation}
\begin{split}
\braket{ab; JM_J | V | cd; JM_J}_{A\,N} =& \int dP\,P^2  \sum_L \,\,\int dpdp'\,p^2p'^2\, \\
&\quad\times 
\sum_{\alpha\,\alpha'}\braket{ab; JM_J | p\alpha, PL}_{N} \braket{p\alpha | V | p'\alpha'}_A \braket{p'\alpha', PL | cd; JM_J}_{N} \, ,
\end{split}
\end{equation}
where the T-coefficients are from Eqs.~\eqref{eqn:tlst_N} and 
$\braket{p\alpha | V | p'\alpha'}_A  \equiv 2 \, \pi(l,S,T) \, \pi(l',S',T') \braket{p\alpha | V | p'\alpha'}$ is the antysymmetrized version of Eqs.~\eqref{eqn:2bmom}.
Because of antisymmetrization, the sums over $\a$ and $\a'$ are limited to odd values of $l+S+T$ and~$l'+S'+T'$.

The equations discussed above can be applied to MS wave functions. CS wave functions can be employed as well, but they need to be first Hankel-transformed to momentum space (Eq.~\eqref{eqn:momtor}), or equivalently the MS potential must be transformed to CS through a Hankel transform:
\begin{equation}
\braket{r\alpha | V | r'\alpha'} = \dfrac{2}{\pi} \,i^{l'-l}\,\int_0^{+\infty} dpdp'\,p^2p'^2\,j_l(pr)\,j_{l'}(p'r')\,\braket{p\alpha | V | p'\alpha'}\,,
\label{eq:Hankel_V2b}
\end{equation}
where the phase $i^{l'-l}$ results from Eq.~\eqref{eqn:Hank_phase} and it is real for parity conserving interactions (for which $l-l'$ is an even number). Eq.~\eqref{eq:Hankel_V2b}
assumes that the new relative and center-of-mass variables ($\vec{r}$, $\vec{R}$) in CS are conjugate to the momenta ($\vec{p}$, $\vec{P}$) of Eq.~\eqref{eqn:2bcoord}. They are related to the single-particle positions through the following change of reference system:
\begin{equation}
\begin{pmatrix}
\vec{r}\\
\vec{R}
\end{pmatrix}
=
\begin{pmatrix}
1 & -1\\
\frac{1}{2} & \frac{1}{2}
\end{pmatrix}
\begin{pmatrix}
\vec{r}_1\\
\vec{r}_2
\end{pmatrix}\,.
\label{eqn:2bcoord_rR}
\end{equation}
The computation of two-body matrix elements in CS follows exactly the same steps discussed above with this new reference system. The only exception that the WC coefficients for the transformation in Eq.~\eqref{eqn:2bwceq} are  substituted by
\begin{equation}
\begin{pmatrix}
s_1 & t_1\\
s_2 & t_2
\end{pmatrix}_{CS}
=
\begin{pmatrix}
\frac{1}{2} & 1\\
-\frac{1}{2} & 1
\end{pmatrix} \, ,
\end{equation}
which follow directly from inverting Eq.~\eqref{eqn:2bcoord_rR}.

\section{Three-body matrix elements}
\label{3b}

In this section we provide a transformation for a change of basis among three-particle states, similarly to the one of Sec.~\ref{Sec:2b_force}, that applies to generic spherical bases.
Let be defined the Jacobi momentum coordinates ($\vec{Q}_{cm}$, $\vec{p}$, $\vec{q}$) as
\begin{equation}
\begin{pmatrix}
\vec{Q}_{cm}\\
\vec{p}\\
\vec{q}
\end{pmatrix}
 = 
 \begin{pmatrix}
1 & 1 & 1\\
\frac{1}{2} & -\frac{1}{2} & 0\\
-\frac{1}{3} & -\frac{1}{3} & \frac{2}{3}
\end{pmatrix}
\begin{pmatrix}
\vec{k}_1\\
\vec{k}_2\\
\vec{k}_3
\end{pmatrix}
\label{eqn:momcoord}
\end{equation}
that represent the center of mass and Jacobi momenta as a function of the single-particle momenta ($\vec{k}_1$, $\vec{k}_2$, $\vec{k}_3$). A general translationally-invariant three-body interaction can be expanded with respect to the momenta $(\vec{k}_1, \vec{k}_2, \vec{k}_3)$,
\begin{equation}
W = \int d\vec{k}_1d\vec{k}_2d\vec{k}_3 \, d\vec{k}'_1d\vec{k}'_2d\vec{k}'_3 \, \ket{\vec{k}_1\vec{k}_2\vec{k}_3} \braket{\vec{k}_1\vec{k}_2\vec{k}_3| W |\vec{k}'_1\vec{k}'_2\vec{k}'_3} \bra{\vec{k}'_1\vec{k}'_2\vec{k}'_3} \, \delta\biggl( \sum_{i=1}^3(\vec{k}_i-\vec{k}'_i)\biggr),
\end{equation}
or it can be equivalently expressed in terms of the momenta ($\vec{Q}_{cm}$, $\vec{p}$, $\vec{q}$):
\begin{equation}
\begin{split}
W &= \int d\vec{Q}_{cm}d\vec{p}d\vec{q}\,d\vec{Q}'_{cm}d\vec{p}'d\vec{q}'\, \ket{\vec{Q}_{cm}\vec{p}\vec{q}} \braket{\vec{Q}_{cm}\vec{p}\vec{q}| W |\vec{Q}'_{cm}\vec{p}'\vec{q}'} \bra{\vec{Q}'_{cm}\vec{p}'\vec{q}'} \, \delta(\vec{Q}_{cm} - \vec{Q}'_{cm})\\
&= \int d\vec{p} d\vec{q} d\vec{p}' d\vec{q}' \, \ket{\vec{p}\vec{q}}\braket{\vec{p}\vec{q} | W | \vec{p}'\vec{q}'} \bra{\vec{p}'\vec{q}'} \otimes \mathds{1}_{\vec{Q}_{cm}},
\label{eqn:cm3b}
\end{split}
\end{equation}
where the center-of-mass contribution has been isolated in the term
\begin{equation}
\mathds{1}_{\vec{Q}_{cm}} = \int d\vec{Q}_{cm}\,\ket{\vec{Q}_{cm}}\bra{\vec{Q}_{cm}}.
\end{equation}
In practice, three-body nuclear interactions that are expressed in terms of Jacobi coordinates, Eq.~\eqref{eqn:cm3b}, are often given in the following $J$- and $T$-coupled partial-wave basis~\cite{glocklefew,Hebeler2015}
\begin{equation}
\begin{split}
 \braket{pq\alpha | W | p'q'\alpha'} \equiv& \braket{pq,[(LS)J,(ls)j]\mathcal{J}M_\mathcal{J},(Tt)\mathcal{T}M_\mathcal{T} | W | p'q',[(L'S')J',(l's')j']\mathcal{J}'M'_\mathcal{J},(T't')\mathcal{T}'M'_\mathcal{T}}
%\\ =& \int_{many\,angles}  \sum_{many\,ms}  \sum_{many\,\t s}  Y()\, Y()\, \cdots CG\; CG \, \cdots \braket{\vec{p}\vec{q} | W | \vec{p}'\vec{q}'}
\,
\label{eqn:3Bmom}
\end{split}
\end{equation}
where the collective index $\alpha$ carries the angular and isospin degrees of freedom similarly to Eq.~\eqref{eqn:twobas} and it is defined as
\begin{equation}
\alpha \equiv \{[(LS)J, (ls)j]\mathcal{J}M_\mathcal{J},(Tt)\mathcal{T}M_\mathcal{T}\} \,.
\label{eqn:3bmomc}
\end{equation}
The quantum numbers $L$, $S$, $J$ and $T$ represent respectively the orbital angular momentum, spin, total angular momentum and isospin of the relative motion of particles 1 and 2 (associate to momentum $\vec{p}$). $l$, $s$, $j$ and $t$ characterize the orbital angular momentum, the spin, the total angular momentum and the isospin of particle 3 relative to the center-of-mass of the other pair of particles (variable $\vec{q}$). The quantum numbers $\mathcal{J}$ and $\mathcal{T}$ represent the intrinsic total angular momentum and the total isospin of the three-body system, while $M_{\mathcal{J}}$ and $M_{\mathcal{T}}$ are their projections along the quantization axis. The parity of state~\eqref{eqn:3bmomc} under spatial inversion is given by $\Pi = (-1)^{L+l}$.

For three-body interactions the antisymmetrization is more cumbersome than the two-body case. For such reason, it is convenient to perform the antisymmetrization directly in the partial-wave basis before projecting the three-body matrix elements in Eq.~\eqref{eqn:3Bmom} to spherical single-particle states.

\subsection{Antisymmetrization}
Let be defined the \emph{cyclic (anti-cyclic) permutation operators} $P_{123}$ ($P_{132}$) as the operators that permute cyclically (anticyclically) a set of 3 particles:
\begin{subequations}
\begin{align}
P_{123}\ket{\a\b\c} &= \ket{\c\a\b},\\
P_{132}\ket{\a\b\c} &= \ket{\b\c\a}.
\end{align}
\end{subequations}
Likewise, the \emph{transposition operator} $T_{ij}$ exchanges the two particles $i$ and $j$ (with $i\neq j\in\{1,2,3\}$).
\begin{subequations}
\begin{align}
T_{12}\ket{\a\b\c} = \ket{\b\a\c},\\
T_{23}\ket{\a\b\c} = \ket{\a\c\b},\\
T_{13}\ket{\a\b\c} = \ket{\c\b\a}.
\end{align}
\end{subequations}
The two operators are connected through the following identities
\begin{align}
P_{123} =& T_{12}T_{23} = T_{23}T_{13}, \\
P_{132} =& T_{13}T_{23} = T_{23}T_{12}.
\end{align}
and the three-body \emph{antisymmetrization operator} can be written as
\begin{equation}
\mathcal{A} \equiv \dfrac{1}{6}(1 - T_{12} - T_{13} - T_{23} + P_{123} + P_{132}) = \dfrac{1}{6}(1-T_{12})(1+P_{123}+P_{132}).
\label{eqn:A3N}
\end{equation}

Because of Eq.~\eqref{eqn:momcoord} and the coupling convention~\eqref{eqn:3Bmom}, the state $\ket{pq\alpha}$ takes the following phase under the inversion of the first two particles
\begin{equation}
T_{12}\ket{pq\alpha} = (-)^{L+S+T}\ket{pq\alpha} \, ,
\end{equation}
which is obtained following the same expansion of Sec.~\ref{Sec:2b_force} and the arguments given below Eq.~\eqref{eq:twob_antisymm}.
The state $\ket{pq\alpha}$ is said to be \emph{partially antisymmetrized} with respect to particles 1 and 2 if
\begin{equation}
(-)^{L+S+T} = -1.
\end{equation}
so that $T_{12}\ket{pq\alpha} = -\ket{pq\alpha}$.
In the following, $\ket{pq\alpha}$ will always be taken to be partially antisymmetric. With this assumption, the matrix elements of the operator in Eq.~\eqref{eqn:A3N} reduce to
\begin{equation}
\begin{split}
\braket{pq\alpha|\mathcal{A}|p'q'\alpha'} =& \dfrac{1}{3}\braket{pq\alpha|1+P_{123}+P_{132}|p'q'\alpha'} \\
=& \dfrac{1}{3}\braket{pq\alpha|1 + T_{12}T_{23} + T_{23}T_{12}|p'q'\alpha'} \\
=& \dfrac{1}{3}\braket{pq\alpha|1 - 2T_{23}|p'q'\alpha'} \, .
\end{split}
\label{eqn:Aant}
\end{equation}

Three-body interactions from ChEFT can always be decomposed in terms of their three \emph{Faddeev components} \cite{glocklefew}
\begin{equation}
W = \sum_{i=1}^3V^{(i)} \, ,
\end{equation}
that are related through cyclic permutations of the three particles and each component $V^{(i)}$ is symmetric under exchange of the two particles $j,k \neq i \in \{1,2,3\}$.  Hence, the interaction can be written in terms of any of the components and appropriate permutation or transposition operators:
\begin{equation}
W = V^{(1)} + V^{(2)} + V^{(3)} = V^{(1)} + P_{123}V^{(1)}P_{123}^{-1} + P_{132}V^{(1)}P_{132}^{-1} = V^{(1)} + T_{12}T_{23}V^{(1)}T_{13}T_{23} + T_{13}T_{23}V^{(1)}T_{12}T_{23}.
\end{equation}
It follows that the product of the 3N interaction operator and the antisymmetrization operator is given by
\begin{equation}
\begin{split}
W \mathcal{A} &= \dfrac{1}{3}(V^{(1)} + P_{123}V^{(1)}P_{123}^{-1} + P_{132}V^{(1)}P_{132}^{-1})(1 + P_{123} + P_{132})\\
&= \dfrac{1}{3}(1+P_{123}+P_{132}) V^{(1)} (1+P_{123}+P_{132}).
\label{eqn:AW}
\end{split}
\end{equation}
One can easily prove that  $W$ and $\mathcal{A}$ commute, so that $W\mathcal{A} = \mathcal{A}\,W$.
When acting on partially antisymmetric states, the matrix element of Eq.~\eqref{eqn:Aant} reads
\begin{equation}
\begin{split}
\braket{pq\alpha|W|p'''q'''\alpha'''}_A ={}& \dfrac{1}{3} \sum_{\alpha' \, \alpha''}\int dp'\,dq'\,dp''\,dq''\,p'^2q'^2p''^2q''^2
\braket{pq\alpha|1-2T_{23}|p'q'\alpha'}\\
& \qquad \qquad \times\braket{p'q'\alpha'|V^{(1)}|p''q''\alpha''}\braket{p''q''\alpha''|1-2T_{23}|p'''q'''\alpha'''},
\end{split}
\label{eq:3Bmon_Antisymm}
\end{equation}
with $\braket{pq\alpha|W|p'''q'''\alpha'''}_A \equiv \braket{pq\alpha|W\mathcal{A}|p'''q'''\alpha'''}$.
The structure of the radial part of the transposition operator is discussed in \ref{T23} and it implies the evaluation of the three-body potential for several instances of momenta $p$ and $q$. For practical implementation in  MS, the three-body matrix elements are often given on a pre-defined mesh of momenta. Thus, an interpolation of the potential is necessary and it can be conveniently performed with the \emph{modified cubic splines} method introduced in Ref.~\cite{glocklespline} to tackle integrals with moving singularities.

\subsection{Regularization}
ChEFT interactions need the application of a regulator in order to separate low-energy and high-energy momenta scales.
In the case of two-nucleon interactions such regulator is typically already included at the level of in the momentum-space matrix elements~\eqref{eqn:2bmom}.
For three-body interactions it is not the case, and the explicit application of a regulator is discussed here. In the following, a non-local MS regulator is used, while a complete explanation of the various type of regulators (non-local, semi-local, local) either in momentum or coordinate space can be found in \cite{Hebeler2021}.
Let define the non-local momentum space regulator as
\begin{equation}
f_\Lambda(\vec{p}, \vec{q}) \equiv \exp{\biggl[-\biggl(\dfrac{\vec{p}^2 + 3/4 \vec{q}^2}{\Lambda^2}\biggr)^n\biggr]} = f_\Lambda(p, q),
\label{eqn:regg}
\end{equation}
where the exponent $n$ typically assumes values in the range from $2$ to $4$.
This formulation of the regulator is particularly convenient because it allows to regularize directly the MS matrix elements in Eq.~\eqref{eqn:3Bmom} through a multiplication with Eq.~\eqref{eqn:regg}:
\begin{equation}
\braket{pq\alpha | V^{(i), \text{reg}} | p'q'\alpha'} = f_\Lambda(p, q)\braket{pq\alpha | V^{(i)} | p'q'\alpha'}f_\Lambda(p', q').
\end{equation}
The structure of the antisymmetrization operator is such that it commutes with respect to any transposition or permutation of the 3 particles:
\begin{equation}
\vec{p}^2 + \dfrac{3}{4}\vec{q}^2 = \dfrac{1}{6}[ (\vec{k}_2-\vec{k}_1)^2 + (\vec{k}_3-\vec{k}_2)^2 + (\vec{k}_1-\vec{k}_3)^2 ].
\end{equation}
Since the antisymmetrization operator can be written as a function of a transposition operator, the regularization can be equivalently performed on antisymmetrized matrix elements
\begin{equation}
\braket{pq\alpha | V^{(i), \text{reg}} | p'q'\alpha'}_A = f_\Lambda(p, q)\braket{pq\alpha | V^{(i)} | p'q'\alpha'}_Af_\Lambda(p', q').
\end{equation}

\subsection{\texorpdfstring{$T$}{T}-coefficient}
This section derives the complete change of basis between the Jacobi MS state that appears in Eq.~\eqref{eqn:3Bmom} and a single-particle $J$-coupled $T$-decoupled three-body state
\begin{equation}
\ket{[(n_\a l_\a j_\a \tau_\a, n_\b l_\b j_\b \tau_\b) J_{12}, n_\c l_\c j_\c \tau_\c]J_{tot}M_{J_{tot}}}
\end{equation}
for a generic spherical basis.
Let first of all factorize the isospin-dependent part
\begin{equation}
\begin{multlined}
\ket{[(n_\a l_\a j_\a \tau_\a, n_\b l_\b j_\b \tau_\b) J_{12}, n_\c l_\c j_\c \tau_\c]J_{tot}M_{J_{tot}}}\\ = \ket{[(n_\a l_\a j_\a, n_\b l_\b j_\b) J_{12}, n_\c l_\c j_\c]J_{tot}M_{J_{tot}}}\otimes\ket{t_\a \tau_\a, t_\b \tau_\b, t_\c \tau_\c}
\label{eqn:3bbas}
\end{multlined}
\end{equation}
and consider only the isospin-independent term. In the following, various intermediate transformations between states are performed.
\vspace{0.2cm}
\\
\underline{STEP 1}
\vspace{-0.6cm}
\begin{equation}
\begin{split}
&\ket{ \{ [ n_\a(l_\a s_\a)j_\a, n_\b(l_\b s_\b)j_\b]J_{12}, n_\c(l_\c s_\c)j_\c \}J_{tot}M_{J_{tot}}}\\
\to
&\ket{ \{ [ n_\a n_\b (l_\a l_\b)\lambda,(s_\a s_\b)S]J_{12}, n_\c(l_\c s_\c)j_\c \}J_{tot}M_{J_{tot}}}
\end{split}
\end{equation}
The $J$-coupling of particles 1 and 2 is turned into an $ls$-coupling\footnote{Here and in the following sections, intermediate quantum number that do not enter explicitly in each transformation will not always be shown. Their implicit presence should be clear from the states in each equation.}:
\begin{equation}
\ket{[(l_\a s_\a)j_\a, (l_\b s_\b)j_\b] J_{12}} = \sum_{\lambda S}\hat{j}_\a\hat{j}_\b\hat{\lambda}\hat{S}
\begin{Bmatrix}
l_\a & s_\a & j_\a\\
l_\b & s_\b & j_\b\\
\lambda & S & J_{12}
\end{Bmatrix}
\ket{[(l_\a l_\b)\lambda\, (s_\a s_\b)S]J_{12}}.
\end{equation}
\vspace{0.2cm}
\\
\underline{STEP 2}
\vspace{-0.6cm}
\begin{equation}
\begin{split}
&\ket{ \{ [ n_\a n_\b (l_\a l_\b)\lambda,(s_\a s_\b)S]J_{12}, n_\c(l_\c s_\c)j_\c \}J_{tot}M_{J_{tot}}}\\
\to
&\ket{ \{ [ Pp(L_{P}L)\lambda,(s_\a s_\b)S]J_{12}, n_\c(l_\c s_\c)j_\c \}J_{tot}M_{J_{tot}}}
\end{split}
\end{equation}
A change of reference system is performed for particles 1 and 2 into their relative and center-of-mass frame, while at the same time the single-particle momenta of particle 1 and 2 are integrated
\begin{equation}
\ket{n_\a n_\b (l_\a l_\b)\lambda} = \int dPdp\,P^2p^2 \sum_{L_PL}\braket{Pp(L_PL) \lambda | n_\a n_\b (l_\a l_\b) \lambda}^{(a)}_{\t_\a \t_\b} \ket{PL_P\,pL,\, \lambda}.
\end{equation}
The WC bracket is associated to the transformation
\begin{equation}
\begin{pmatrix}
\vec{k}_1\\
\vec{k}_2
\end{pmatrix}
=
\begin{pmatrix}
\frac{1}{2} & 1\\
\frac{1}{2} & -1
\end{pmatrix}
\begin{pmatrix}
\vec{P}\\
\vec{p}
\end{pmatrix}
\label{eqn:wca}
\end{equation}
and the associated matrix of WC coefficients is
\begin{equation}
\begin{pmatrix}
s_1 & t_1\\
s_2 & t_2
\end{pmatrix}_{MS}^a
=
\begin{pmatrix}
\frac{1}{2} & 1\\
\frac{1}{2} & -1
\end{pmatrix},
\end{equation}
where the superscript $a$ is used to distinguish the WC matrix from the one entering at step 5 below.
\vspace{0.2cm}
\\
\underline{STEP 3}
\vspace{-0.4cm}
\begin{equation}
\begin{split}
&\ket{ \{ [ Pp(L_{P}L)\lambda,(s_\a s_\b)S]J_{12}, n_\c(l_\c s_\c)j_\c \}J_{tot}M_{J_{tot}}}\\
\to
&\ket{ \{ [ PL_P, p(LS)J]J_{12}, n_\c(l_\c s_\c)j_\c \}J_{tot}M_{J_{tot}}}
\end{split}
\end{equation}
For particles 1 and 2, the angular momenta $L$, $L_P$  and $S$ are recoupled to have the \emph{total relative} angular momentum $J$ as the intermediate quantum number:
\begin{equation}
\begin{split}
\ket{[(L_PL)\lambda, S]J_{12}} &= \sum_J \braket{[L_P,(LS)J]J_{12} | [(L_PL)\lambda, S]J_{12}} \ket{[L_P,(LS)J]J_{12}}\\
&= \sum_J (-)^{L_P+L+S+J_{12}} \hat{\lambda} \hat{J}
\begin{Bmatrix}
L_P & L & \lambda\\
S & J_{12} & J
\end{Bmatrix}
\ket{[L_P,(LS)J]J_{12}}.
\end{split}
\end{equation}
\vspace{0.2cm}
\\
\underline{STEP 4}
\vspace{-0.4cm}
\begin{equation}
\begin{split}
&\ket{ \{ [ PL_P, p(LS)J]J_{12}, n_\c(l_\c s_\c)j_\c \}J_{tot}M_{J_{tot}}}\\
\to
&\ket{ \{ Pn_\c(L_Pl_\c)\Lambda, p[(LS)J,s_\c]X \}J_{tot}M_{J_{tot}}}
\end{split}
\end{equation}
A $9j$ symbol is used to change the structure of the coupling scheme of $J_{tot}$:
\begin{equation}
\begin{split}
&\ket{\{[L_P,(LS)J]J_{12}, (l_\c s_\c)j_\c\}J_{tot}} \\
&\qquad \qquad \qquad = \sum_{\Lambda X} \braket{ \{ (L_Pl_\c)\Lambda, [(LS)J,s_\c]X \}J_{tot} | \{[L_P,(LS)J]J_{12}, (l_\c s_\c)j_\c\}J_{tot} }
\ket{\{ (L_Pl_\c)\Lambda, [(LS)J,s_\c]X \}J_{tot}}\\
&\qquad \qquad \qquad = \sum_{\Lambda X} \hat{J}_{12} \hat{j}_\c \hat{\Lambda} \hat{X}
\begin{Bmatrix}
L_P & J & J_{12}\\
l_\c & s_\c & j_\c\\
\Lambda & X & J_{tot}
\end{Bmatrix}
\ket{\{ (L_Pl_\c)\Lambda, [(LS)J,s_\c]X \}J_{tot}}\\
&\qquad \qquad \qquad = \sum_{\Lambda X} \hat{J}_{12} \hat{j}_\c \hat{\Lambda} \hat{X}
\begin{Bmatrix}
l_\c & L_P & \Lambda\\
j_\c & J_{12} & J_{tot}\\
s_\c & J & X
\end{Bmatrix}
\ket{\{ (L_Pl_\c)\Lambda, [(LS)J,s_\c]X \}J_{tot}},
\end{split}
\end{equation}
where at the last line the properties of invariance under exchange of rows and columns and swap of rows of the 9$j$-symbols have been used.
\vspace{0.2cm}
\\
\underline{STEP 5}
\vspace{-0.4cm}
\begin{equation}
\begin{split}
&\ket{ \{ Pn_\c(L_Pl_\c)\Lambda, p[(LS)J,s_\c]X \}J_{tot}M_{J_{tot}}}\\
\to
&\ket{ \{ Q_{cm}q(l_{cm}l)\Lambda, p[(LS)J,s_\c]X \}J_{tot}M_{J_{tot}}}
\end{split}
\end{equation}
A second change of reference system is performed, transforming the coordinates of the center-of-mass of the first two particles (1 and 2) and particle 3 in the total center-of-mass and the Jacobi coordinate between the center-of-mass of particles 1 and 2 and the third particle
\begin{equation}
\begin{dcases}
\vec{q} = \dfrac{2}{3}\biggl[ \vec{k}_3 - \dfrac{1}{2}(\vec{k}_1 + \vec{k}_2) \biggr] = \dfrac{2}{3}\biggl[ \vec{k}_3 - \dfrac{1}{2}\vec{P} \biggr]\\
\vec{Q}_{cm} = \vec{k}_1 + \vec{k}_2 + \vec{k}_3 = \vec{P} + \vec{k}_3
\end{dcases}\, ,
\label{eq:qQ_vsk3P}
\end{equation}
\begin{equation}
\ket{Pn_\c(L_p l_\c)\Lambda} = \int dQ_{cm}dq\,Q_{cm}^2q^2\sum_{l_{cm}\,l} \braket{Q_{cm}q(l_{cm}l)\Lambda|Pn_\c(L_p l_\c)\Lambda}^{(b)}_{\tau_\c} \ket{Q_{cm}q(l_{cm}l)\Lambda}.
\end{equation}
The mixed WC bracket (see~\ref{coeff_MixedWC}) is the one associated with the inverse transformation of Eq.~\eqref{eq:qQ_vsk3P}
\begin{equation}
\begin{pmatrix}
\vec{P}\\
\vec{k}_3
\end{pmatrix}
=
\begin{pmatrix}
\frac{2}{3} & -1\\
\frac{1}{3} & 1
\end{pmatrix}
\begin{pmatrix}
\vec{Q}_{cm}\\
\vec{q}
\end{pmatrix}
\label{eqn:wcb}
\end{equation}
and the associated matrix of WC coefficients is
\begin{equation}
\begin{pmatrix}
s_1 & t_1\\
s_2 & t_2
\end{pmatrix}^b_{MS}
=
\begin{pmatrix}
\frac{2}{3} & -1\\
\frac{1}{3} & 1
\end{pmatrix},
\label{eq:WCcoeffs_3B_b_Pk3}
\end{equation}
where the label $b$ is used to distinguish between the WC matrices that appear in this procedure.
\vspace{0.2cm}
\\
\underline{STEP 6}
\vspace{-0.6cm}
\begin{equation}
\begin{split}
&\ket{ \{ Q_{cm}q(l_{cm}l)\Lambda, p[(LS)J,s_\c]X \}J_{tot}M_{J_{tot}}}\\
\to
&\ket{ \{ Q_{cm}l_{cm}, qp[l,[(LS)J,s_\c]X]\mathcal{J} \}J_{tot}M_{J_{tot}}}
\end{split}
\end{equation}
A $6j$ is used to change again the internal coupling of $J_{tot}$:

\begin{equation}
\begin{split}
\ket{[(l_{cm}l)\Lambda, X]J_{tot}} &= \sum_{\mathcal{J}}\braket{[l_{cm},(lX)\mathcal{J}]J_{tot}|[(l_{cm}l)\Lambda, X]J_{tot}} \ket{[l_{cm},(lX)\mathcal{J}]J_{tot}}\\
&= \sum_{\mathcal{J}} (-)^{l_{cm}+l+X+J_{tot}} \hat{\Lambda}\hat{\mathcal{J}}
\begin{Bmatrix}
l & \Lambda & l_{cm}\\
J_{tot} & \mathcal{J} & X
\end{Bmatrix}
\ket{[l_{cm},(lX)\mathcal{J}]J_{tot}}.
\end{split}
\end{equation}
\vspace{0.2cm}
\\
\underline{STEP 7}
\vspace{-0.6cm}
\begin{equation}
\begin{split}
&\ket{ \{ Q_{cm}l_{cm}, qp[l,[(LS)J,s_\c]X]\mathcal{J} \}J_{tot}M_{J_{tot}}}\\
\to
&\ket{ \{ Q_{cm}l_{cm}, pq[(LS_\c)J, (ls)j]\mathcal{J} \}J_{tot}M_{J_{tot}}}
\end{split}
\end{equation}
The coupling of $\mathcal{J}$ is changed to give the final $j$ coupling in the MS basis:

\begin{equation}
\begin{split}
\ket{[l, (Js_\c)X]\mathcal{J}} &= \sum_j \braket{[(ls_\c)j, J]\mathcal{J}|[l, (Js_\c)X]\mathcal{J}} \ket{[(ls_\c)j, J]\mathcal{J}}\\
&= \sum_j (-)^{l+j+X+J} \hat{j}\hat{X}
\begin{Bmatrix}
l & s_\c & j\\
J & \mathcal{J} & X
\end{Bmatrix}
\ket{[J, (ls_\c)j]\mathcal{J}}.
\end{split}
\end{equation}
\vspace{0.4cm}
\\
\underline{STEP 8}
\vspace{-0.2cm}
\begin{equation}
\ket{t_\a \tau_\a, t_\b \tau_\b, t_\c \tau_\c} \to
\ket{[(t_\a t_\b)T, t_\c] \mathcal{T}M_{\mathcal{T}}}
\end{equation}
The isospin is recoupled
\begin{equation}
\ket{t_\a \tau_\a, t_\b \tau_\b, t_\c \tau_\c} = \sum_{T\,\mathcal{T}} \braket{t_\a \tau_\a t_\b \tau_\b | TM_T} \braket{TM_T t_\c \tau_\c | \mathcal{T}M_{\mathcal{T}}} \ket{[(t_\a t_\b)T, t_\c] \mathcal{T}M_{\mathcal{T}}} \, ,
\end{equation}
where the sums over $M_T$ and $M_{\mathcal{T}}$ drop because of the constraint from charge conservation.

To find the basis transformation coefficient one proceeds similarly to the case of two-body interactions.
The steps 1 to 8 discussed above are applied in sequence to expand Eq.~\eqref{eqn:3bbas} on states $\ket{[Q_{cm}l_{cm}, p q \alpha]J_{tot}}$ of given center of mass momentum and partial waves~\eqref{eqn:3bmomc}.
Multiplying to the left with state the state $\bra{[Q'_{cm}l'_{cm}, p'q'\alpha'] J_{tot}}$ and exploiting the orthonormality relation\, 
\begin{equation}
\braket{[Q'_{cm}l'_{cm}, {p'q'\alpha'}]J_{tot}|[{Q_{cm}l_{cm}}, {pq\alpha}]J_{tot}} = \dfrac{\delta(Q_{cm}-Q'_{cm})}{Q_{cm}^2} \delta_{l_{cm}l'_{cm}} \dfrac{\delta(p-p')}{p^2} \delta_{LL'}\delta_{SS'}\delta_{JJ'}\delta_{TT'}\dfrac{\delta{(q-q')}}{q^2}\delta_{ll'}\delta_{jj'}\delta_{\mathcal{J}\mathcal{J}'}\delta_{\mathcal{T}\mathcal{T}'}\,,
\end{equation}
the following expression for the three-body $T$-coefficient is found:
\begin{equation}
\begin{split}
T^{3B} \equiv& \braket{[Q_{cm}l_{cm}, pq\alpha]J_{tot} | [(ab)J_{12}c]J_{tot}}\\
=& \int dPP^2 \sum_{\lambda L_P}\sum_{\Lambda X} \braket{t_\a \tau_\a t_\b \tau_\b | TM_T} \braket{T M_T t_\c \tau_\c | \mathcal{T}M_\mathcal{T}} \hat{j}_\a\hat{j}_\b\hat{\lambda}\hat{S}
\begin{Bmatrix}
l_\a & s_\a & j_\a\\
l_\b & s_\b & j_\b\\
\lambda & S & J_{12}
\end{Bmatrix}\\
& \qquad \times  \braket{P p\, (L_P L) \lambda | n_\a n_\b \, (l_\a l_\b) \lambda}^{(a)}_{\t_\a \t_\b} (-)^{L_P+L+S+J_{12}} \hat{\lambda} \hat{J}
\begin{Bmatrix}
L_P & L & \lambda\\
S & J_{12} & J
\end{Bmatrix}\\
& \qquad \times \hat{J}_{12} \hat{j}_\c \hat{\Lambda} \hat{X}
\begin{Bmatrix}
l_\c & L_P & \Lambda\\
j_\c & J_{12} & J_{tot}\\
s_\c & J & X
\end{Bmatrix} \braket{Q_{cm}q(l_{cm}l)\Lambda|Pn_\c(L_p l_\c)\Lambda}^{(b)}_{\tau_\c}\\
& \qquad \times (-)^{l_{cm}+l+X+J_{tot}} \hat{\Lambda}\hat{\mathcal{J}}
\begin{Bmatrix}
l & \Lambda & l_{cm}\\
J_{tot} & \mathcal{J} & X
\end{Bmatrix} (-)^{l+j+J+X} \hat{j}\hat{X}
\begin{Bmatrix}
l & s_\c & j\\
J & \mathcal{J} & X
\end{Bmatrix} \,.
\end{split}
\label{eqn:Tc}
\end{equation}
Eq.~\eqref{eqn:Tc} differs from the already known expression for the three-body $T$-coefficient~\cite{Miyagi} since it projects a MS plane-wave basis to a generic isospin-dependent spherical basis. Hence, it is not limited to HO states.
Eq.~\eqref{eqn:Tc} can be further simplified via the use of a Wigner $12j$-symbol of the first kind \cite{Wig12j} (see~\ref{angmom_12j})
\begin{equation}
\begin{split}
& \begin{Bmatrix}
J && L_P && \Lambda && l_{cm} &\\
& J_{12} && l_\c && l && \mathcal{J}\\
J_{tot} && j_\c && s_\c && j &
\end{Bmatrix}_1\\
&\qquad \qquad = (-)^{J-\Lambda-J_{tot}+s_\c} \sum_X \hat{X}^2
\begin{Bmatrix}
l_\c & L_P & \Lambda\\
j_\c & J_{12} & J_{tot}\\
s_\c & J & X
\end{Bmatrix}
\begin{Bmatrix}
l & \Lambda & l_{cm}\\
J_{tot} & \mathcal{J} & X
\end{Bmatrix}
\begin{Bmatrix}
l & s_\c & j\\
J & \mathcal{J} & X
\end{Bmatrix}.
\end{split}
\label{eqn:12jo}
\end{equation}
As all the Wigner $3nj$ symbols, the $12j$ symbol can be decomposed in a summation over a single index of a product of $6j$ symbols. Such decomposition reads
\begin{equation}
\begin{split}
& \begin{Bmatrix}
J && L_P && \Lambda && l_{cm} &\\
& J_{12} && l_\c && l && \mathcal{J}\\
J_{tot} && j_\c && s_\c && j &
\end{Bmatrix}_1\\
&\qquad \qquad = \sum_x (-)^{Y-x}\hat{x}^2
\begin{Bmatrix}
J & L_P & J_{12}\\
j_\c & J_{tot} & x
\end{Bmatrix}
\begin{Bmatrix}
L_P & \Lambda & l_\c\\
s_\c & j_\c & x
\end{Bmatrix}
\begin{Bmatrix}
\Lambda & l_{cm} & l\\
j & s_\c & x
\end{Bmatrix}
\begin{Bmatrix}
l_{cm} & J_{tot} & \mathcal{J}\\
J & j & x
\end{Bmatrix},
\end{split}
\label{eqn:12jf6j}
\end{equation}
where $Y$ is given by the sum of all the angular momenta appearing in the $12j$ symbol:
\begin{equation}
Y \equiv J + L_P + \Lambda + l_{cm} + J_{12} + l_\c + l + \mathcal{J} + J_{tot} + j_\c + s_\c + j.
\end{equation}
The expression in Eq.~\eqref{eqn:12jf6j} allows for an efficient implementation of the $12j$ symbols through the precalculation of the $6j$ coefficients. With these simplifications, the $T$-coefficient can be re-written in a compact expression
\begin{equation}
\begin{split}
T^{3B} =& \braket{[Q_{cm}l_{cm}, pq\alpha]J_{tot} | [(ab)J_{12}c]J_{tot}}\\
=& (-)^{S+L-l_{cm}+J_{12}+j-s_\c} \hat{j}_\a \hat{j}_\b \hat{S}\hat{J}\hat{J}_{12}\hat{j}_\c\hat{\mathcal{J}}\hat{j}\sum_\lambda \hat\lambda^2
\begin{Bmatrix}
l_\a & s_\a & j_\a\\
l_\b & s_\b & j_\b\\
\lambda & S & J_{12}
\end{Bmatrix}
\sum_{L_P} (-)^{L_P}
\begin{Bmatrix}
L_P & L & \lambda\\
S & J_{12} & J
\end{Bmatrix}\\
&\qquad  \times\int dP\,P^2 \braket{P p \, (L_P L) \lambda | n_\a n_\b\, (l_\a l_\b) \lambda}^{(a)}_{\t_\a \t_\b} \sum_\Lambda (-)^{\Lambda} \hat{\Lambda}^2
\braket{Q_{cm}q(l_{cm}l)\Lambda|Pn_\c(L_p l_\c)\Lambda}^{(b)}_{\tau_\c}\\
&\qquad \times\begin{Bmatrix}
J && L_P && \Lambda && l_{cm} &\\
& J_{12} && l_\c && l && \mathcal{J}\\
J_{tot} && j_\c && s_\c && j &
\end{Bmatrix}_1 \braket{t_\a \tau_\a t_\b \tau_\b | TM_T} \braket{T M_T t_\c \tau_\c | \mathcal{T}M_\mathcal{T}},
\end{split}
\label{eqn:Tcoeff3}
\end{equation}
where the limits of integration over $P$ are constrained by the values of $Q_{cm}$ and $q$ through the mixed WC bracket.
The integral over $P$ can be converted in an integral over the cosine $x_b$ of the angle between $Q_{cm}$ and $q$,
\begin{equation}
\begin{split}
T^{3B} =& \braket{[Q_{cm}l_{cm}, pq\alpha]J_{tot} | [(ab)J_{12}c]J_{tot}}\\
=& (-)^{S+L-l_{cm}+J_{12}+j-s_\c} \hat{j}_\a \hat{j}_\b \hat{S}\hat{J}\hat{J}_{12}\hat{j}_\c\hat{\mathcal{J}}\hat{j}\sum_\lambda \hat\lambda^2
\begin{Bmatrix}
l_\a & s_\a & j_\a\\
l_\b & s_\b & j_\b\\
\lambda & S & J_{12}
\end{Bmatrix}
\sum_{L_P} (-)^{L_P}
\begin{Bmatrix}
L_P & L & \lambda\\
S & J_{12} & J
\end{Bmatrix}\\
&\qquad\times 8\pi^2\int_{-1}^1 dx_b \braket{PL_PpL; \lambda | n_\a l_\a n_\b l_\b; \lambda}^{(a)}_{\t_\a \t_\b}\\
&\qquad\times \sum_\Lambda (-)^{\Lambda} \hat{\Lambda}^2 \;
 A(x_b, \dfrac{Q_{cm}}{q}, l_{cm}, l, L_p, l_\c, \Lambda; s_1^b, t_1^b, s_2^b, t_2^b) 
 \;\, \tilde\phi_{n_{\c}l_{\c}\tau_\c}(\tilde{k}_3)\\
&\qquad \times\begin{Bmatrix}
J && L_P && \Lambda && l_{cm} &\\
& J_{12} && l_\c && l && \mathcal{J}\\
J_{tot} && j_\c && s_\c && j &
\end{Bmatrix}_1
 \braket{t_\a \tau_\a t_\b \tau_\b | TM_T} \braket{T M_T t_\c \tau_\c | \mathcal{T}M_\mathcal{T}} \,,
\end{split}
\label{eqn:Tcoeff}
\end{equation}
where $\tilde\phi_{n_{\c}l_{\c}\tau_\c}(k)$ is the radial single-particle wave function~\eqref{eqn:psi_sp_spher_MS} in momentum space and the values
\begin{equation}
\begin{split}
\tilde{k}_3 = \abs{t^b_2} \, q \sqrt{1+2x_b y_b z_b+(y_b z_b)^2}
\qquad\qquad \hbox{and} \qquad\qquad
P = \abs{t_1^b} \, q \sqrt{1 + 2x_by_b + y_b^2}
\end{split}
\label{eq:k3_tilde}
\end{equation}
 depend on the integration variable $x_b$ and are constrained by Eqs.~\eqref{eqn:r1} and~\eqref{eqn:r2} with coefficients~\eqref{eq:WCcoeffs_3B_b_Pk3}.
In Eqs.~\eqref{eqn:Tcoeff} and ~\eqref{eq:k3_tilde}, the subscript (superscript) $b$ denotes quantities relative to the mixed WC coefficient for transformation~\eqref{eqn:wcb} and the definition of the variables $x_b$, $y_b$ and $z_b$ is given in details in \ref{coeff}.
Finally, the $J$-coupled $T$-decoupled three-body matrix element is given by
\begin{equation}
\begin{split}
&\braket{[(ab)J_{12}c]J_{tot}|W|[(de)J'_{12}f]J_{tot}} =\\
&\qquad\qquad= \int dQ_{cm}Q_{cm}^2\,\sum_{l_{cm}}\int dp\,dq\,p^2q^2 \int dp'\,dq'\,p'^2q'^2\sum_{\alpha \,\alpha'} \braket{[(ab)J_{12}c]J_{tot} | [Q_{cm}l_{cm}, pq\alpha]J_{tot}} \\
&\qquad\qquad\qquad\qquad\times\braket{pq\a | \hat{W} | p'q'\a'}_A \, \braket{[Q_{cm}l_{cm}, p'q'\alpha']J_{tot} | [(de)J'_{12}f]J_{tot}}
\end{split}
\label{eq:3B_final_transform}
\end{equation}
where the independence of the three-body potential from the center-of-mass of the three-body system (Eq.~\eqref{eqn:cm3b}) has been exploited. Eq.~\eqref{eq:3B_final_transform} extends the transformation used in Ref.~\cite{Miyagi} to any spherical single-particle basis.

As for the two-body interactions, it can be useful to define CS three-body matrix elements to be used with CS regulators or wave functions. In this case the new Jacobi coordinates ($\vec{R}_{cm}$, $\vec{r}$, $\vec{s}$) read
\begin{equation}
\begin{pmatrix}
\vec{R}_{cm}\\
\vec{r}\\
\vec{s}
\end{pmatrix}
 = 
 \begin{pmatrix}
\frac{1}{3} & \frac{1}{3} & \frac{1}{3}\\
1 & -1 & 0\\
-\frac{1}{2} & -\frac{1}{2} & 1
\end{pmatrix}
\begin{pmatrix}
\vec{r}_1\\
\vec{r}_2\\
\vec{r}_3
\end{pmatrix} 
\label{eqn:coord3b3b}
\end{equation}
and are conjugate to ($\vec{Q}_{cm}$, $\vec{p}$, $\vec{q}$) from Eq.~\eqref{eqn:momcoord}.
When the matrix elements of the interaction are known in MS (as in Eqs.~\eqref{eq:3Bmon_Antisymm} or ~\eqref{eqn:3Bmom}), the corresponding relative CS can be obtained directly from a quadruple Hankel transform:
\begin{equation}
\begin{split}
\braket{rs\alpha | W | r's'\alpha'} = \dfrac{4}{\pi^2} \,i^{L'+l'-L-l}\, \int_0^{+\infty} dpdqdp'dq'\,p^2q^2p'^2q'^2\,j_L(pr)j_l(qs)j_{L'}(p'r')j_{l'}(q's')\braket{pq\alpha | W | p'q'\alpha'} \,.
\end{split}
\end{equation}
All the steps for the computation of the $T$-coefficients discussed in this section remain the same for CS with the exception the WC coefficients for the transformations in  Eqs.~\eqref{eqn:wca} and \eqref{eqn:wcb} must be substituted by
\begin{equation}
\begin{pmatrix}
s_1 & t_1\\
s_2 & t_2
\end{pmatrix}^a_{CS}=
\begin{pmatrix}
1 & \frac{1}{2}\\
1 & -\frac{1}{2}
\end{pmatrix}
\end{equation}
and
\begin{equation}
\begin{pmatrix}
s_1 & t_1\\
s_2 & t_2
\end{pmatrix}^b_{CS}=
\begin{pmatrix}
1 & -\frac{1}{3}\\
1 & \frac{2}{3}
\end{pmatrix} \, .
\end{equation}

\section{Conclusions}
\label{conclusions}
A complete analytical derivation of matrix elements of a realistic nuclear Hamiltonian expanded on a generic spherical single-particle basis has been presented, including up to three-body interactions. This calculation requires to use WC brackets, which differently from the Moshinsky brackets are no longer diagonal in major oscillator shells. To do this, the vector bracket from \cite{BB} has been generalized to the case of a generic change of reference system and an efficient expression of the WC coefficient which is computationally more convenient than the one from \cite{WC} has been obtained. The full expansion of two- and three-body matrix elements on a generic single-particle basis constitutes a novelty for what concerns the handling of the radial part of the equations. Also, as for the three-body sector, a new scheme of angular momenta couplings that allows to re-obtain the expression for the three-body $T$-coefficient from \cite{Miyagi} has been presented.
While these formulas are written for the specific case of nuclear interactions, they can be easily adapted to the case of other many-body systems with rotationally invariant Hamiltonians.
The working equations presented in this work are suitable for implementation on high-performance computers. Hence, this work should pave the way to systematic investigations of the possible advantages of bases different from the traditional HO states in the study of bulk and spectroscopic properties of atomic nuclei. On the other hand, the implementation of these analytic expressions in numerical calculations represents a challenge.
In particular, the dependence of the WC coefficients from continuous variables (momenta or positions) tends to increase both the computational time and the storage requirements, so that new computational strategies are to be designed.  Work in this direction is currently underway.

\section*{Acknowledgements}
A. S. is supported by the European Union's Horizon 2020 research and innovation program under grant agreement No 800945 - NUMERICS - H2020-MSCA-COFUND-2017.
The authors thank V. Som\`{a} for valuable discussions and for reviewing the manuscript.

\appendix

\section{The spherical components of vectors}
\label{sphcomp}
Let us recall how to express vectors in \emph{spherical components} \cite{Dev} since it is easier to perform rotations in this basis.
Consider a generic vector $\vec{A}$ expressed in Cartesian coordinates $\{\vec{e}_x, \vec{e}_y, \vec{e}_z\}$:
\begin{equation}
\vec{A} = A_x\vec{e}_x + A_y\vec{e}_y + A_z\vec{e}_z.
\end{equation}
One defines a spherical basis $\{\vec\epsilon_{1-1}, \vec\epsilon_{10}, \vec\epsilon_{11}\}$ related to the Cartesian one through the relations
\begin{equation}
\vec\epsilon_{1-1} = \dfrac{\vec{e}_x - i\vec{e}_y}{\sqrt{2}}, \qquad\vec\epsilon_{10} = \vec{e}_z, \qquad\vec\epsilon_{11} = -\dfrac{\vec{e}_x + i\vec{e}_y}{\sqrt{2}}.
\end{equation}
The vector $\vec{A}$ can be expressed in this basis as
\begin{equation}
\begin{split}
\vec{A} &= -A_{1\,-1}\vec\epsilon_{1 1} + A_{10}\vec\epsilon_{10} - A_{11}\vec\epsilon_{1\,-1}\\
&= \sum_{\mu=-1,0,1}(-1)^\mu A_{1\mu}\vec\epsilon_{1\,-\mu},
\end{split}
\end{equation}
where the components $A_{1-1}$, $A_{10}$ and $A_{11}$ read
\begin{equation}
A_{1\,-1} = \dfrac{A_x - iA_y}{\sqrt{2}}, \qquad A_{10} = A_z, \qquad A_{11} = -\dfrac{A_x + iA_y}{\sqrt{2}} \, ,
\end{equation}
and the complex conjugate of $A_{1\mu}$ is
\begin{equation}
A_{1\mu}^* = (-1)^\mu A_{1-\mu} \, .
\end{equation}
A generic vector $\vec{r}$ in coordinate or in momentum space can be expressed in spherical components
\begin{equation}
\vec{r} = -r_{1\,-1}\vec\epsilon_{11} + r_{10}\vec\epsilon_{10} - r_{11}\vec\epsilon_{1\,-1},
\end{equation}
with
\begin{equation}
r_{1\,-1} = \dfrac{x-iy}{\sqrt{2}}, \qquad r_{10} = z, \qquad r_{11} = -\dfrac{x+iy}{\sqrt{2}}.
\end{equation}
Since the Cartesian coordinates ($x$, $y$, $z$) are expressed as a function of the spherical coordinates ($r$, $\vartheta$, $\phi$) by
\begin{equation}
x = r\sin\vartheta\cos\phi, \qquad y = r\sin\vartheta\sin\phi, \qquad z = r\cos\vartheta \,,
\end{equation}
the spherical components of the vector can then be written as
\begin{subequations}
\begin{align}
r_{1-1} &= \dfrac{r}{\sqrt{2}}\sin\vartheta e^{-i\phi} = \sqrt{\dfrac{4\pi}{3}}\,r\,Y_{1-1}(\hat{r}) \, ,\\
r_{10} &= r\cos\vartheta = \sqrt{\dfrac{4\pi}{3}}\,r\,Y_{10}(\hat{r}) \, ,\\
r_{11} &= -\dfrac{r}{\sqrt{2}}\sin\vartheta e^{i\phi} = \sqrt{\dfrac{4\pi}{3}}\,r\,Y_{11}(\hat{r}) \, ,
\end{align}
\end{subequations}
where $Y_{1\mu}$ are order-1 spherical harmonics
\begin{subequations}
\begin{align}
Y_{1-1}(\hat{r}) &= \sqrt{\dfrac{3}{8\pi}}\sin\vartheta e^{-i\phi} \, ,\\
Y_{10}(\hat{r}) &= \sqrt{\dfrac{3}{4\pi}}\cos\vartheta \, ,\\
Y_{11}(\hat{r}) &= -\sqrt{\dfrac{3}{8\pi}}\sin\vartheta e^{i\phi} \, .
\end{align}
\end{subequations}
The vector $\vec{r}$ can then be expressed in spherical components
\begin{equation}
\vec{r} = \sqrt{\dfrac{4\pi}{3}}r\sum_{\mu}(-1)^\mu Y_{1\mu}\vec\epsilon_{1-\mu} \, .
\label{eqn:r_sph}
\end{equation}
It is then easy to write the scalar product of two vectors $\vec{A}$ and $\vec{B}$ as
\begin{equation}
\begin{split}
\vec{A}\cdot\vec{B} &= A_xB_x + A_yB_y + A_zB_z\\
&= \sum_{\mu}(-1)^\mu A_{1\mu} B_{1-\mu}\\
&= \sum_{\mu} A_{1\mu}B_{1\mu}^* \, .
\end{split}
\label{eqn:prod_sph}
\end{equation}

\section{Phases in the Fourier transform between Harmonic Oscillator wave functions in coordinate and momentum space}
\label{HOmomcoord}
A generic three-dimensional wave function in coordinate space $\Psi(\vec{r}) = \phi_{nl}(r)Y_{lm_l}(\hat{r})$ can be Fourier-transformed to its MS representation by means of the plane-wave expansion:
\begin{equation}
e^{i\vec{k}\cdot\vec{r}} = 4\pi \sum_{l\,m_l}i^{l}j_{l}(kr)Y_{l\,m_l}(\hat{k})Y_{l\,m_l}^*(\hat{r}).
\label{eqn:pw}
\end{equation}
The complete transformation is obtained from Eq.~\eqref{eqn:pw} as follows:
\begin{equation}
\begin{split}
\widetilde\Psi(\vec{k}) &= \mathcal{F}\{ \Psi(\vec{r}) \} (\vec{k})\\
&= \mathcal{F}\{ \phi_{nl}(r)Y_{lm_l}(\hat{r}) \} (\vec{k})\\
&= \dfrac 1{\sqrt{(2\pi)^3}} \int d^3\vec{r}\,e^{-i\vec{k}\cdot\vec{r}}\phi_{nl}(r)Y_{lm_l}(\hat{r})\\
&= \dfrac {4\pi}{\sqrt{(2\pi)^3}} \sum_{l'm_l'} Y_{l'm_l'}(\hat{k})(-i)^{l'}\int d^3\vec{r}\,j_{l'}(kr)Y_{l'm_l'}^*(\hat{r})Y_{lm_l}(\hat{r})\phi_{nl}(r)\\
&= \sqrt{\dfrac 2 \pi}\sum_{l'm_l'}Y_{l'm_l'}(\hat{k})(-i)^{l'}\delta_{ll'}\delta_{m_lm_l'}\int_0^{+\infty} dr\,r^2 j_l(kr)\phi_{nl}(r)\\
&= (-i)^l \,\hat{\phi}_{nl}(k) \, Y_{l m_l }(\hat{k}),
\end{split}
\label{eqn:complexphase}
\end{equation}
with
\begin{equation}
\hat{\phi}_{nl}(k) = \sqrt{\dfrac 2 \pi} \int_0^{+\infty} dr\,r^2j_l(kr)\phi_{nl}(r).
\label{eqn:phik}
\end{equation}
Eq.~\eqref{eqn:complexphase} then shows an additional imaginary phase $(-i)^l$ that must be explicitly taken into account when performing a Hankel transform on the radial part of a total wave function.

In the case of the HO wave function, their radial components in CS and MS are related through Eq.~\eqref{eqn:complexphase} and their analytical expressions read
\begin{equation}
\phi_{nl}(r) = \sqrt{\dfrac{2 \, (n!)}{\Gamma(n+l+3/2)\,b^3}}\,\biggl( \dfrac{r}{b} \biggr)^l e^{-\frac{1}{2}(\frac{r}{b})^2}L_n^{l+1/2}\biggl( \dfrac{r^2}{b^2} \biggr)
\label{eqn:phir}
\end{equation}
and
\begin{equation}
\hat{\phi}_{nl}(k) = (-)^n\sqrt{\dfrac{2\,(n!)\,b^3}{\Gamma(n+l+3/2)}} \, (kb)^l e^{-\frac{1}{2}(kb)^2} L_n^{l+1/2}((kb)^2),
\label{eqn:phik2}
\end{equation}
where $\Gamma$ represents the Gamma function, $L$ the Laguerre polynomial and $b = \sqrt{\hbar/(m\omega)}$ the oscillator length.

\section{Angular momenta algebra}
\label{angmom}
Here, we collect the results from angular momentum  algebra that have been exploited in the derivation of the main text and give specific definitions for the $12j$ coefficients.  Several well known relations among $3j$, $6j$ and $9j$ symbols are also included for completeness.
We use the convention
\begin{equation}
\hat{j} = \sqrt{2j+1}
\label{eqn:def_hatj}
\end{equation}
throughout this manuscript.
Several useful angular momentum identities are collected in Ref.~\cite{Talmi} while more advanced results are taken from \cite{Var}.

\subsection{Clebsch-Gordan coefficients}
Let $j_1$ and $j_2$ be two angular momenta with projections $m_1$ and $m_2$ along the quantization axis. The \emph{Clebsch-Gordan} coefficient expresses the probability amplitude that $j_1$ and $j_2$ are coupled to a third angular momentum $j_3$ with projection $m_3$ and is represented as
\begin{equation}
\braket{j_1m_1j_2m_2 | j_3m_3}.
\end{equation}
The Clebsch-Gordan coefficient is non-vanishing only if the three angular momenta $\{j_1j_2j_3\}$ satisfy the \emph{triangular inequality}:
\begin{itemize}
\item
$j_1 < j_2+j_3, \quad j_2 < j_3 + j_1, \quad j_3 < j_1 + j_2 \qquad  \Longrightarrow \qquad \abs{j_1-j_2} \leq j_3 \leq j_1+j_2$\, ,
\item
$j_1+j_2+j_3 \qquad\text{always integer}$\, ,
\end{itemize}
and if the following condition on the projections of angular momenta is satisfied:
\begin{equation}
m_1+m_2=m_3 \, .
\label{eqn:msumrule}
\end{equation}
The Clebsch-Gordan coefficient is self-adjoint:
\begin{equation}
\braket{j_1m_1j_2m_2 | j_3m_3} = \braket{j_3m_3 | j_1m_1j_2m_2}
\end{equation}
and it satisfies the following exchange symmetries:
\begin{subequations}
\label{eqn:cgall}
\begin{align}
\braket{j_1m_1\,j_2m_2|j_3m_3}
&= (-1)^{j_1+j_2-j_3} \braket{j_1\, (-m_1)\;j_2\, (-m_2) | j_3 \, (-m_3)}
\label{eqn:cg1}
\\
&= (-1)^{j_1+j_2-j_3} \braket{j_2m_2\,j_1m_1 | j_3 m_3}\\
&= (-1)^{j_1-m_1} \dfrac{\hat{j_3}}{\hat{j_2}}\braket{j_1m_1\; j_3 \, (-m_3) | j_2\,(-m_2)}.
\label{eqn:cg3}
\end{align}
\end{subequations}
The quantum numbers entering a non-vanishing Clebsch-Gordan coefficient satisfy the following two relations
\begin{equation}
(-1)^{2(j_i-m_i)} = 1, \quad \forall \, i \in \{1,2,3\},
\end{equation}
\begin{equation}
(-1)^{2(j_1+j_2+j_3)} = 1.
\end{equation}
The \emph{orthonormality relation} of the Clebsch-Gordan coefficients, sometimes also referred to as \emph{unitarity relation}, reads
\begin{equation}
\sum_{m_1m_2} \braket{j_1m_1j_2m_2|j_3m_3} \braket{j_1m_1j_2m_2|j_3'm_3'} = \delta_{j_3j_3'}\delta_{m_3m_3'}.
\label{eqn:CGort}
\end{equation}

\subsection{\texorpdfstring{$3j$}{3j} symbol}
The Wigner $3j$ symbol is defined from the Clebsch-Gordan coefficient as
\begin{equation}
\begin{pmatrix}
j_1 & j_2 & j_3\\
m_1 & m_2 & m_3
\label{eqn:3jtoclebsch}
\end{pmatrix}
= (-1)^{j_1-j_2-m_3}\dfrac 1{\hat{j}_3} \braket{j_1m_1j_2m_2 | j_3(-m_3)},
\end{equation}
while the inverse relation is
\begin{equation}
\label{eqn:GCvs3j}
\braket{j_1m_1j_2m_2 | j_3m_3} = (-1)^{-j_1+j_2-m_3}\hat{j}_3
\begin{pmatrix}
j_1 & j_2 & j_3\\
m_1 & m_2 & -m_3
\end{pmatrix}.
\end{equation}
The $3j$ symbols are more symmetric objects than Clebsch-Gordan coefficients.
A $3j$ symbol is invariant under even permutations of its columns
\begin{equation}
\begin{pmatrix}
j_1 & j_2 & j_3\\
m_1 & m_2 & m_3
\end{pmatrix}
=
\begin{pmatrix}
j_2 & j_3 & j_1\\
m_2 & m_3 & m_1
\end{pmatrix}
=
\begin{pmatrix}
j_3 & j_1 & j_2\\
m_3 & m_1 & m_2
\end{pmatrix},
\end{equation}
while any odd permutation of the columns introduces an additional phase:
\begin{equation}
\begin{pmatrix}
j_1 & j_2 & j_3\\
m_1 & m_2 & m_3
\end{pmatrix}
= (-1)^{j_1+j_2+j_3}
\begin{pmatrix}
j_2 & j_1 & j_3\\
m_2 & m_1 & m_3
\end{pmatrix}
= (-1)^{j_1+j_2+j_3}
\begin{pmatrix}
j_1 & j_3 & j_2\\
m_1 & m_3 & m_2
\end{pmatrix}
= (-1)^{j_1+j_2+j_3}
\begin{pmatrix}
j_3 & j_2 & j_1\\
m_3 & m_2 & m_1
\end{pmatrix}
\, .
\end{equation}
The same phase is involved in \emph{time reversal} transformations:
\begin{equation}
\begin{pmatrix}
j_1 & j_2 & j_3\\
-m_1 & -m_2 & -m_3
\end{pmatrix}
 = (-1)^{j_1+j_2+j_3}
\begin{pmatrix}
j_1 & j_2 & j_3\\
m_1 & m_2 & m_3
\end{pmatrix}.
\end{equation}
The $3j$ symbols appear in evaluating of the Slater integrals over three spherical harmonics:
\begin{equation}
\begin{split}
\braket{l_1m_1 | Y_{l_2m_2} | l_3m_3} &=
\int d\Omega\, Y_{l_1m_1}^*(\Omega)Y_{l_2m_2}(\Omega)Y_{l_3m_3}(\Omega)\\
&= \sqrt{\dfrac{1}{4\pi}} \hat{l}_1\hat{l}_2\hat{l}_3
(-1)^{-m_1}
\begin{pmatrix}
l_2 & l_3 & l_1\\
0 & 0 & 0
\end{pmatrix}
\begin{pmatrix}
l_2 & l_3 & l_1\\
m_2 & m_3 & -m_1
\end{pmatrix}\\
&= \sqrt{\dfrac{1}{4\pi}} \dfrac{\hat{l}_2\hat{l}_3}{\hat{l}_1}
\braket{l_3 0 \, l_2 0|l_1 0} \braket{l_3 m_3 \, l_2 m_2|l_1 m_1},
\end{split}
\label{eqn:gaunt}
\end{equation}
which follows directly from the so called~\emph{Gaunt formula}.

\subsection{\texorpdfstring{$6j$}{6j} symbol}
The Wigner $6j$ symbols are denoted by curly braces and appear when coupling three different angular momenta $j_1$, $j_2$ and $j_3$ to a total momentum $j$. The probability amplitude among different possible coupling reads
\begin{equation}
\braket{[(j_1j_2)j_{12},j_3]jm | [j_1, (j_2j_3)j_{23}] j'm'} = \delta_{jj'}\delta_{mm'}(-1)^{j_1+j_2+j_3+j}\hat{j}_{12}\hat{j}_{23}
\begin{Bmatrix}
j_1 & j_2 & j_{12}\\
j_3 & j & j_{23}
\end{Bmatrix}
\label{eqn:6jdef}
\end{equation}
where we have used the notation $(j_a j_b)j_c$ for the coupling of momenta $j_a$ and $j_b$ according to the Clebsch-Gordan coefficient $\braket{j_a m_a \, j_b m_b|j_c m_c}$ (and similarly for ``$[,]$'').
From Eqs.~\eqref{eqn:cgall} and~\eqref{eqn:6jdef} it follows that
\begin{equation}
\braket{[(j_1 j_2)j_{12},j_3]jm | [(j_1j_3)j_{13},j_2]j'm'} = \delta_{jj'}\delta_{mm'} (-1)^{j_2+j_3+j_{12}+j_{13}}\hat{j}_{12}\hat{j}_{13}
\begin{Bmatrix}
j_2 & j_1 & j_{12}\\
j_3 & j & j_{13}
\end{Bmatrix}\,,
\end{equation}
\begin{equation}
\braket{[j_1,(j_2j_3)j_{23}]jm | [(j_1j_3)j_{13},j_2]j'm'} = \delta_{jj'}\delta_{mm'} (-1)^{j_1+j+j_{23}}\hat{j}_{13}\hat{j}_{23}
\begin{Bmatrix}
j_1 & j_3 & j_{13}\\
j_2 & j & j_{23}
\end{Bmatrix}\,.
\end{equation}
The $6j$ symbol is defined as the sum of four $3j$ symbols
\begin{equation}
\begin{gathered}
\begin{Bmatrix}
j_1 & j_2 & j_3\\
j_1' & j_2' & j_3'
\end{Bmatrix}
= \sum_{\substack{m_1m_2m_3\\m'_1m'_2m'_3}}(-1)^{j_1'+j_2'+j_3'+m_1'+m_2'+m_3'}\\
\begin{pmatrix}
j_1 & j_2 & j_3\\
m_1 & m_2 & m_3
\end{pmatrix}
\begin{pmatrix}
j_1 & j_2' & j_3'\\
m_1 & m_2' & -m_3'
\end{pmatrix}
\begin{pmatrix}
j_1' & j_2 & j_3'\\
-m_1' & m_2 & m_3'
\end{pmatrix}
\begin{pmatrix}
j_1' & j_2' & j_3\\
m_1' & -m_2' & m_3
\end{pmatrix}.
\end{gathered}
\label{eqn:6jfrom3j}
\end{equation}
It is invariant under any permutation of the columns
\begin{equation}
\begin{Bmatrix}
j_1 & j_2 & j_3\\
j_4 & j_5 & j_6
\end{Bmatrix}
=
\begin{Bmatrix}
j_2 & j_1 & j_3\\
j_5 & j_4 & j_6
\end{Bmatrix}
=
\begin{Bmatrix}
j_1 & j_3 & j_2\\
j_4 & j_6 & j_5
\end{Bmatrix},
\end{equation}
and under the exchange of upper and lower arguments in any pair of columns
\begin{equation}
\begin{Bmatrix}
j_1 & j_2 & j_3\\
j_4 & j_5 & j_6
\end{Bmatrix}
=
\begin{Bmatrix}
j_4 & j_5 & j_3\\
j_1 & j_2 & j_6
\end{Bmatrix}
=
\begin{Bmatrix}
j_1 & j_5 & j_6\\
j_4 & j_2 & j_3
\end{Bmatrix}.
\end{equation}
The following two equations relate $3j$ and $6j$ symbols~\cite{Talmi}:
\begin{equation}
\begin{gathered}
\begin{pmatrix}
j_1 & j_2 & j_3\\
m_1 & m_2 & m_3
\end{pmatrix}
\begin{Bmatrix}
j_1 & j_2 & j_3\\
j_1' & j_2' & j_3'
\end{Bmatrix} \\=
\sum_{m_1'm_2'm_3'} (-1)^{j_1'+j_2'+j_3'+m_1'+m_2'+m_3'}
\begin{pmatrix}
j_1 & j_2' & j_3'\\
m_1 & m_2' & -m_3'
\end{pmatrix}
\begin{pmatrix}
j_1' & j_2 & j_3'\\
-m_1' & m_2 & m_3'
\end{pmatrix}
\begin{pmatrix}
j_1' & j_2' & j_3\\
m_1' & -m_2' & m_3
\end{pmatrix},
\end{gathered}
\end{equation}

\begin{equation}
%\begin{multlined}
\begin{pmatrix}
l & l' & k\\
0 & 0 & 0
\end{pmatrix}
\begin{Bmatrix}
l & l' & k\\
j' & j & 1/2
\end{Bmatrix}
= -\dfrac{1}{2}[1 + (-1)^{l+l'+k}] \hat{l}^{-1}\hat{l'}^{-1}
\begin{pmatrix}
j & j' & k\\
1/2 & -1/2 & 0
\end{pmatrix}.
%\end{multlined}
\end{equation}
These identities can be easily rewritten by substituting the $3j$ symbols with Clebsch-Gordan coefficients by means of Eq.~\eqref{eqn:3jtoclebsch}:
\begin{equation}
\begin{split}
\braket{j_1  j_2 m_1 m_2 | j_3 m_3}
\begin{Bmatrix}
j_1 & j_2 & j_3\\
j_1' & j_2' & j_3'
\end{Bmatrix}=&
\sum_{m_1' \,m_2'\, m_3'} \dfrac{(-1)^{-j_1'-j_2'+j_3'-m_3'+m_2-m_1}}{\hat{j'}_3^2 } \, \braket{j_1m_1j_2'm_2' | j_3'm_3'}\\
&\qquad\qquad \times\braket{j_1'-\!m_1'j_2m_2 | j_3'-\!m_3'}\braket{j_1'm_1'j_2'-\!m_2' | j_3 m_3},
\end{split}
\label{eqn:3j6j}
\end{equation}

\begin{equation}
\begin{multlined}
\braket{l0l'0|k0}
\begin{Bmatrix}
l & l' & k\\
j' & j & 1/2
\end{Bmatrix}
= -\dfrac{[1+(-1)^{l+l'+k}]}{2}   \dfrac{(-1)^{j-j'+l'-l}}{\hat{l} \; \hat{l'}}\braket{j \dfrac 12 \, j' -\!\dfrac 12  |k0}.
\label{eqn:3j6j2}
\end{multlined}
\end{equation}

\subsection{\texorpdfstring{$9j$}{9j} symbol}
The Wigner $9j$ symbol appears in the evaluation of the probability amplitude
\begin{equation}
\begin{multlined}
\braket{[(j_1 \, j_2)j_{12}, (j_3\,j_4)j_{34}] j m | [(j_1\,j_3)j_{13},(j_2\,j_4)j_{24}] j' m'} = \delta_{jj'}\delta_{mm'}  \hat{j}_{12} \hat{j}_{34} \hat{j}_{13} \hat{j}_{24} 
\begin{Bmatrix}
j_1 & j_2 & j_{12}\\
j_3 & j_4 & j_{34}\\
j_{13} & j_{24} & j
\end{Bmatrix}.
\end{multlined}
\end{equation}
The $9j$ symbol is invariant under reflection about either diagonal as well as even permutations of its rows or columns:
\begin{equation}
\begin{Bmatrix}
j_1 & j_2 & j_3\\
j_4 & j_5 & j_6\\
j_7 & j_8 & j_9
\end{Bmatrix}
=
\begin{Bmatrix}
j_1 & j_4 & j_7\\
j_2 & j_5 & j_8\\
j_3 & j_6 & j_9
\end{Bmatrix}
=
\begin{Bmatrix}
j_9 & j_6 & j_3\\
j_8 & j_5 & j_2\\
j_7 & j_4 & j_1
\end{Bmatrix}
=
\begin{Bmatrix}
j_7 & j_4 & j_1\\
j_9 & j_6 & j_3\\
j_8 & j_5 & j_2
\end{Bmatrix}.
\end{equation}
An odd permutation of the columns or of the rows yields
\begin{equation}
\begin{Bmatrix}
j_1 & j_2 & j_3\\
j_4 & j_5 & j_6\\
j_7 & j_8 & j_9
\end{Bmatrix}
=
(-1)^S
\begin{Bmatrix}
j_4 & j_5 & j_6\\
j_1 & j_2 & j_3\\
j_7 & j_8 & j_9
\end{Bmatrix}
=
(-1)^S
\begin{Bmatrix}
j_2 & j_1 & j_3\\
j_5 & j_4 & j_6\\
j_8 & j_7 & j_9
\end{Bmatrix},
\end{equation}
with $S = \sum_{i=1}^9j_i$.
The Wigner $9j$ symbol is related to the Wigner $6j$ symbol by the following relation
\begin{equation}
\begin{Bmatrix}
j_1 & j_2 & j_3\\
j_4 & j_5 & j_6\\
j_7 & j_8 & j_9
\end{Bmatrix}
=\sum_x(-1)^{2x}\hat{x}^2
\begin{Bmatrix}
j_1 & j_4 & j_7\\
j_8 & j_9 & x
\end{Bmatrix}
\begin{Bmatrix}
j_2 & j_5 & j_8\\
j_4 & x & j_6
\end{Bmatrix}
\begin{Bmatrix}
j_3 & j_6 & j_9\\
x & j_1 & j_2
\end{Bmatrix}.
\end{equation}
\subsection{\texorpdfstring{$12j$}{12j} symbol}
\label{angmom_12j}

The Wigner $12j$ symbol \cite{Wig12j} of the first kind appears in the coupling of 5 angular momenta:
\begin{equation}
\begin{gathered}
\ket{[b_{12},\, [ b_{23},\, [ b_{34},\, (b_{41}\,a_1)c_4\,]c_3\,]c_2\,]c_1} \\
= (-1)^{2a_1} \sum_{a_2a_3a_4} \hat{a}_2\hat{a}_3\hat{a}_4\hat{c}_2\hat{c}_3\hat{c}_4
\begin{Bmatrix}
a_1 && a_2 && a_3 && a_4 &\\
& b_{12} && b_{23} && b_{34} && b_{41}\\
c_1 && c_2 && c_3 && c_4 &
\end{Bmatrix}_1 
\ket{[\,[\,[(a_1 b_{12})a_2, \, b_{23}]a_3,\, b_{34}]a_4,\, b_{41}]c_1}\, .
\end{gathered}
\end{equation}
The $12j$ coefficients can be expanded in terms of $6j$ coefficients:
\begin{equation}
\begin{gathered}
\begin{Bmatrix}
a_1 && a_2 && a_3 && a_4 &\\
& b_{12} && b_{23} && b_{34} && b_{41}\\
c_1 && c_2 && c_3 && c_4 &
\end{Bmatrix}_1\\
=
\sum_{x}(-)^{S-x}\hat{x}^2
\begin{Bmatrix}
a_1 & a_2 & b_{12}\\
c_2 & c_1 & x
\end{Bmatrix}
\begin{Bmatrix}
a_2 & a_3 & b_{23}\\
c_3 & c_2 & x
\end{Bmatrix}
\begin{Bmatrix}
a_3 & a_4 & b_{34}\\
c_4 & c_3 & x
\end{Bmatrix}
\begin{Bmatrix}
a_4 & c_1 & b_{41}\\
a_1 & c_4 & x
\end{Bmatrix}
\end{gathered}
\end{equation}
with $S = \sum_{i=1}^4 (a_i+c_i)+b_{12}+b_{23}+b_{34}+b_{41}$. The summed momenta $x$ is integer (semi-integer) if $S$ is integer (semi-integer). The $12j$ symbol can also be related to a summation over $9j$ and $6j$ symbols
\begin{equation}
\begin{gathered}
\begin{Bmatrix}
a_1 && a_2 && a_3 && a_4 &\\
& b_{12} && b_{23} && b_{34} && b_{41}\\
c_1 && c_2 && c_3 && c_4 &
\end{Bmatrix}_1\\
= (-1)^{a_1-a_3-c_1+c_3}\sum_x \hat{x}^2
\begin{Bmatrix}
b_{23} & a_2 & a_3\\
c_2 & b_{12} & c_1\\
c_3 & a_1 & x
\end{Bmatrix}
\begin{Bmatrix}
b_{34} & a_3 & a_4\\
c_1 & b_{41} & x
\end{Bmatrix}
\begin{Bmatrix}
b_{34} & c_3 & c_4\\
a_1 & b_{41} & x
\end{Bmatrix},
\end{gathered}
\end{equation}
and is invariant under cyclic or anti-cyclic permutations of the columns $\{a_1, b_{12}, c_1\}$, $\{a_2, b_{23}, c_2\}$, $\{a_3, b_{34}, c_3\}$ and $\{a_4, b_{41}, c_4\}$ and under exchange of the first and the third row.
The $12j$ symbol of the first kind satisfies 8 triangular inequalities
\begin{equation}
\{a_1 b_{12} a_2\}, \{a_2 b_{23} a_3\}, \{a_3 b_{34} a_4\}, \{a_4 b_{41} c_1\},
\{c_1 b_{12} c_2\}, \{c_2 b_{23} c_3\}, \{c_3 b_{34} c_4\}, \{c_4 b_{41} a_1\}
\end{equation}
and 2 \emph{tetragonal inequalities}
\begin{equation}
\{a_1 c_1 a_3 c_3\}, \{a_2 c_2 a_4 c_4\}.
\end{equation}
A set of 4 angular momenta $\{j_1, j_2, j_3, j_4\}$ is said to satisfy a tetragonal inequality if it satisfies the following properties:
\begin{itemize}
\item
$j_1+j_2+j_3+j_4$ is an integer
\item
$j_1 \le j_2+j_3+j_4$,\,\, $j_2 \le j_3+j_4+j_1$,\,\, $j_3 \le j_4+j_1+j_2$,\,\, $j_4 \le j_1+j_2+j_3$
\end{itemize}
The inequalities at the second point can be re-written in a compact way as
\begin{equation}
\dfrac{\abs{j_3-j_4}+\abs{j_2-j_3}+\abs{j_2-j_4}-j_2-j_3-j_4}{3} \le j_1 \le j_2+j_3+j_4.
\end{equation}

\section{Coefficients for general changes of coordinates}
\label{coeff}

We demonstrate the derivation of the coefficients for the change of coordinates of two-particle states. The results reported in this appendix are valid for both momentum and coordinate space. Consider the general linear transformation
\begin{equation}
\begin{dcases}
\vec{r}_1 = s_1\vec{r} + t_1\vec{R}\\
\vec{r}_2 = s_2\vec{r} + t_2\vec{R} \, ,
\end{dcases}
\label{eqn:first}
\end{equation}
where the vectors $\vec{r}_1$, $\vec{r}_2$, $\vec{r}$ and $\vec{R}$ can live either in coordinate or in momentum space. The transformation~\eqref{eqn:first} implies that the vectors $\vec{r}_1$ and $\vec{r}_2$ live on the same plane as $\vec{r}$ and $\vec{R}$. Due to rotational invariance angular coefficients will depend only on the angle between any two vectors and on the ratio of their length.
Thus, it is useful to follow Ref.~\cite{WC} and define the quantities
\begin{equation}
x \equiv \cos(\widehat{rR}) = \dfrac{\vec{r}\cdot\vec{R}}{\abs{rR}},\qquad\quad
y = \dfrac{s_1r}{t_1R},\qquad\quad
z = \dfrac{s_2t_1}{s_1t_2} \,,
\label{eqn:xyz_vars}
\end{equation}
from which one can also write the magnitudes of $r_1$ and $r_2$ as
\begin{align}
r_1 =& \abs{t_1}R\sqrt{1+2xy+y^2} \, ,
\label{eqn:r1} \\
r_2 =& \abs{t_2}R\sqrt{1+2xyz+y^2z^2} \, .
\label{eqn:r2}
\end{align}

\subsection{Vector bracket}
To obtain the transformation among states coupled in angular momentum one starts by defining the ket
\begin{equation}
\label{eqn:I_def}
\begin{split}
\ket{I} &\equiv \ket{rlRL,\lambda\mu}\\
&= \sum_{m_lm_L}\braket{lm_lLm_L | \lambda\mu} \ket{rlm_lRLm_L}\\
&= \sum_{m_lm_L}\braket{lm_lLm_L | \lambda\mu}\int d\hat{r}d\hat{R}\,\, Y_{lm_l}(\hat{r})Y_{Lm_L}(\hat{R})\ket{\vec{r}\vec{R}}\,,
\end{split}
\end{equation}
where $\lambda$ results from the coupling of $l$ and $L$ and $\mu$ is the projection along the axis of quantization. Eq.~\eqref{eqn:I_def} follows directly from the representation of a (spherical) single-particle basis state and of its angular wave function:
\begin{equation}
\braket{\vec{r}|\psi} = \phi_{nl}(r)Y_{lm_l}(\hat{r}),
\end{equation}
\begin{equation}
\braket{\hat{r}|rlm_l} = \ket{r}Y_{lm_l}(\hat{r}).
\end{equation}
Similarly, we consider the angular momentum representation of a state $\ket{\pvec{r}_1\pvec{r}_2}$:
\begin{equation}
\label{eqn:II_def}
\ket{II} \equiv \ket{r_1l_1r_2l_2, \lambda\mu} \, , %= \ket{r_1r_2(l_1l_2)\lambda\mu},
\end{equation}
where $l_1$ and $l_2$ are coupled to $\lambda$.
The vector bracket associated to the transformation~\eqref{eqn:first} is given by the overlap
\begin{equation}
\begin{split}
\braket{I|II} &= \braket{rR(lL)\lambda\mu | r_1r_2(l_1l_2)\lambda\mu}\\
&= \int d\pvec{r}''d\pvec{R}''d\pvec{r}_1'd\pvec{r}_2'\,\,\braket{I|\pvec{r}''\vec{R}''}\braket{\pvec{r}''\pvec{R}'' | \pvec{r}_1'\pvec{r}_2'}\braket{\pvec{r}_1'\pvec{r}_2' | II},
\label{eqn:III}
\end{split}
\end{equation}
where the brackets are given by
\begin{align}
\label{eq:r1r2rR}
\braket{\pvec{r}''\pvec{R}'' | \pvec{r}_1'\pvec{r}_2'} &= \delta(\pvec{r}_1'-s_1\pvec{r}''-t_1\pvec{R}'')\delta(\pvec{r_2}'-s_2\pvec{r}''-t_2\pvec{R}'')\,,\\
\label{eq:IrR}
\braket{I|\pvec{r}''\vec{R}''} &= \sum_{m_lm_L}\braket{lm_lLm_L | \lambda\mu} \int d\hat{r}d\hat{R}\,Y_{lm_l}^*(\hat{r})Y_{Lm_L}^*(\hat{R})\braket{\pvec{r}\pvec{R} | \pvec{r}''\pvec{R}''} \notag \\
&= \sum_{m_lm_L}\braket{lm_lLm_L | \lambda\mu} Y_{lm_l}^*(\hat{r}'')Y_{Lm_L}^*(\hat{R}'')\dfrac{\delta{(r-r'')}}{r^2}\dfrac{\delta{(R-R'')}}{R^2}\,,
\end{align}
with
\begin{equation}
\braket{\pvec{r}\pvec{R} | \pvec{r}''\pvec{R}''} = \dfrac{\delta{(r-r'')}}{r^2}\dfrac{\delta{(R-R'')}}{R^2}\delta{(\hat{r}-\hat{r}'')}\delta{(\hat{R}-\hat{R}'')}.
\end{equation}
For the state~\eqref{eqn:II_def} one obtains:
\begin{equation}
\label{eq:r1r2II}
\braket{\pvec{r}_1'\pvec{r}_2' | II} = \sum_{m_1m_2}\braket{l_1m_1l_2m_2|\lambda\mu}Y_{l_1m_1}(\hat{r}_1')Y_{l_2m_2}(\hat{r}_2')\dfrac{\delta{(r_1-r_1')}}{r_1^2}\dfrac{\delta{(r_2-r_2')}}{r_2^2}\,.
\end{equation}
Combining Eqs.~\eqref{eq:r1r2rR},~\eqref{eq:IrR} and~\eqref{eq:r1r2II} allows to simplify Eq.~\eqref{eqn:III} into
\begin{equation}
\begin{gathered}
\braket{I|II} = \int d\pvec{r}''d\pvec{R}''d\pvec{r}'_1d\pvec{r}'_2\,\,\dfrac{\delta{(r-r'')}}{r^2}\dfrac{\delta{(R-R'')}}{R^2}\dfrac{\delta{(r_1-r_1')}}{r_1^2}\dfrac{\delta{(r_2-r_2')}}{r_2^2}A_{I,II}(x'',r''/R'')\\
\times\,\,\delta{(\pvec{r}_1'-s_1\pvec{r}''-t_1\pvec{R}'')}\delta{(\pvec{r}_2'-s_2\pvec{r}''-t_2\pvec{R}'')} \, ,
\end{gathered}
\label{eqn:III2}
\end{equation}
where the so-called \emph{angular bracket} $A_{I,II}$ is defined as
\begin{equation}
A_{I,II}(x'',r''/R'') \equiv \biggl[ \sum_{m_l\,m_L}\braket{lm_lLm_L|\lambda\mu} Y^*_{lm_l}(\hat{r}'')Y^*_{Lm_L}(\hat{R}'') \biggr] \biggl[ \sum_{m_1\,m_2}\braket{l_1m_1l_2m_2|\lambda\mu}Y_{l_1m_1}(\hat{r}'_1)Y_{l_2m_2}(\hat{r}'_2) \biggr] 
\label{eqn:Ab}
\end{equation}
and depends implicitly on the angular momenta and the coefficients of transformation~\eqref{eqn:first}.
The transformation properties of states $\ket{I}$ and $\ket{II}$ under rotation imply that their overlap~\eqref{eqn:III} is diagonal in $\lambda$ and $\mu$ and it is independent on the value of $\mu$.  This allows to rewrite 
Eq.~\eqref{eqn:Ab} in the form
\begin{equation}
\begin{split}
A_{I,II}(x'',r''/R'') = \dfrac{1}{2\lambda+1} &\sum_\mu\biggl[ \sum_{m_l\,m_L}\braket{lm_lLm_L|\lambda\mu} Y^*_{lm_l}(\hat{r}'')Y^*_{Lm_L}(\hat{R}'') \biggr]\\
&\qquad \times\biggl[ \sum_{m_1\,m_2}\braket{l_1m_1l_2m_2|\lambda\mu}Y_{l_1m_1}(\hat{r}'_1)Y_{l_2m_2}(\hat{r}'_2) \biggr] \,,
\end{split}
\label{eqn:AIx}
\end{equation}
which is manifestly invariant under rotation. The angular bracket can then be seen as a scalar quantity that can be shown to depend only on the two quantities $x''\equiv\cos{(\widehat{r''R''})}$ and $r''/R''$. The cosine of $\widehat{r''R''}$ can be obtained from \eqref{eqn:first} exploiting the scalar products $\pvec{r}_1\cdot\pvec{r}_1$ or $\pvec{r}_2\cdot\pvec{r}_2$:
\begin{equation}
\cos{(\widehat{r''R''})} = \dfrac{r_1'^2-s_1^2r''^2-t_1^2R''^2}{2s_1t_1r''R''} = \dfrac{r_2'^2-s_2^2r''^2-t_2^2R''^2}{2s_2t_2r''R''}.
\label{eqn:x}
\end{equation}

When integrating in Eq.~\eqref{eqn:III2} over $\pvec{r}'_1$ and $\pvec{r}'_2$, the two deltas $\delta(r_1-r'_1)$ and $\delta(r_2-r'_2)$ fix the values of $r'_1$ and $r'_2$ according to Eqs.~\eqref{eqn:r1} and~\eqref{eqn:r2}:
\begin{equation}
{r}_1'=\sqrt{(s_1\pvec{r}''+t_1\pvec{R}'')^2}=f_1(\cos{(\widehat{r''R''})},r'',R'')\,,
\end{equation}
\begin{equation}
{r}_2'=\sqrt{(s_2\pvec{r}''+t_2\pvec{R}'')^2}=f_2(\cos{(\widehat{r''R''})},r'',R'')\,,
\end{equation}
where we put in evidence that their values depend only on the magnitudes $r''$, $R''$ and the angle $x''$ among them.
It follows that
\begin{equation}
\braket{I|II} = \int d\pvec{r}''d\pvec{R}''\,\,\dfrac{\delta{(r-r'')}}{r^2}\dfrac{\delta{(R-R'')}}{R^2}\dfrac{\delta{(r_1-r_1')}}{r_1^2}\dfrac{\delta{(r_2-r_2')}}{r_2^2}A_{I,II}(x'',r''/R'') \,.
\end{equation}
An integration over the two solid angles $\hat{r}$ and $\hat{R}$ can then be performed, with the notation $d\hat{r} \equiv d\Omega_r = d\varphi_rd\cos\vartheta_r$.
Given two vectors $\vec{r}$ and $\vec{R}$ in the three-dimensional space and a generic function $f(x)$, with $x = \cos{\widehat{rR}} = \cos\vartheta_x$, one finds
\begin{equation}
\int d\hat{r}d\hat{R}\, f(x) = 8\pi^2\int d(\cos{\vartheta_x}) \, f(x) = 8\pi^2\int_{-1}^1dx\,f(x),
\end{equation}
and therefore
\begin{equation}
\braket{I|II} = 8\pi^2 \int dr'' dR''\, r''^2 R''^2 \int dx'' \, \dfrac{\delta{(r-r'')}\delta{(R-R'')}}{r^2R^2} \dfrac{\delta{(r_1-r'_1)}\delta{(r_2-r'_2)}}{r_1^2r_2^2} A_{I,II}(x'',r''/R'').
\label{eqn:III3}
\end{equation}
To evaluate Eq.~\eqref{eqn:III3} we need to express $r'_1$ and $r'_2$ in terms of the integration variables and perform an appropriate transformation of the last two Dirac delta. We start considering
\begin{equation}
\delta{(r_1-r'_1)} = \delta \left(r_1 - \sqrt{s_1^2r''^2 + t_1^2R''^2 + 2s_1t_1r''R''x}\right) \, .
\end{equation}
Exploiting the composition property of the Dirac delta for continuously differentiable functions
\begin{equation}
\delta(g(x)) = \sum_i\dfrac{\delta(x-x_i)}{\abs{g'(x_i)}},
\end{equation}
one can rewrite the Dirac delta as
\begin{equation}
\delta{(r_1^2-r_1'^{2})} = \dfrac{1}{2r_1'} [ \delta(r_1-r'_1) + \delta(r_1+r'_1) ],
\end{equation}
where $\delta(r_1+r'_1) = 0$ since $r_1, r'_1>0$.
Hence,
\begin{equation}
\delta(r_1-r'_1) = 2r_1\delta(r_1^2 - r_1'^2)\,,
\end{equation}
\begin{equation}
\delta(r_2-r'_2) = 2r_2\delta(r_2^2 - r_2'^2)\,,
\end{equation}
and
\begin{equation}
\braket{I|II} = \dfrac{8\pi^2}{r_1^2r_2^2} \int dr'' dR'' dx'' \, 4r_1r_2 \delta(r_1^2-r_1'^2)\delta(r_2^2-r_2'^2)\delta(R-R'')\delta(r-r'') A_{I,II}(x'',r''/R'')\, .
\end{equation}
We then manipulate again the two Dirac deltas for performing the integral over
$x$:
\begin{equation}
\label{eqn:delta_x}
\begin{split}
\delta(r_1^2 - r_1'^2) &= \delta(r_1^2 - (s_1^2r''^2 + t_1^2R''^2 + 2s_1t_1r''R''x''))\\
&= \dfrac{1}{2\abs{s_1t_1}r''R''} \delta{\biggl(x'' - \dfrac{r_1^2 - s_1^2r''^2 - t_1^2R''^2}{2s_1t_1r''R''}\biggr)} \,,
\end{split}
\end{equation}
and
\begin{equation}
\begin{split}
\delta(r_2^2 - r_2'^2) &= \delta\biggl(r_2^2 - s_2^2r''^2 - t_2^2R''^2 - 2s_2t_2r''R''\dfrac{r_1^2-s_1^2r''^2-t_1^2R''^2}{2s_1t_1r''R''}\biggr)\\
&= \abs{s_1t_1} \delta{(-s_2t_2r_1^2 + s_1t_1r_2^2 + (s_2t_2s_1^2 - s_1t_1s_2^2)r''^2+(s_2t_2t_1^2-s_1t_1t_2^2)R''^2)}\\
&= \abs{s_1t_1} \delta{(w)}
\end{split}
\end{equation}
with
\begin{equation}
  w \equiv s_2t_2r_1^2 - s_1t_1r_2^2 + (t_1s_2-t_2s_1)(s_1s_2r''^2-t_1t_2R''^2) \, .
\end{equation}
Finally, Eq.~\eqref{eqn:III3} reduces to
 \begin{equation}
\braket{I|II} = 16\pi^2\dfrac{1}{rRr_1r_2} A_{I,II}(x,r/R) \, \delta(w) \, \theta{(1-x^2)},
\label{eqn:VB}
\end{equation}
where $x=\cos(\widehat{rR})$ is evaluated according to Eq.~\eqref{eqn:x}, the Heaviside function $\theta$ imposes the constraint that the integral over Eq.~\eqref{eqn:delta_x} vanishes unless $-1\leq \cos(\widehat{rR}) \leq 1$ and the full dependency of the function $A_{I,II}$ is
\begin{equation}
 A_{I,II}  =  A_{I,II}(x, r/R, l, L, l_1, l_2, \lambda; s_1, t_1, s_2, t_2).
\end{equation}
Eq.~\eqref{eqn:VB} is referred to as the \emph{vector bracket} and has been obtained in the case of a specific change of coordinates in \cite{KungKuoRatcliff, VBPolls}.
Following the notation of Ref.~\cite{BB}, this coefficient will be simply denoted as $A_{I,II}(x)$.

\subsection{Angular bracket \texorpdfstring{$A_{I, II}(x)$}{A(I II)(x)}}
\begin{figure}
\centering 
\includegraphics[width=0.4\textwidth]{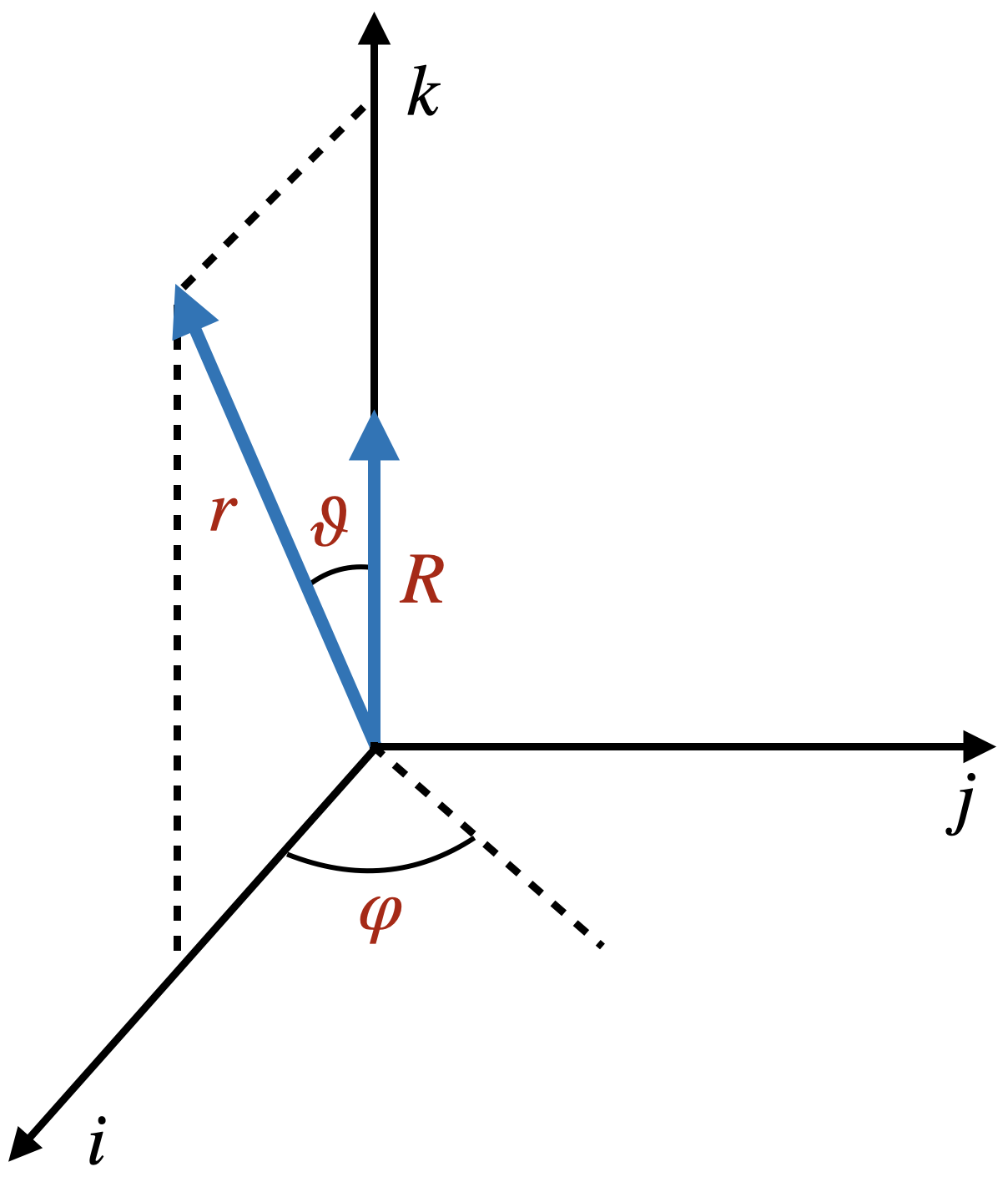}	
\caption{Reference frame used for the angular bracket} 
\label{fig:coord}%
\end{figure}
The rotational invariance of $A_{I,II}$ can be exploited to arrange the vectors $\vec{r}$, $\vec{R}$, $\vec{r}_1$ and $\vec{r}_2$ in space so that the final expression for the angular bracket is easy to evaluate numerically~\cite{BB}. 
As shown in Fig.~\ref{fig:coord}, we chose a reference frame $(\hat{\bf i},\hat{\bf j},\hat{\bf k})$ such that all vectors lie on the $j=0$ plane, $\vec{R}$ lies along the  axis $\hat{\bf k}$ and $\vec{r}$ is on the $i>0$ half plane.
Consider now the general expression for $A_{I,II}(x)$
\begin{equation}
\begin{split}
A_{I,II}(x) = \dfrac{1}{2\lambda+1} & \sum_\mu \biggl[ \sum_{mM} \braket{lmLM|\lambda\mu}Y^*_{lm}(\hat{r})Y^*_{LM}(\hat{R}) \biggr]\\ 
& \qquad \times\biggl[ \sum_{m_1m_2} \braket{l_1m_1l_2m_2|\lambda\mu}Y_{l_1m_1}(\hat{r}_1)Y_{l_2m_2}(\hat{r}_2) \biggr].
\end{split}
\end{equation}
From the specific disposition of the vectors in Fig.~\ref{fig:coord} it follows that $\vartheta_R = 0$, so that the spherical harmonic $Y_{LM}(\hat{R})$ is just
\begin{equation}
Y_{LM}(0,0) = \delta_{M0}\sqrt{\dfrac{2L+1}{4\pi}}.
\end{equation}
and it is independent of $\varphi_R = 0$. On the other hand $\vartheta_r = \arccos{x}$ and $\varphi_r = 0$.

Let consider now the value of the cosines of the angles between $r_1$ and $R$ and $r_2$ and $R$.
\begin{equation}
\vec{r}_i\cdot\vec{R} = s_i \vec{r}\cdot\vec{R} + t_i R^2 = s_irRx + t_iR^2 = r_iR\cos{\vartheta_i}
\end{equation}
\begin{equation}
\Longrightarrow\qquad \cos{\vartheta_i} = \dfrac{s_irx+t_iR}{r_i}\qquad\quad \text{for}\,\,i = 1,2
\end{equation}
which can be shown to be only a function of $r/R$, $x$, $y$ and $z$, simply by substituting Eqs.~\eqref{eqn:r1} and \eqref{eqn:r2}.
Since $\vec{R}$ has no components along the axis $\hat{\bf i}$, Eq.~\eqref{eqn:first} implies that
\begin{equation}
\varphi_i = 
\begin{dcases}
0\qquad\text{if}\,\,s_i>0\\
\pi\qquad\text{if}\,\,s_i<0 \,.
\end{dcases}
\end{equation}
Recalling the property of the spherical harmonics
\begin{equation}
Y_{lm}(\vartheta, \pi+\varphi) = (-)^mY_{lm}(\vartheta,\varphi)\,,
\end{equation}
it is useful to define the quantities $c_1$ and $c_2$ such that
\begin{equation}
c_i = 
\begin{dcases}
0\qquad\,\,\,\,\text{if}\,\,s_i>0\\
m_i\qquad\text{if}\,\,s_i<0
\end{dcases}\qquad \text{for}\; \,i=1,2 \,.
\end{equation}
Eventually, the expression found for the angular bracket $A_{I, II}(x)$ is particularly convenient for the implementation since it includes only two summations:
\begin{equation}
\begin{split}
A_{I,II}(x) =& \dfrac{1}{2\lambda+1}\sum_\mu \braket{l\mu L0 | \lambda\mu} Y^*_{l\mu}(\vartheta, 0) \, Y_{L0}^*(0,0)\\
&\qquad \times\sum_{m_1} \braket{l_1m_1l_2(\mu-m_1) | \lambda\mu} Y_{l_1m_1}(\vartheta_1,0) \, Y_{l_2m_2}(\vartheta_2,0) \, (-)^{c_1+c_2}.
\label{eqn:Ax}
\end{split}
\end{equation}

\subsection{Wong-Clement bracket}
\label{coeff_WC}
The Wong-Clement brackets are used to express the wave functions of two particles with radial states $\ket{n_1\,l_1}$ and $\ket{n_2\,l_2}$ in terms of new coordinates $r$ and $R$. They are obtained by a double-integration of~\eqref{eqn:VB}:
\begin{equation}
\braket{rR(lL)\lambda | n_1n_2(l_1l_2)\lambda} = \int dr_1dr_2\,r_1^2r_2^2 \phi_{n_1l_1}(r_1)\phi_{n_2l_2}(r_2) \braket{rR(lL)\lambda | r_1r_2(l_1l_2)\lambda},
\label{eqn:WCc}
\end{equation}
where $\phi_{n_1l_1}(r)$ and $\phi_{n_2l_2}(r)$ are orthonormal radial wave functions. Eq.~\eqref{eqn:WCc} represents the so called \emph{Wong-Clement} bracket \cite{WC}, which can be seen as a generalization of the coefficients obtained in \cite{BaymanKallio}. The $s_1$, $s_2$, $t_1$ and $t_2$ will be referred to as the \emph{Wong-Clement coefficients}.
Exploiting Eq.~\eqref{eqn:VB}, the delta over $w$ simplifies the integration over $r_2$:
\begin{equation}
\label{eqn:dw_WC}
\delta(w) = \dfrac{1}{2r_2\abs{s_1t_1}} \delta{\biggl( r_2 - \sqrt{\biggl|{\dfrac{s_2t_2r_1^2 + (t_1s_2-t_2s_1)(s_1s_2r^2-t_1t_2R^2)}{s_1t_1}}\biggr|} \biggr)}.
\end{equation}
The other integration can be simplified by the change of variable
\begin{equation}
dr_1 = \dfrac{s_1t_1rR}{r_1}dx \,.
\end{equation}
From Eq.~\eqref{eqn:r1}, \eqref{eqn:r2} and \eqref{eqn:x}, one finds that if $s_1t_1>0$, the integration over $x$ is on the range $[-1,1]$ while if $s_1t_1<0$ it is on the range $[1,-1]$. However, the change of integration sign is absorbed by removing the absolute value in the denominator of Eq.~\eqref{eqn:dw_WC}.
Simplifying all the terms, the final expression for the WC bracket is then obtained
\begin{equation}
\braket{rR(lL)\lambda | n_1n_2(l_1l_2)\lambda} = 8\pi^2\int_{-1}^1 dx\,\phi_{n_1l_1}(r_1)\phi_{n_2l_2}(r_2)A_{I,II}(x) \,,
\end{equation}
with $r_1$ and~$r_2$ determined by Eqs.~\eqref{eqn:r1} and~\eqref{eqn:r2}, respectively.
The WC bracket can include a dependency on the isospin through the radial part of the wave functions. In general it can be written as
\begin{equation}
\braket{rR(lL)\lambda | n_1n_2(l_1l_2)\lambda}_{\tau_1 \tau_2}
\equiv \braket{rR(lL)\lambda, \tau_1\tau_2 | n_1n_2(l_1l_2)\lambda, \tau_1 \tau_2}
=8\pi^2\int_{-1}^1 dx\,\phi_{n_1l_1\tau_1}(r_1)\phi_{n_2l_2\tau_2}(r_2)A_{I,II}(x).
\end{equation}

\subsection{Mixed Wong-Clement bracket}
\label{coeff_MixedWC}
A single integration of Eq.~\eqref{eqn:VB} over the radial state of one particle gives the \emph{mixed Wong-Clement bracket}:
\begin{equation}
\braket{rR(lL)\lambda | r_1n_2(l_1l_2)\lambda} = \int dr_2\,r_2^2\, \phi_{n_2l_2}(r_2) \braket{rR(lL)\lambda | r_1r_2(l_1l_2)\lambda}.
\label{eqn:WCmix}
\end{equation}
Proceeding in analogy to the previous section, one finds
\begin{equation}
\braket{rR(lL)\lambda | r_1n_2(l_1l_2)\lambda} = 8\pi^2\dfrac{1}{\abs{s_1t_1}rRr_1} \vartheta{(1-x^2)}A_{I,II}(x)\phi_{n_2l_2}(\tilde{r}_2),
\end{equation}
where the argument of the radial wave function
\begin{equation}
\tilde{r}_2 = \sqrt{\dfrac{s_2t_2r_1^2 + (t_1s_2-t_2s_1)(s_1s_2r^2-t_1t_2R^2)}{s_1t_1}}
\label{eq:tld_r2_MixWC} 
\end{equation}
is constrained by of the Dirac delta $\delta(\omega)$. Notice that substituting Eq.~\eqref{eqn:r1} into~\eqref{eq:tld_r2_MixWC} recovers Eq.~\eqref{eqn:r2} exactly.
Furthermore, the Heaviside theta constrains the value of $x$.
As for the WC bracket, an explicit dependency on the isospin can be included in the bracket through the wave function and Eq.~\eqref{eqn:WCmix} becomes
\begin{equation}
\braket{rR(lL)\lambda | r_1n_2(l_1l_2)\lambda}_{\tau_2} = \int dr_2\,r_2^2\, \phi_{n_2l_2\tau_2}(r_2) \braket{rR(lL)\lambda | r_1r_2(l_1l_2)\lambda}.
\end{equation}

\section{Transposition operator \texorpdfstring{$T_{23}$}{T23}}
\label{T23}
Let us consider the following three definitions for three-particles states:
\begin{align}
\ket{pq\alpha} &\equiv \ket{pq, \, [(LS)J,(ls)j]\mathcal{J}M_\mathcal{J}, \,(Tt)\mathcal{T}M_\mathcal{T}} \,, \\
\ket{pq\beta} &\equiv \ket{pq,\, [(L\,l)\mathcal{L},(Ss)\mathcal{S}]\mathcal{J}M_\mathcal{J},\, (Tt)\mathcal{T}M_\mathcal{T}}\, , \\
\ket{pq\gamma} &\equiv \ket{pq,\, (L\,l)\mathcal{L}M_\mathcal{L},\, (Ss)\mathcal{S}M_\mathcal{S},\, (Tt)\mathcal{T}M_\m{T}}\, .
\end{align}
The transformations between the coupling schemes $\alpha$, $\beta$ and $\gamma$ read
\begin{equation}
\ket{pq\alpha} = \sum_{\m{L}\m{S}} \hat{\m{L}}\hat{\m{S}}\hat{{J}}\hat{{j}}
\begin{Bmatrix}
L & S & J\\
l & s & j\\
\m{L} & \m{S} & \m{J}
\end{Bmatrix}
\ket{pq\beta},
\end{equation}
\begin{equation}
\ket{pq\beta} = \sum_{M_\m{L}M_\m{S}} \braket{\m{L}M_{\m{L}}\m{S}M_{\m{S}}|\m{J}M_{\m{J}}} \ket{pq\gamma}
\end{equation}
and the state $\ket{pq\alpha}$ can therefore be written as
\begin{equation}
\begin{split}
\ket{pq\alpha} &= \sum_{\m{L}\m{S}M_\m{L}M_\m{S}} \hat{\m{L}}\hat{\m{S}}\hat{{J}}\hat{{j}}
\begin{Bmatrix}
L & S & J\\
l & s & j\\
\m{L} & \m{S} & \m{J}
\end{Bmatrix}
\braket{\m{L}M_{\m{L}}\m{S}M_{\m{S}}|\m{J}M_{\m{J}}}
\ket{pq\gamma}\\
&= \sum_{\m{L}\m{S}M_\m{L}M_\m{S}} \hat{\m{L}}\hat{\m{S}}\hat{{J}}\hat{{j}}
\begin{Bmatrix}
L & S & J\\
l & s & j\\
\m{L} & \m{S} & \m{J}
\end{Bmatrix}
\braket{\m{L}M_{\m{L}}\m{S}M_{\m{S}}|\m{J}M_{\m{J}}}\\
&\qquad\qquad\times\ket{pq,(L\,l)\m{L}M_{\m{L}}}\otimes \ket{(Ss)\m{S}M_{\m{S}}}
\otimes \ket{(Tt)\m{T}M_{\m{T}}}.
\end{split}
\end{equation}
The expectation value of the transposition operator $T_{23}$ over the states $\ket{pq\alpha}$ reads
\begin{equation}
\begin{split}
\braket{pq\a|T_{23}|p'q'\a'} =& \sum_{\m{L}M_{\m{L}}\m{S}M_{\m{S}}\m{L'}M_{\m{L}}'\m{S'}M_{\m{S}}'}
\hat{\m{L}}\hat{\m{S}}\hat{{J}}\hat{{j}}\hat{\m{L}}'\hat{\m{S}}'\hat{{J}}'\hat{{j}}' \\
&\times
\begin{Bmatrix}
L & S & J\\
l & s & j\\
\m{L} & \m{S} & \m{J}
\end{Bmatrix}
\braket{\m{L}M_{\m{L}}\m{S}M_{\m{S}}|\m{J}M_{\m{J}}}
\begin{Bmatrix}
L' & S' & J'\\
l & s' & j'\\
\m{L}' & \m{S}' & \m{J}'
\end{Bmatrix}
\braket{\m{L}'M_{\m{L}}'\m{S}'M_{\m{S}}'|\m{J}'M_{\m{J}}'}\\
&\times \braket{pq,(L\,l)\m{L}M_{\m{L}}|T_{23}|p'q',(L'l')\m{L}'M_{\m{L}}'}\\
&\times \braket{(Ss)\m{S}M_{\m{S}}|T_{23}|(S's')\m{S}'M_{\m{S}}'} \braket{(Tt)\m{T}M_{\m{T}}|T_{23}|(T't')\m{T}'M_{\m{T}}'}.
\end{split}
\end{equation}
The coefficient $\braket{pq,(L\,l)\m{L}M_{\m{L}}|T_{23}|p'q',(L'l')\m{L}'M_{\m{L}}'}$ 
is proportional to $\delta_{\m{L}\m{L}'}\delta_{M_\m{L}M'_\m{L}}$ and it is simply given by the vector bracket in Eq.~\eqref{eqn:VB} evaluated for an appropriate transformation between the $\vec{p}$ and~$\vec{q}$ and the result of inverting particles 2 and 3 on $\vec{p}'$ and~$\vec{q}'$.
For fermions, $s = t = 1/2$, the spin and isospin contributions can be simplified as follows~\cite{glocklefew, Nav1999, Nav2007}:
\begin{equation}
\braket{(Ss)\m{S}M_{\m{S}}|T_{23}|(S's')\m{S}'M_{\m{S}}'} = \delta_{\m{S} \ \m{S}'}\delta_{M_\m{S} \, M'_\m{S}} (-1)^{1+S+S'}\, \hat{S}\, \hat{S}' \,
\begin{Bmatrix}
1/2 & 1/2 & S\\
1/2 & \m{S} & S'
\end{Bmatrix}
\end{equation}
and
\begin{equation}
\label{eqn:T23_IsoSp}
\braket{(Tt)\m{T}M_{\m{T}}|T_{23}|(T't')\m{T}'M_{\m{T}}'} = \delta_{\m{T} \m{T}'}\delta_{M_\m{T} M'_\m{T}} (-1)^{1+T+T'}\, \hat{T} \, \hat{T}' \,
\begin{Bmatrix}
1/2 & 1/2 & T\\
1/2 & \m{T} & T'
\end{Bmatrix}.
\end{equation}
Eventually, a term $\delta_{\m{J}\m{J}'}\delta_{M_\m{J}M_\m{J}'}$ arises from the orthogonality of the Clebsch-Gordan coefficients. Then everything simplifies into
\begin{equation}
\begin{split}
\braket{pq\a|T_{23}|p'q'\a'} =& \sum_{\m{L}\m{S}} \delta_{\m{J}\m{J}'}\delta_{M_\m{J}M_\m{J}'} \delta_{\mathcal{T}\mathcal{T}'}\delta_{M_{\mathcal{T}}M_{\mathcal{T}'}} (-)^{1+S+S'} (-)^{1+T+T'} \hat{\m{L}}^2 \, \hat{\m{S}}^2 \, \hat{{J}}' \hat{{j}}' \, \hat{{J}} \hat{{j}} \, \hat{S}\hat{S}' \, \hat{T}\hat{T}'\\
&\times
\begin{Bmatrix}
L & S & J\\
l & 1/2 & j\\
\m{L} & \m{S} & \m{J}
\end{Bmatrix}
\begin{Bmatrix}
L' & S' & J'\\
l & 1/2 & j\\
\m{L} & \m{S} & \m{J}
\end{Bmatrix}\\
&\times\braket{pq,(L\,l)\m{L} | T_{23} | p'q',(L'l')\m{L}}
\begin{Bmatrix}
1/2 & 1/2 & S\\
1/2 & \m{S} & S'
\end{Bmatrix}
\begin{Bmatrix}
1/2 & 1/2 & T\\
1/2 & \m{T} & T'
\end{Bmatrix}.
\end{split}
\label{eqn:isos}
\end{equation}

\subsection*{Radial component of the transposition operator}
The bracket of the operator $T_{23}$ over momentum states is given by
\begin{equation}
\braket{\vec{p} \vec{q} |T_{23}| \vec{p}' \vec{q}'} = \delta(\vec{p}' - \tilde{\vec{p}}) \delta(\vec{q}' - \tilde{\vec{q}}) \, ,
\label{eqn:pqT23p1q1_a}
\end{equation}
where the vectors $\tilde{\vec{p}}$ and $\tilde{\vec{q}}$ result from by applying the transposition operator to the left side, $\ket{\tilde{\vec{p}} \tilde{\vec{q}}}=T_{23} \ket{\vec{p} \vec{q}}$, exploiting its property of being self-adjoint. Inverting Eq.~\eqref{eqn:momcoord} to find the single-particle momenta, one finds
\begin{equation}
\begin{pmatrix}
\vec{k}_1\\
\vec{k}_2\\
\vec{k}_3
\end{pmatrix}
 = 
 \begin{pmatrix}
\frac{1}{3} &  1 & -\frac{1}{2}\\
\frac{1}{3} & -1 & -\frac{1}{2} \\
\frac{1}{3} &  0 &  1
\end{pmatrix}
\begin{pmatrix}
\vec{Q}_{cm}\\
\vec{p}\\
\vec{q}
\end{pmatrix},
\label{eqn:momcoord_inv}
\end{equation}
and reconstructing $\tilde{\vec{p}}$ and $\tilde{\vec{q}}$ after permuting $\vec{k}_2$ and $\vec{k}_3$
\begin{equation}
\begin{pmatrix}
\vec{Q}_{cm}\\
\tilde{\vec{p}}\\
\tilde{\vec{q}}
\end{pmatrix}
=
 \begin{pmatrix}
1 & 1 & 1 \\
\frac{1}{2}  & -\frac{1}{2} & 0 \\
-\frac{1}{3} & -\frac{1}{3} & \frac{2}{3} 
\end{pmatrix}
\begin{pmatrix}
\vec{k}_1\\
\vec{k}_3\\
\vec{k}_2
\end{pmatrix}
 = 
 \begin{pmatrix}
1  &  0 & 0\\
0 & \frac{1}{2} &     -\frac{3}{4} \\
0 & -1 & -\frac{1}{2}
\end{pmatrix}
\begin{pmatrix}
\vec{Q}_{cm}\\
\vec{p}\\
\vec{q}
\end{pmatrix} \, ,
\label{eqn:momT23}
\end{equation}
which shows explicitly that $T_{23}$ does not affect the center of mass motion. Thus, Eq.~\eqref{eqn:pqT23p1q1_a} becomes
\begin{equation}
\braket{\vec{p} \vec{q} |T_{23}| \vec{p}' \vec{q}'} =
\delta (\vec{p}' -s_1 \vec{p} - t_1 \vec{q} ) \,
\delta (\vec{q}' -s_2 \vec{p} - t_2 \vec{q})
\label{eqn:pqT23p1q1_b} \,,
\end{equation}
with the following change of coordinates:
\begin{equation}
\begin{pmatrix}
s_1 & t_1\\
s_2 & t_2
\end{pmatrix}
=
\begin{pmatrix}
\frac{1}{2} & -\frac{3}{4}\\
-1 & -\frac{1}{2}
\end{pmatrix}_{MS} \,.
\end{equation}
Comparing to Eq.~\eqref{eq:r1r2rR} and performing the angular momentum coupling as in Eq.~\eqref{eqn:III}, the radial part of Eq.~\eqref{eqn:pqT23p1q1_a} reduces to the \emph{vector bracket}~\eqref{eqn:VB}:
\begin{equation}
\begin{split}
\braket{pq,(L\,l)\m{L} |T_{23}| p'q',(L'l')\m{L}'} =&
\braket{pq(L\,l)\m{L} | p'q'(L'l')\m{L}'} \\
=&   16\pi^2\dfrac{1}{p \, q \, p' \, q'} A_{I,II}(x,p/q) \, \delta(w) \, \theta{(1-x^2)} \,,
\end{split}
\label{eqn:pqT23p1q1_VB} 
\end{equation}
where $x=\cos(\widehat{\vec{p} \vec{q}})$ is computed from $p$, $q$ and $p'$ and the Dirac delta imposes that only a mesh of values of $p'$ and $q'$ is geometrically allowed changing the angle between the vectors $\vec{p}'$ and $\vec{q}'$.

When working in CS [see Eq.~\eqref{eqn:coord3b3b}] the whole derivation for the matrix elements of the transposition operator $T_{23}$ remains valid up to Eq.~\eqref{eqn:isos} but the vector bracket~\eqref{eqn:pqT23p1q1_VB} must then be associated to the following WC transformation coefficients:
\begin{equation}
\begin{pmatrix}
s_1 & t_1\\
s_2 & t_2
\end{pmatrix}=
\begin{pmatrix}
\frac{1}{2} & -1\\
-\frac{3}{4} & -\frac{1}{2}
\end{pmatrix}_{CS}\,.
\end{equation}

\subsection*{Spin/isospin contributions to the transposition operator}
The spin and isospin contributions in Eq.~\eqref{eqn:isos} have the same structure and can be obtained starting from a decoupling of the angular momenta. We report here the derivation of the isospin term as the spin case is exactly equivalent.
\begin{equation}
\label{eqn:T23tt}
\begin{split}
&\braket{[(t_\a t_\b)T_{\a\b},t_\c]\m{T}M_\m{T} | T_{23} | [(t_\d t_\e)T_{\d\e},t_\f]\m{T}M_\m{T}} \\
&\qquad \qquad = \sum_{\substack{\tau_\a\tau_\b\\M_{T_{\a\b}}\tau_\c}} \braket{t_\a \tau_\a t_\b\tau_\b | T_{\a\b}M_{T_{\a\b}}} \braket{T_{\a\b}M_{T_{\a\b}}t_\c \tau_\c | \m{T}M_\m{T}} \sum_{\substack{\tau_\d\tau_\e\\M_{T_{\d\e}}\tau_\f}} \braket{t_\d \tau_\d t_\e\tau_\e | T_{\d\e}M_{T_{\d\e}}} \braket{T_{\d\e}M_{T_{\d\e}}t_\f \tau_\f | \m{T}M_\m{T}} \\
& \qquad \qquad \qquad \qquad \qquad \times \braket{t_\a \tau_\a, t_\b \tau_\b, t_\c \tau_\c | T_{23} | t_\d \tau_\d, t_\e \tau_\e, t_\f \tau_\f} \, .
\end{split}
\end{equation}
Having decoupled all the angular momenta, the particles on the right-hand side of the latter bracket can now be exchanged
\begin{equation}
\begin{split}
\braket{t_\a \tau_\a, t_\b \tau_\b, t_\c \tau_\c | T_{23} | t_\d \tau_\d, t_\e \tau_\e, t_\f \tau_\f} &=  \braket{t_\a \tau_\a, t_\b \tau_\b, t_\c \tau_\c | t_\d \tau_\d, t_\f \tau_\f, t_\e \tau_\e}\\
&= \delta_{t_\a t_\d}\delta_{\tau_\a\tau_\d}\delta_{t_\b t_\f}\delta_{\tau_\b\tau_\f}\delta_{t_\c t_\e}\delta_{\tau_\c\tau_\e}.
\end{split}
\end{equation}
Eq.~\eqref{eqn:T23tt} becomes
\begin{equation}
\begin{split}
&\braket{[(t_\a t_\b)T_{\a\b},t_\c]\m{T}M_\m{T} | T_{23} | [(t_\d t_\e)T_{\d\e},t_\f]\m{T}M_\m{T}}\\
&\qquad= \dfrac{\delta_{t_\a t_\d} \delta_{t_\b t_\f} \delta_{t_\c t_\e}}{2 \m{T}+1}
\sum_{\substack{\tau_\a\;\tau_\b\;\tau_\c \\ M_{T_{\a\b}} \,M_{T_{\d\e}} \,M_\m{T} }} \braket{t_\a\tau_\a t_\b\tau_\b | T_{\a\b}M_{T_{\a\b}}} \braket{T_{\a\b}M_{T_{\a\b}}t_\c \tau_\c | \m{T}M_\m{T}} \, \braket{t_\a\tau_\a t_\c\tau_\c | T_{\d\e}M_{T_{\d\e}}} \braket{T_{\d\e}M_{T_{\d\e}}t_\b \tau_\b | \m{T}M_\m{T}} \\
&\qquad= \dfrac{\delta_{t_\a t_\d} \delta_{t_\b t_\f} \delta_{t_\c t_\e}}{2 \m{T}+1} \sum_{\substack{\tau_\a\;\tau_\b\;\tau_\c \\ M_{T_{\a\b}} \,M_{T_{\d\e}} \,M_\m{T} }}  \hat{\m{T}}^2\, \hat{T}_{\a\b}\, \hat{T}_{\d\e} 
 (-1)^{-t_\a+t_\b-M_{T_{\a\b}}} (-1)^{-T_{\a\b}+t_\c-M_\m{T}}
 (-1)^{-t_\a+t_\c-M_{T_{\d\e}}} (-1)^{-T_{\d\e}+t_\b-M_\m{T}} \\
&\qquad\qquad\qquad\times
\begin{pmatrix}
t_\a & t_\b & T_{\a\b}\\
\tau_\a & \tau_\b & -M_{T_{\a\b}}
\end{pmatrix}
\begin{pmatrix}
T_{\a\b}     & t_\c   & \m{T} \\
M_{T_{\a\b}} &\tau_\c & -M_\m{T}
\end{pmatrix}
\begin{pmatrix}
t_\a & t_\c & T_{\d\e}\\
\tau_\a & \tau_\c & -M_{T_{\d\e}}
\end{pmatrix}
\begin{pmatrix}
 T_{\d\e}     & t_\b    & \m{T}\\
 M_{T_{\d\e}} & \tau_\b & -M_\m{T}
\end{pmatrix}\, ,
\end{split}
\end{equation}
where we have used Eq.~\eqref{eqn:GCvs3j} and the independence of the bracket from $M_T$ due to rotational symmetry.
The final summation over $3j$ symbols can be simplified with Eq.~\eqref{eqn:6jfrom3j}:
\begin{equation}
\braket{[(t_\a t_\b)T_{\a\b},t_\c]\m{T}M_\m{T} | T_{23} | [(t_\d t_\e)T_{\d\e},t_\f]\m{T}M_\m{T}} = \delta_{t_\a t_\d} \delta_{t_\b t_\f} \delta_{t_\c t_\e}(-1)^{t_\a+t_\c+T_{\a\b} + T_{\d\e}} \hat{T}_{\a\b} \hat{T}_{\d\e}
\begin{Bmatrix}
t_\b & t_\a  &T_{\a\b}\\
t_\c & \m{T} & T_{\d\e}
\end{Bmatrix}\, .
\end{equation}
Specializing to the case of isospin $t=1/2$ one recovers Eq.~\eqref{eqn:T23_IsoSp}.

\section{Alternative version of the three-body \texorpdfstring{$T$}{T}-coefficient}
\label{Tcoeff2}
In the following, an alternative version of the three-body $T$-coefficient presented in Sec.~\ref{3b} is presented. The angular momenta couplings required to carry out this transformation are already known \cite{nogga2006} and only the radial part has been changed to allow the possibility of using a generic spherical basis. Only the isospin independent part is showed, since the transformation of the isospin-dependent state is unchanged. Step 1, 2 and 5 are the same as in Sec.~\ref{3b} but they are repeated here for clarity.
\vspace{0.2cm}
\\
\underline{STEP 1}
\vspace{-0.4cm}
\begin{equation}
\begin{split}
&\ket{ \{ [ n_\a(l_\a s_\a)j_\a, n_\b(l_\b s_\b)j_\b]J_{12}, n_\c(l_\c s_\c)j_\c \}J_{tot}M_{J_{tot}}}\\
\to
&\ket{ \{ [ n_\a n_\b (l_\a l_\b)\lambda,(s_\a s_\b)S]J_{12}, n_\c(l_\c s_\c)j_\c \}J_{tot}M_{J_{tot}}}
\end{split}
\end{equation}
The $j$ coupling of particles 1 and 2 is changed to $ls$-coupling:
\begin{equation}
\ket{[(l_\a s_\a)j_\a, (l_\b s_\b)j_\b] J_{12}} = \sum_{\lambda\,S} \hat{j}_\a\hat{j}_\b\hat{\lambda}\hat{S}
\begin{Bmatrix}
l_\a & s_\a & j_\a\\
l_\b & s_\b & j_\b\\
\lambda & S & J_{12}
\end{Bmatrix}
\ket{[(l_\a l_\b)\lambda, (s_\a s_\b)S]J_{12}}.
\end{equation}
\vspace{0.2cm}
\\
\underline{STEP 2}
\vspace{-0.4cm}
\begin{equation}
\begin{split}
&\ket{ \{ [ n_\a n_\b (l_\a l_\b)\lambda,(s_\a s_\b)S]J_{12}, n_\c(l_\c s_\c)j_\c \}J_{tot}M_{J_{tot}}}\\
\to
&\ket{ \{ [ Pp(L_{P}L)\lambda,(s_\a s_\b)S]J_{12}, n_\c(l_\c s_\c)j_\c \}J_{tot}M_{J_{tot}}}
\end{split}
\end{equation}
A change of reference system is performed for particles 1 and 2 into their relative and center-of-mass frame and at the same time the single-particle momenta of particle 1 and 2 are integrated:
\begin{equation}
\ket{n_\a n_\b (l_\a l_\b)\lambda} = \int dPdp\,P^2p^2 \sum_{L_P\,L}\braket{Pp(L_PL) \lambda | n_\a n_\b (l_\a l_\b) \lambda}^{(a)}_{\t_\a \t_\b} \ket{PL_P\, pL,\, \lambda}.
\end{equation}
The WC bracket is associated to the transformation
\begin{equation}
\begin{pmatrix}
\vec{k}_1\\
\vec{k}_2
\end{pmatrix}
=
\begin{pmatrix}
\frac{1}{2} & 1\\
\frac{1}{2} & -1
\end{pmatrix}
\begin{pmatrix}
\vec{P}\\
\vec{p}
\end{pmatrix} \,,
\end{equation}
and matrix of WC coefficients for the transformation is then
\begin{equation}
\begin{pmatrix}
s_1 & t_1\\
s_2 & t_2
\end{pmatrix}^a_{MS}
=
\begin{pmatrix}
\frac{1}{2} & 1\\
\frac{1}{2} & -1
\end{pmatrix}\,.
\end{equation}
\vspace{0.2cm}
\\
\underline{STEP 3}
\vspace{-0.4cm}
\begin{equation}
\begin{split}
&\ket{ \{ [ Pp(L_{P}L)\lambda,(s_\a s_\b)S]J_{12}, n_\c(l_\c s_\c)j_\c \}J_{tot}M_{J_{tot}}}\\
\to
&\ket{ \{ [ Pp(L_{P}L)\lambda, n_\c l_\c]L_3, [(s_\a s_\b)S s_\c]\mathcal{S} \}J_{tot}M_{J_{tot}}}
\end{split}
\end{equation}
The coupling scheme of $J_{tot}$ is changed again from $J-$ to $ls$-coupling:
\begin{equation}
\ket{[(\lambda S)J_{12},(l_\c s_\c)j_\c]J_{tot}} = \sum_{L_3 \, \mathcal{S}} \hat{J}_{12}\hat{j}_\c\hat{L}_3\hat{\mathcal{S}}
\begin{Bmatrix}
\lambda & S & J_{12}\\
l_\c & s_\c & j_\c\\
L_3 & \mathcal{S} & J_{tot}
\end{Bmatrix}
\ket{[(\lambda l_\c)L_3, (S s_\c)\mathcal{S}]J_{tot}}.
\end{equation}
\vspace{0.2cm}
\\
\underline{STEP 4}
\vspace{-0.4cm}
\begin{equation}
\begin{split}
&\ket{ \{ [ Pp(L_{P}L)\lambda, n_\c l_\c]L_3, [(s_\a s_\b)S, s_\c]\mathcal{S} \}J_{tot}M_{J_{tot}}}\\
\to
&\ket{ \{ [ P n_\c p(L_Pl_\c)\Lambda L]L_3, [(s_\a s_\b)S, s_\c]\mathcal{S} \}J_{tot}M_{J_{tot}}}
\end{split}
\end{equation}
The coupling of $L_3$ is changed through a $6j$ coefficient:
\begin{equation}
\begin{split}
\ket{[(L_PL)\lambda, l_\c]L_3} &= \sum_\Lambda \braket{[(L_Pl_\c)\Lambda, L]L_3|[(L_PL)\lambda, l_\c]L_3} \ket{[(L_Pl_\c)\Lambda, L] L_3}\\
&= \sum_\Lambda (-)^{L+l_\c+\lambda+\Lambda}\hat{\lambda}\hat{\Lambda}
\begin{Bmatrix}
L & L_P & \lambda\\
l_\c & L_3 & \Lambda
\end{Bmatrix}
\ket{[(L_Pl_\c)\Lambda, L] L_3}.
\end{split}
\end{equation}
\vspace{0.2cm}
\\
\underline{STEP 5}
\vspace{-0.4cm}
\begin{equation}
\begin{split}
&\ket{ \{ [ P n_\c p (L_P l_\c)\Lambda, L]L_3, [(s_\a s_\b)S, s_\c]\mathcal{S} \}J_{tot}M_{J_{tot}}}\\
\to
&\ket{ \{ [ Q_{cm}q(l_{cm}l)\Lambda, pL]L_3, [(s_\a s_\b)S, s_\c]\mathcal{S} \}J_{tot}M_{J_{tot}}}
\end{split}
\end{equation}
A second change of reference system is performed, transforming the coordinates of the relative and center-of-mass of the first two particles (1 and 2) and particle 3 in the total center-of-mass and the relative coordinate between the center-of-mass of particles 1 and 2 and the third particle
\begin{equation}
\begin{dcases}
\vec{q} = \dfrac{2}{3}\biggl[ \vec{k}_3 - \dfrac{1}{2}(\vec{k}_1 + \vec{k}_2) \biggr] = \dfrac{2}{3}\biggl[ \vec{k}_3 - \dfrac{1}{2}\vec{P} \biggr]\\
\vec{Q}_{cm} = \vec{k}_1 + \vec{k}_2 + \vec{k}_3 = \vec{P} + \vec{k}_3
\end{dcases}
\end{equation}
\begin{equation}
\ket{Pn_\c(L_p l_\c)\Lambda} = \int dQ_{cm}dq\,Q_{cm}^2q^2\sum_{l_{cm}\,l} \braket{Q_{cm}q(l_{cm}\,l)\Lambda|Pn_\c(L_p l_\c)\Lambda}^{(b)}_{\t_\c} \ket{Q_{cm}q(l_{cm}\,l)\Lambda}.
\end{equation}
The mixed WC bracket (Eq.~\eqref{eqn:WCmix}) represents the transformation
\begin{equation}
\begin{pmatrix}
\vec{P}\\
\vec{k}_3
\end{pmatrix}
=
\begin{pmatrix}
\frac{2}{3} & -1\\
\frac{1}{3} & 1
\end{pmatrix}
\begin{pmatrix}
\vec{Q}_{cm}\\
\vec{q}
\end{pmatrix}.
\end{equation}
The matrix of WC coefficients associated to the transformation is then
\begin{equation}
\begin{pmatrix}
s_1 & t_1\\
s_2 & t_2
\end{pmatrix}^b_{MS}
=
\begin{pmatrix}
\frac{2}{3} & -1\\
\frac{1}{3} & 1
\end{pmatrix}.
\end{equation}
\vspace{0.2cm}
\\
\underline{STEP 6}
\vspace{-0.4cm}
\begin{equation}
\begin{split}
&\ket{ \{ [ Q_{cm}q(l_{cm}l)\Lambda, pL]L_3, [(s_\a s_\b)S, s_\c]\mathcal{S} \}J_{tot}M_{J_{tot}}}\\
%\to & \ket{ \{ [ Q_{cm}\,q\,p\,l_{cm},(lL)\mathcal{L}]L_3, [(s_\a s_\b)S, s_\c]\mathcal{S} \}J_{tot}M_{J_{tot}}}\\
\to
&\ket{ \{ [ Q_{cm}\,p\,q\,l_{cm},(L\,l)\mathcal{L}]L_3, [(s_\a s_\b)S, s_\c]\mathcal{S} \}J_{tot}M_{J_{tot}}}
\end{split}
\end{equation}
The angular momenta $L$ and $l$ are recoupled:
\begin{equation}
\begin{split}
\ket{[(l_{cm}\,l)\Lambda, L]L_3} &= \sum_{\mathcal{L}} \braket{[l_{cm},(lL)\mathcal{L}]L_3 | [(l_{cm}\,l)\Lambda, L]L_3} \ket{[l_{cm},(lL)\mathcal{L}]L_3}\\
&= \sum_{\mathcal{L}} (-)^{l_{cm}+L_3+\mathcal{L}}\hat\Lambda \hat{\mathcal{L}}
\begin{Bmatrix}
l_{cm} & l & \Lambda\\
L & L_3 & \mathcal{L}
\end{Bmatrix}
\ket{[l_{cm},(L\,l)\mathcal{L}]L_3}\, .
\end{split}
\end{equation}
\vspace{0.2cm}
\\
\underline{STEP 7}
\vspace{-0.4cm}
\begin{equation}
\begin{split}
&\ket{ \{ [ Q_{cm}\,p\,q\,l_{cm},(L\,l)\mathcal{L}]L_3, [(s_\a s_\b)S, s_\c]\mathcal{S} \}J_{tot}M_{J_{tot}}}\\
\to
&\ket{ \{ Q_{cm}l_{cm},\,[p\,q(L\,l)\mathcal{L}, (S s_\c)\mathcal{S}]\mathcal{J}  \}J_{tot}M_{J_{tot}}}
\end{split}
\end{equation}
The orbital angular momenta scheme is changed again, coupling $\mathcal{L}$ and $\mathcal{S}$ in $\mathcal{J}$:
\begin{equation}
\begin{split}
\ket{[(l_{cm}\mathcal{L})L_3, \mathcal{S}]J_{tot}} &= \sum_{\mathcal{J}} \braket{[l_{cm},(\mathcal{L}\mathcal{S})\mathcal{J}]J_{tot} | [(l_{cm}\mathcal{L})L_3,\mathcal{S}]J_{tot}} \ket{[l_{cm},(\mathcal{L}\mathcal{S})\mathcal{J}]J_{tot}}\\
&= \sum_{\mathcal{J}} (-)^{{l_{cm}}+\mathcal{L}+\mathcal{S}+J_{tot}} \hat{L}_3\hat{\mathcal{J}}
\begin{Bmatrix}
l_{cm} & \mathcal{L} & L_3\\
\mathcal{S} & J_{tot} & \mathcal{J}
\end{Bmatrix}
\ket{[l_{cm},(\mathcal{L}\mathcal{S})\mathcal{J}]J_{tot}}.
\end{split}
\end{equation}
\vspace{0.2cm}
\\
\underline{STEP 8}
\vspace{-0.4cm}
\begin{equation}
\begin{split}
&\ket{ \{ Q_{cm}l_{cm},\,[p\,q(L\,l)\mathcal{L},  (S s_\c)\mathcal{S}]\mathcal{J}  \}J_{tot}M_{J_{tot}}}\\
\to
&\ket{ \{ Q_{cm}l_{cm},\,[p\,q(LS)J, (ls_\c)j]\mathcal{J}  \}J_{tot}M_{J_{tot}}}
\end{split}
\end{equation}
Another change of couplings in the internal structure of $\mathcal{J}$ is performed:
\begin{equation}
\ket{[(L\,l)\mathcal{L},(Ss_\c)\mathcal{S}]\mathcal{J}} = \sum_{Jj} \hat{\mathcal{L}}\hat{\mathcal{S}}\hat{J}\hat{j}
\begin{Bmatrix}
L & l & \mathcal{L}\\
S & s_\c & \mathcal{S}\\
J & j & \mathcal{J}
\end{Bmatrix}
\ket{[(LS)J,(ls_\c)j]\mathcal{J}}.
\end{equation}
The complete change of basis is
\begin{equation}
\begin{split}
\ket{[(ab)J_{12}c]J_{tot}} =& \int dP\,dp\,dQ_{cm}\,dq\,\sum_{T\,\mathcal{T}} \sum_{\lambda\,S} \sum_{L_P\,L} \sum_{L_3\,\mathcal{S}} \sum_{\Lambda} \sum_{l_{cm}\,l}\sum_{\mathcal{L}} \sum_{\mathcal{J}} \sum_{J\,j}\\
&\qquad\times \braket{t_\a \tau_\a t_\b \tau_\b | TM_T} \braket{T M_T t_\c \tau_\c | \mathcal{T}M_\mathcal{T}}\\
&\qquad\times \hat{j}_\a \hat{j}_\b \hat{\lambda} \hat{S}
\begin{Bmatrix}
l_\a & s_\a & j_\a\\
l_\b & s_\b & j_\b\\
\lambda & S & J_{12}
\end{Bmatrix}
 P^2\, p^2\, \braket{Pp(L_PL)\lambda | n_\a n_\b (l_\a l_\b) \lambda}^{(a)}_{\t_\a \t_\b}\\
&\qquad\times \hat{J}_{12} \hat{j}_\c \hat{L}_3 \hat{\mathcal{S}}
\begin{Bmatrix}
\lambda & S & J_{12}\\
l_\c & s_\c & j_\c\\
L_3 & \mathcal{S} & \mathcal{J}
\end{Bmatrix}
 (-)^{L+l_\c+\lambda+\Lambda} \hat{\lambda} \hat{\Lambda}
\begin{Bmatrix}
L & L_P & \lambda\\
l_\c & L_3 & \Lambda
\end{Bmatrix}\\
&\qquad\times Q_{cm}^2\,q^2\,\braket{Q_{cm}q(l_{cm} l) \Lambda | P n_\c (L_P l_\c) \Lambda}^{(b)}_{\t_\c}\\
&\qquad\times (-)^{l_{cm}+L_3+\mathcal{L}} \hat{\Lambda}\hat{\mathcal{L}}
\begin{Bmatrix}
l_{cm} & l & \Lambda \\
L & L_3 & \mathcal{L}
\end{Bmatrix} \\
&\qquad\times (-)^{l_{cm}+\mathcal{L}+\mathcal{S}+J_{tot}}
\hat{L}_3 \hat{\mathcal{J}}
\begin{Bmatrix}
l_{cm} & \mathcal{L} & L_3\\
\mathcal{S} & J_{tot} & \mathcal{J}
\end{Bmatrix}  \\
&\qquad\times \hat{\mathcal{L}}\hat{\mathcal{S}}\hat{J}\hat{j}
\begin{Bmatrix}
L & l & \mathcal{L}\\
S & s_\c & \mathcal{S}\\
J & j & \mathcal{J}
\end{Bmatrix}
\ket{[Q_{cm}l_{cm}, pq\alpha]J_{tot}}.
\end{split}
\end{equation}
Eventually the three-body $T$-coefficient reads
\begin{equation}
\begin{split}
T^{3B} &= \braket{[Q_{cm}l_{cm}, pq\alpha]J_{tot} | [(ab)J_{12}c]J_{tot}}\\
&= \int dP P^2\, \sum_{\lambda\, L_P}\sum_{L_3\, \mathcal{S}} \sum_{\Lambda\, \mathcal{L}}
(-)^{l_\c+\lambda+\Lambda+L+\mathcal{S}+J_{tot}-L_3}
 \hat{j}_\a \hat{j}_\b \hat{\lambda} \hat{S}
\begin{Bmatrix}
l_\a & s_\a & j_\a\\
l_\b & s_\b & j_\b\\
\lambda & S & J_{12}
\end{Bmatrix}\\
&\qquad\times\braket{Pp(L_PL)\lambda | n_\a n_\b (l_\a l_\b) \lambda}^{(a)}_{\t_\a \t_\b}
 \hat{J}_{12} \hat{j}_\c \hat{L}_3 \hat{\mathcal{S}}
\begin{Bmatrix}
\lambda & S & J_{12}\\
l_\c & s_\c & j_\c\\
L_3 & \mathcal{S} & J_{tot}
\end{Bmatrix}
\hat{\lambda} \hat{\Lambda}
\begin{Bmatrix}
L & L_P & \lambda\\
l_\c & L_3 & \Lambda
\end{Bmatrix}\\
&\qquad\times\braket{Q_{cm}q(l_{cm}l) \Lambda | P n_\c (L_Pl_\c) \Lambda}^{(b)}_{\t_\c}
 \hat{\Lambda}\hat{\mathcal{L}}
\begin{Bmatrix}
l_{cm} & l & \Lambda\\
L & L_3 & \mathcal{L}
\end{Bmatrix}
\hat{L}_3\hat{\mathcal{J}}
\begin{Bmatrix}
l_{cm} & \mathcal{L} & L_3\\
\mathcal{S} & J_{tot} & \mathcal{J}
\end{Bmatrix}\\
&\qquad\times\hat{\mathcal{L}}\hat{\mathcal{S}}\hat{J}\hat{j}
\begin{Bmatrix}
L & l & \mathcal{L}\\
S & s_\c & \mathcal{S}\\
J & j & \mathcal{J}
\end{Bmatrix}
 \braket{t_\a \tau_\a t_\b \tau_\b | TM_T} \braket{T M_T t_\c \tau_\c | \mathcal{T}M_\mathcal{T}}.
\end{split}
\label{eqn:Tcoeff2}
\end{equation}

\bibliography{bibliography.bib}

\end{document}